北京航空航天大學

# 硕士学位论文

# SSS-1P姿态确定与控制子系统的软件设计与开发

作者姓名　侯赛因

学科专业　空间技术应用

指导教师　孙亮

培养院系　国际学院

# Software Design and Development of Attitude Determination and Control Subsystem for SSS-1P

A Dissertation Submitted for the Degree of Master

**Candidate: Amir Hossein Alikhah Mishamandani**

**Supervisor: Dr. Sun Liang**

School of Astronautics

Beihang University, Beijing, China

**中图分类号：V475. 1**

**论文编号：10006LS1725241**

# 硕 士 学 位 论 文

# SSS-1P 姿态确定与控制子系统的软件设计与开发

| | | | |
|---|---|---|---|
| 作者姓名 | 侯赛因 | 申请学位级别 | 工程硕士 |
| 指导教师姓名 | 孙亮 | 职称 | 讲师 |
| 学科专业 | 空间技术应用 - 微卫星应用 | 研究方向 | 姿态确定和控制系统 |
| 学习时间自 | 2017 年 9 月 12 日 | 起至 | 2019 年 3 月 27 日止 |
| 论文提交日期 | 2019 年 5 月 15 日 | 论文答辩日期 | 2019 年 5 月 27 日 |
| 学位授予单位 | 北京航空航天大学 | 学位授予日期 | 年 月 日 |

# 关于学位论文的独创性声明

本人郑重声明：所呈交的论文是本人在指导教师指导下独立进行研究工作所取得的成果，论文中有关资料和数据是实事求是的。尽我所知，除文中已经加以标注和致谢外，本论文不包含其他人已经发表或撰写的研究成果，也不包含本人或他人为获得北京航空航天大学或其它教育机构的学位或学历证书而使用过的材料。与我一同工作的同志对研究所做的任何贡献均已在论文中作出了明确的说明。

若有不实之处，本人愿意承担相关法律责任。

学位论文作者签名：__________　　　日期：　　年　月　日

# 学位论文使用授权书

本人完全同意北京航空航天大学有权使用本学位论文（包括但不限于其印刷版和电子版），使用方式包括但不限于：保留学位论文，按规定向国家有关部门（机构）送交学位论文，以学术交流为目的赠送和交换学位论文，允许学位论文被查阅、借阅和复印，将学位论文的全部或部分内容编入有关数据库进行检索，采用影印、缩印或其他复制手段保存学位论文。

保密学位论文在解密后的使用授权同上。

学位论文作者签名：______　　　日期：　　年　月　日

指导教师签名：________　　　日期：　　年　月　日

# 摘要

亚太空间合作组织大学小卫星项目是由亚太空间合作组织（APSCO）发起，由APSCO各成员国大学联合研制的三颗小卫星，其中包括1颗微小卫星（SSS-1）和2颗纳星（SSS-2A 和 SSS-2B）,并共同完成在轨技术验证、星间通信、空间科学探测及空间遥感等应用任务。SSS-1P 是 SSS-1 的先导星，用于 SSS-1 微小卫星的全功能飞行验证。本论文针对 SSS-1P 的姿态确定与控制分系统，进行控制系统软件的设计与开发，具有重要的实际工程意义。

首先，依据任务需求和技术指标，确定控制策略和硬件配置；设计合理的姿态确定及姿态控制算法；建立姿控系统数学模型，并通过数值仿真，分析扰动影响，验证算法的可行性。

然后，根据姿控分系统的设计方案，确定姿控软件的功能、性能、接口、数据和环境等需求，以确保软件的完整性、明确性、一致性和可验证性。在 C 语言环境下，对姿态确定与控制分系统的软件进行模块化设计与开发。软件调用层次简单，具有可扩展性；接口实现标准化设计，具有可维护性。通过与MATLAB动力学仿真结果进行对比，验证软件设计（SIL: software in loop）的正确性。

最后，基于 MATLAB 搭建运动模拟器（仿真计算机），实现敏感器、执行机构和卫星动力学数学模型等功能；将设计的软件移植至星上计算机，实现控制器功能；完成控制器闭路试验（PIL: processer in loop），通过对比，验证软件设计的正确性。

# Abstract

The SSS-1P Test micro-satellite is a student satellite program at Beijing University of Aeronautics and Astronautics (BUAA) that is part of the Student Small Satellite (SSS) Program run by the Asia-Pacific Space Cooperation Organization (APSCO). The Attitude Determination and Control System (ADCS) is one of the critical boards of any satellite, specially the micro-satellites. The ADCS is the bridge linking sensors data to actuators by several computationally complex algorithms such as Extended Kalman Filter (EKF), Detumbling and Tumbling controller and so on. In this thesis, I will focus on the implementation of each of the required algorithms and then make a whole integrated package. This thesis focuses on developing the first version of the on-board computer software for ADCS of SSS-1P and required preliminary tests to assure the minimum performance of the provided package with Software-In-the-Loop (SIL) and Processor-In-the-Loop (PIL) tests.

In this thesis, I will introduce the fundamental dynamic and kinematic non-linear equations. The linearization for a specific prerational condition and the assumptions to achieve the linearization will be introduced. The model will be linearized for a specific operating condition for the design and development of the controllers. The disturbances including atmospheric drag, solar radiation and so on are also introduced with regarding to the operating Altitude (500 ~ 600 km). The brief introduction of different magnetic field models and their differences will be introduced as well. In addition, the SGP4 predictive model to estimate the position and the velocity of the satellite discussed.

In addition, the thesis describes the software development phase, some additional algorithms that I implemented but they are not a part of mathematical models, such as system manager, time manager and so on with pseudo code. The architecture of the software is also described in this section.

Finally, the numerical simulations are analysed and the simple compression between tests results from PIL, SIL, and MATLAB simulation are presented. This section will presents the validation and the trustworthy of the final package.

# Contents

# List of Figures

# List of Tables

# 1. Introduction

## 1.1 Student Small Satellite (SSS-1P)

The SSS-1 Satellite is a student satellite program at Beijing University of Aeronautics and Astronautics (BUAA) that is part of the Student Small Satellite (SSS) Program run by the Asia-Pacific Space Cooperation Organization (APSCO). The SSS project includes three satellites: one primary micro-satellite named SSS-1, two companion nano-satellites named SSS-2A and SSS-2B. The SSS-1 is a 30kg nano-satellite with dimension of 320mm×320mm×650mm, whereas the SSS-2A and SSS-2B are 3.0kg nano-satellite with 3U CubeSat structure design. The three satellites construct a constellation as shown in Figure 1-1.

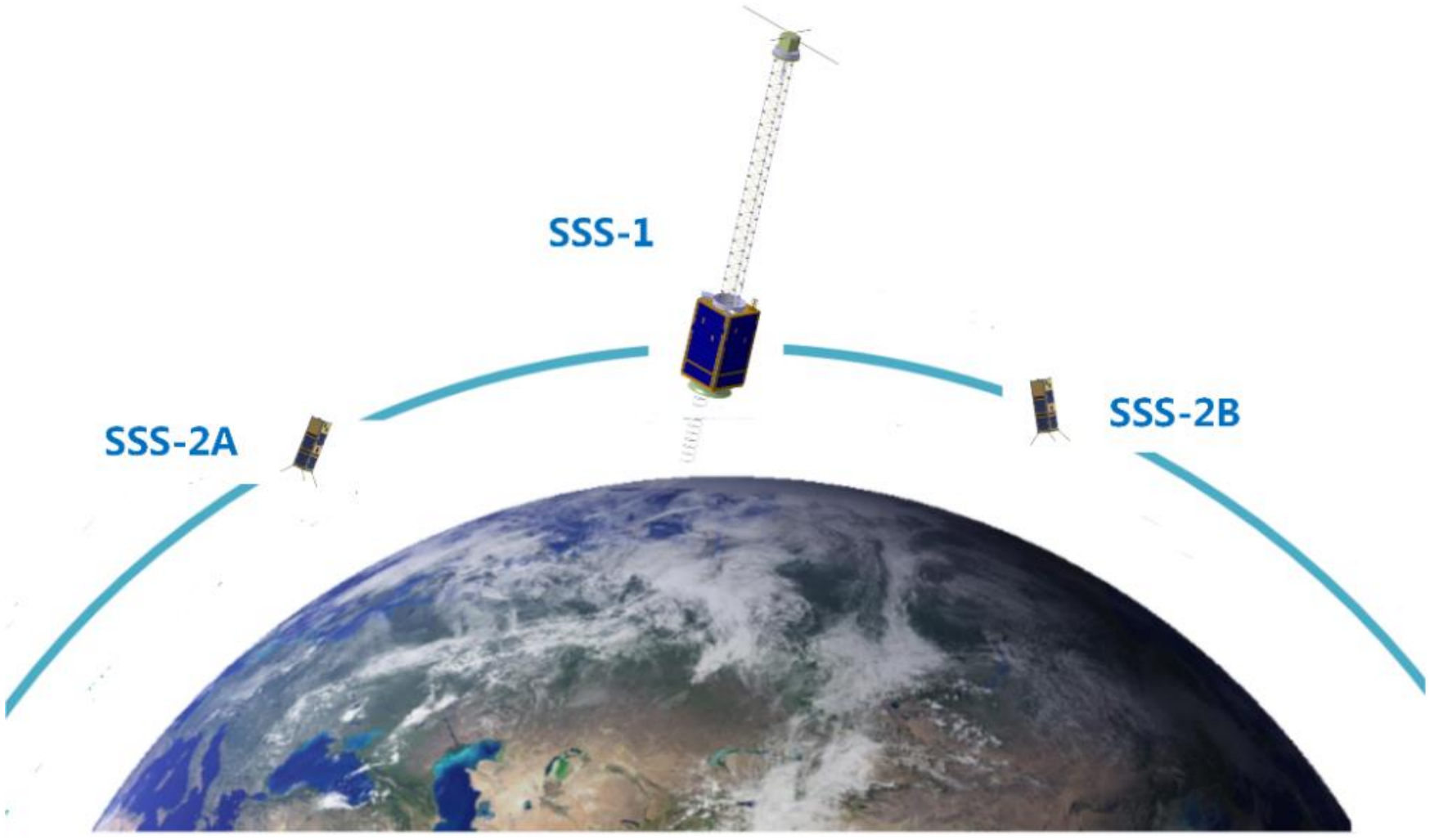

Figure 1-1 The SSS platform

The SSS-1P Test Satellite is a student satellite program at Beijing University of Aeronautics and Astronautics (BUAA) that is supposed to be lunched prior to SSS-1 satellite to verify the reliability of the designed hardware, software and mission. The technical specifications of the ADCS in active mode are listed as follows [1]:

- Three axes stabilized
- Attitude determination accuracy
- Angular error: 0.5° (3σ)
- Angular velocity error: 0.001°/s(3σ)
- Attitude Control Accuracy
- Angular error: 1° (3σ)
- Angular velocity error: 0.01°/s (3σ)

The SSS-1 micro-satellite platform is a rectangular box with the size of no more than 350mm×350 mm×650mm and the mass of less than 30kg. SSS-1 is a micro-satellite with the configuration of a main-sat.

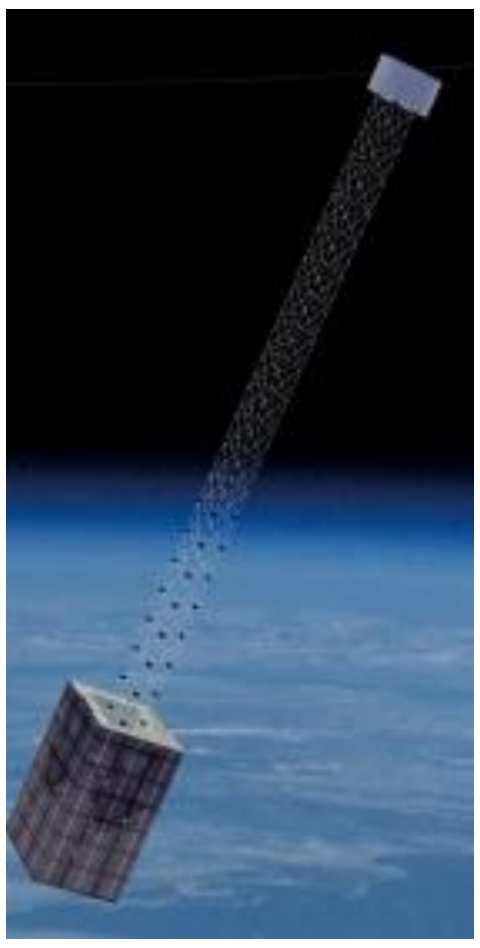

Figure 1-4 SSS-1 platform

The components of each subsystem are designed to be assembled on different panel. The satellite equipment layout is shown as below.

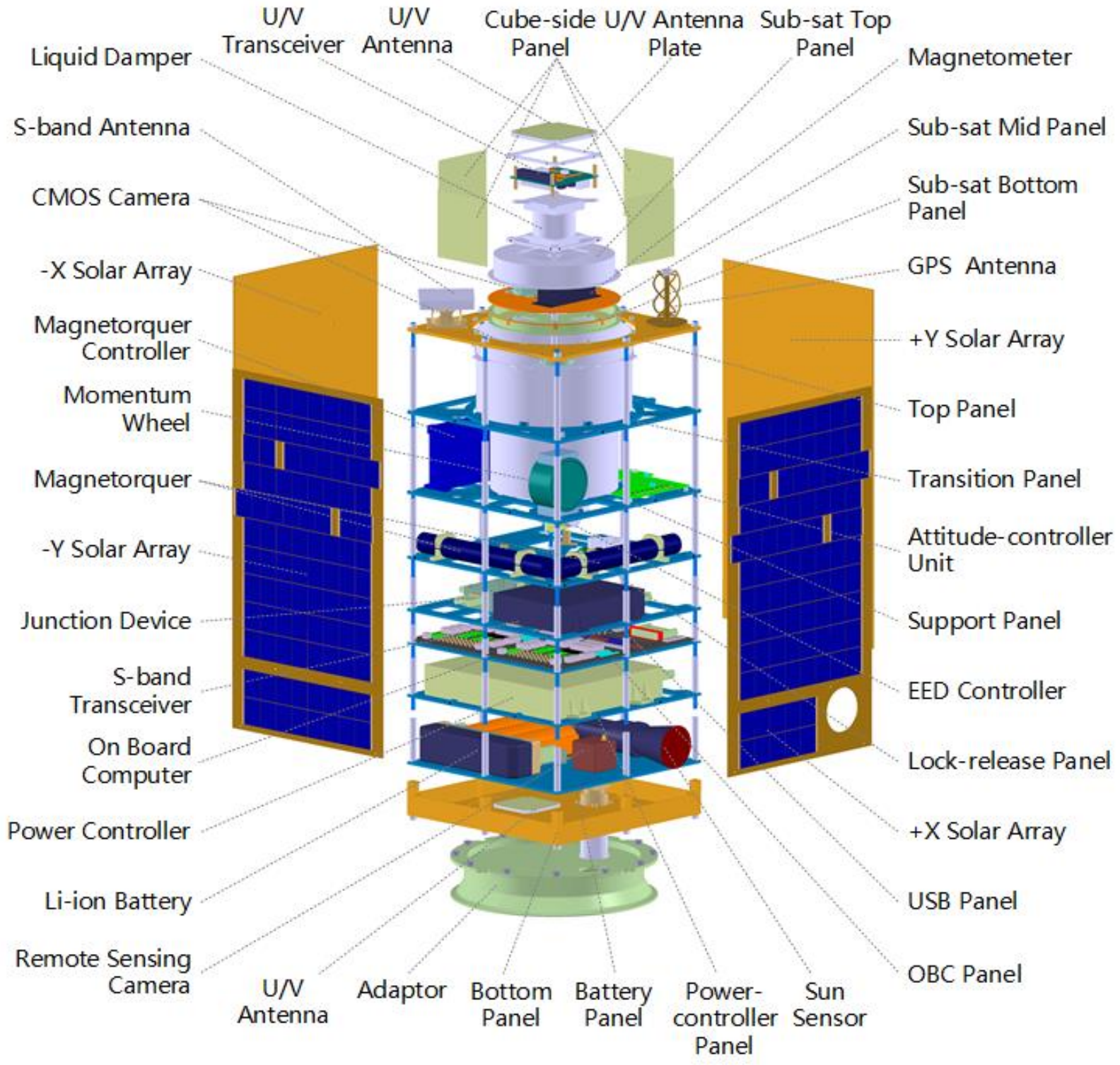

Figure 1-5 Equipment layout

ADCS is required to provide the in-orbit attitude control and determination functions. The attitude control function includes the stabilization and detumbling control of the satellite to the prerequisite value, while the attitude determination function is to facilitate the estimated attitude information. ADCS also operates and records the pertinent data after the separation of satellite from the launcher. The ADCS requirements that allocated by system engineering based on the mission of the SSS-1P satellite are indicated in the table supplied.

Table 1-1 The ADCS Requirements

| Item | Statement |
|---|---|
| SSS1-ACS-0010 | During nominal operations the satellite shall be three-axis stabilized. |
| SSS1-ACS-0020 | The nominal pointing attitude of the satellite shall be Nadir pointing. |
| SSS1-ACS-0030 | The accuracy of the on-board and on-ground estimations of attitude, angular rate and orbital position shall be derived from the payload requirements and from the satellite operation requirements. |
| SSS1-ACS-0040 | ACS shall provide pointing control accuracy of better than 1 degree in optional testing mode, and better than 5 degrees in normal mode; Pointing stability better than 0.01degree/sec in optional testing mode, and better than 0.1degree/sec in normal mode. |
| SSS1-ACS-0050 | The ACS shall provide the necessary data to OBDH to define the orbital position and the attitude information at all times |
| SSS1-ACS-0060 | The ACS shall provide sufficient information to allow the ground segment to reconstruct the attitude and to determine attitude control loop characteristics |
| SSS1-ACS-0070 | The ACS shall provide sufficient data to monitor its configuration, health and operation through telemetry. |

The requirements on ACS clearly demand the three axes stabilized system. Three different kinds of sensors are used to observe the motion of the satellite and two different kinds of actuators are used to control its motion. The sensor and actuator types of the SSS-1 ACS have already been mentioned in make list document. The detailed list is given here below:

- 1 IMU, ADIS16467-1 (outputs gyroscope, accelerometer, and temperature)
- 2 sun sensors, SSOCD60
- 2 magnetometers (MGM) HMR2300R— measuring all 3 axes—
- electronics unit (ACU)
- One GPS receiver, piNAV-NG
- 3 magneto torquers (MGT) with and a common MGT electronic unit
- 1 reaction wheels (RW) in y axis with a common RW assembly electronic unit

When the satellite is in a course pointing mode, the MGT electronics is powered and prepared desire accuracy. Proper torques for precise fine pointing are applied by means of using the Reaction Wheels (RW). If the reaction wheels run into saturation, the MGTs are used to desaturate the RWs.

Figure 1.4 indicates the relations between the control unit, attitude determination unit, and different ADCS sensors and actuators.

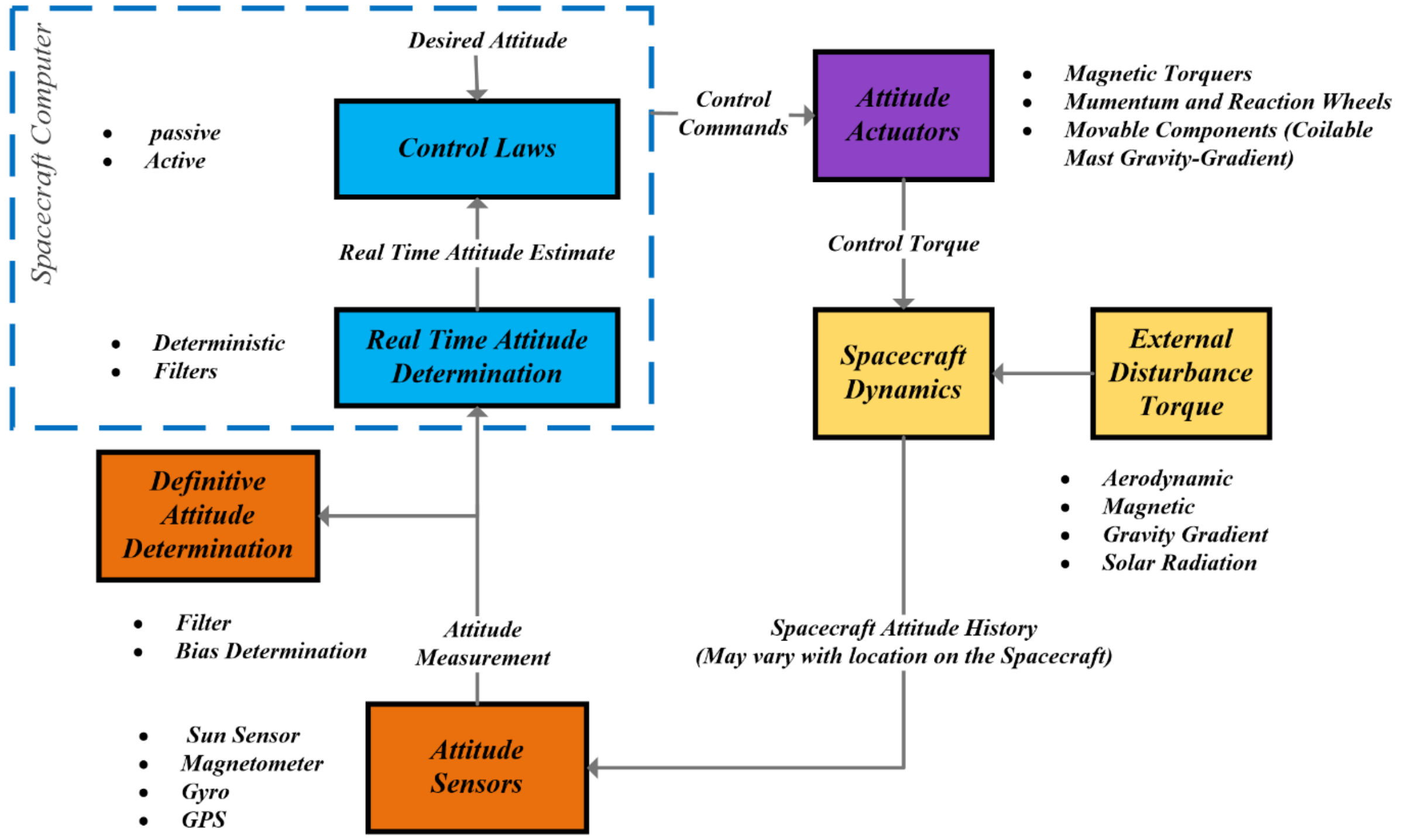

Figure 1-6 Schematic of the SSS-1P's ADCS configuration.

In order to design the ADCS the total part of this schematic should be carried out completely. It's clear that the various parts of this configuration have remarkable impact on the rest of the system. The main part of this systems can be classified as dynamics and kinematic of satellite, determination of external torques, accurate modeling of sensors and actuators, control modes and attitude determination and control (ADCS) unit.

## Hardware Specification

During the design process there were several hardware available. After a wide discussion with experts and analyzing the mission the hardware selected and their specification explained in this chapter.

### The Sun Sensor

The sun sensor selected for this project is SSOC-A60. The maximum precision of this sensor is $0.1°$ which is totally suitable for our mission as based on the project specification the sun sensor should have $0.5°$ precision which it means the requirements also passed as well. The main characteristic of this sensor presented in the table below.

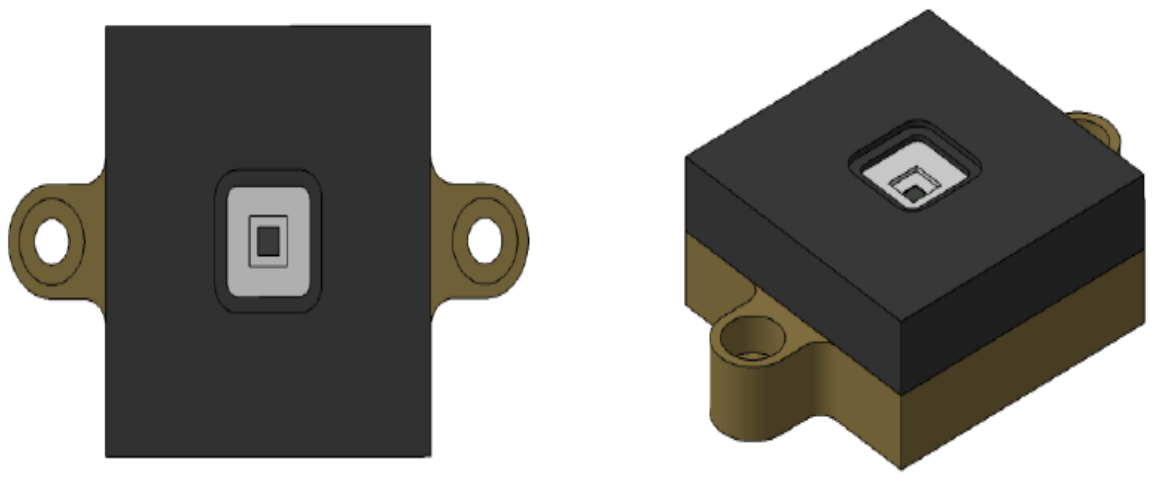

*Figure 1-7 The Sun Sensor SSOC-A60.*

*Table 1-2 SSOC-A60 Specification.*

| | |
|---|---|
| Field of View (FOV) | ±60° |
| Accuracy in FOV | 0.3° (highest 0.05°) |
| Supply voltage | 5V |
| Average consumption | 36mW |
| Communication interface | AD |
| Weight | 25g |
| Dimensions | 30mm*30mm*12mm |
| Operating temperature | -40°C~+85°C |

**The magnetometer**

The selected magnetometer is Honeywell's HMR2300 triaxial magnetometer. Attitude determination requires an accuracy of up to 0.5°. According to the accuracy of 0.5°, the magnetic field measurement accuracy of the magnetometer is required to be better than 500 nT which based on specification this magnetometer accuracy is 120 nT so it passes the minimum requirements.

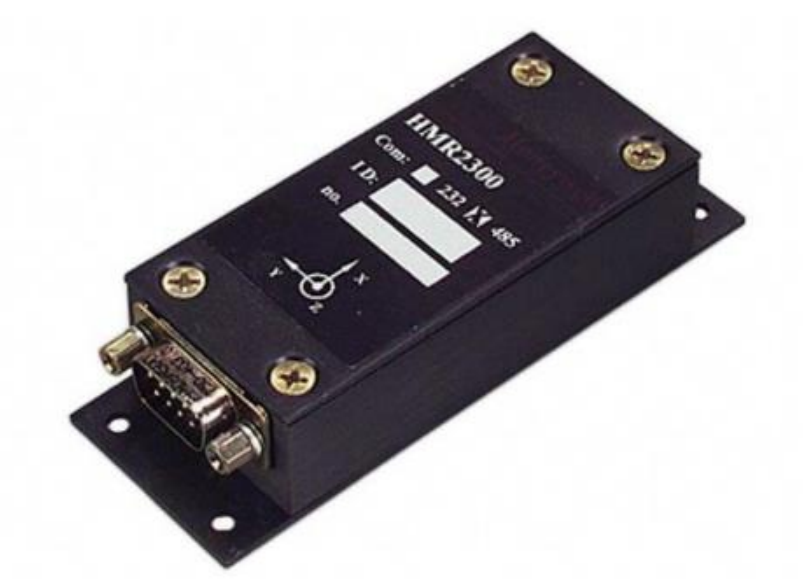

*Figure 1-8 Honeywell's HMR2300 triaxial magnetomete.*

### The Gyroscope

The selected device as the gyroscope for the project is ADIS 16445. Inertial sensors have high measurement accuracy. In general, gyro drift is in the range of 0.003°/h~1°/h. This range meets the project requirements as well.

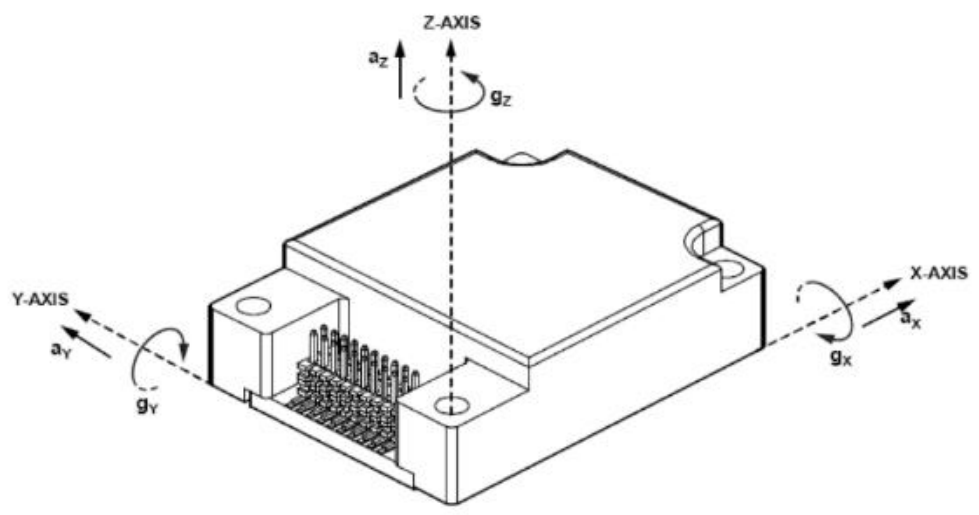

*Figure 1-9 ADIS 16445 Specification.*

*Table 1-4 ADIS 16445 Specification.*

| | |
|---|---|
| Range | $\pm 250°/s$ |
| Sensitivity | $0.0025°/s$ |
| Noise characteristics | $0.22°/s$ |
| Operating voltage | 3.3V |
| Power consumption | 0.25W |
| Communication Interface | SPI |
| Weight | 15g |
| Dimensions | 38mm*24.5mm*11mm |
| Operating temperature | -40℃~+85℃ |

### The Magnetorquer

Torque can be produced to change the attitude of a satellite by generating a magnetic moment. This moment can be produced by a current flowing in a coil wire. Pulse Width Modulation (PWM) is a well-tested method of controlling the current in an inductive circuit. The coil can either have an air core, or a ferrite core that is made from a special type of magnetic alloy of low permanence and good flux linearity. The magnetic torque device selected in this scheme is a customized product composed of three magnetic rods and an associated magnetic torque controller.

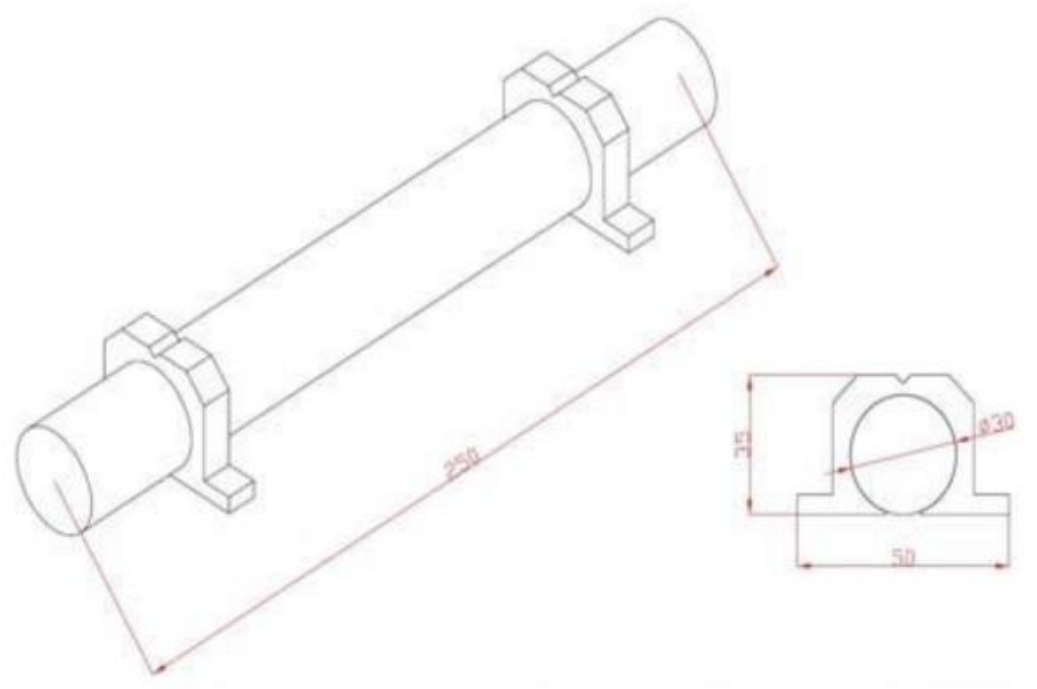

*Figure 1-10 Magnetic bar custom product outline diagram.*

*Table 1-5 The Magnetorquer Specification.*

| | |
|---|---|
| Max. magnetic moment | ≥12A·$m2@9V$ |
| Max. current | 100mA |
| Operating voltage | 9~12.5V |
| Max. power consumption | <5W |
| Communication interface | RS422@9600bps |
| Total mass | <2.8kg |
| Magnet size | 250mm*35mm*50mm |
| Controller size | 220mm*94mm*30mm |
| Operating temperature | -20℃~+65℃ |

**The Reaction Wheels**

Simply, they are wheels that are driven by motors with the purpose of introducing momentum into the satellite system. Because momentum must be conserved in a closed system, the addition of this momentum causes the satellite to rotate. Thus, applying torque to the system via reaction wheels is a means of rotating the satellite along the axis of rotation of the wheel. With three wheels mounted orthogonally, a rotation about any arbitrary axis is achievable with the correct combination of torque to the three wheels. The momentum wheel can provide offset angular momentum in the direction of the satellite's pitch axis, which acts as a fixed axis. In addition, it can provide angular momentum exchange on the pitch axis to speed up the attitude control of the axis. The choice of offset momentum size mainly considers the influence of satellite nutation and precession. According to the demand analysis of the angular momentum, the maximum angular momentum of the momentum wheel should not be less than "0.052N·m·s". In addition, the maximum output torque, constant/peak power consumption and other indicators need to be considered. According to the results of previous investigations, the V70 momentum wheel provided by the China's National University of Science and Technology was selected for this program.

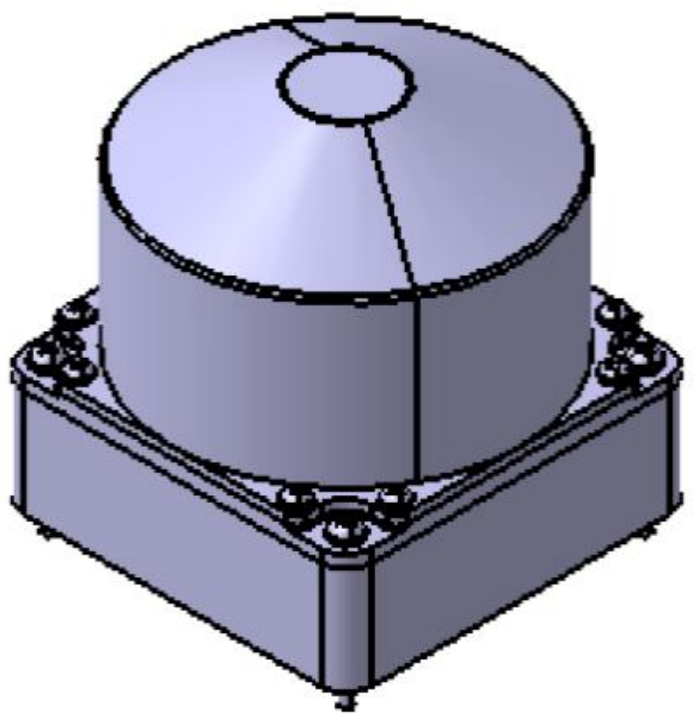

*Figure 1-11 The V70 Momentum Wheel.*

*Table 1-6 The Magnetorquer Specification.*

| | |
|---|---|
| Range of rotation | -6000rpm~6000rpm |
| Moment of inertia | $1.067\times10-4kg\cdot m2$ |
| Max. output torque | 0.01N·m |
| Max. angular momentum | 0.067N·m·s |
| Constant power consumption | <0.8W@0rpm<br><1.5W@3000rpm<br><2.4W@6000rpm |
| Peak power consumption | 6W |
| Acceleration performance | 45s |
| Speed accuracy | 0.5rpm |
| Operating voltage | 10~14V |
| Communication interface | CAN@500kbps |
| Weight | 0.45kg |
| Dimensions | 70mm*70mm*73mm |
| Operating temperature | -20℃~+50℃ |

However the design of the ADCS parts and the scheme above are changing a basic configuration is considered and the development of the software started and the next sub-section will explain the research direction of this thesis.

## 1.2 Research Direction

This thesis focuses on developing the mathematical model of the satellite. The outcome of the mathematical modeling are non-linear equations. The non-linear equations are difficult to underestand, analyze and the developed simultion model is of the high complexity.

Furthermore, the design of controllers and filters will be very complicating. Therefore,

we will simplify the equations by linearizing them and then we proceed to design the filters and control algorithms for them.

The next step was developing the MATLAB and Simulink model. The designed filters and control algorithms are also implemented in the simulation model. The response of the developed model must meet the minimum stability criteria and mission requirements.

The disturbances and the orbital equations and the dynamic model is developed and the closed loop system is provided in matlab. The developed model, tested successfully and the task went to the next phase.

The algorithm design and development for the on-board computer code had multiple challenges such as some Simulink toolboxes are not applicable while developing the C code such as laplas integrals. Even more, some matlab functions are not easy to implement in C language such as algorithms to calculate the matrices operations such as determinant, inverse, Singular Value Decomposition, four equations and three unknown and so on. Several numerical approximations and numerical methods must be implemented in the design process. For instance, the four unknown and three unknown uses defines objects in matlab and this is not convertible by matlab to C code. Therefore, to tackle this issue least square method used, however, the nature of four equations and three unknown is in a way that it might not have a solution at all that should be considered in the future versions and the failure analysis of the ADCS package.

The ADCS software has also some requirements such as compatibility with linux and limitations set by the processing power of selected processor of on-board computer or the maximum variable size and the storage. In addition, the package must be created in C language.

The developed C based project includes three types of packages. The first type of packages are the ones could be used for free and under MIT or public GNU/GNU2 licence from github. The second types of the packages are the one created by myself based on the developed algortihm based on the mathematical equations and numerical methods. The last type of the paclages are the one created in MATLAB as a function and then converted to the C code using MEX toolbox of MATLAB.

Each package is tested seperately with several methods such as generating data with random distribution or available data sets on the internet. The paclages all verified to work with no problem. The results of each package such as (World Magnetic Model, IGRF and so on) in case of availability was tested with the same matlab package.

After completing the whole integrated package the results were compared with the matlab model and verified the cases. To achieve this, several extra packages such as dynamical model and orbital model developed to be able to implement a SIL test.

The last section of the thesis is also dedicated to PIL test and the results obtained from the tests. Finally, the simple comparission between outcomes from different packages in different simulation and tests.

## 1.3 Literature Review

A satellite project is a complex system, which involves different disciplines of science and engineering, such as space physics, material science, communications, mechanical, electrical and computer engineering. Due to the advance of information technology, Micro-Electromechanical Systems (MEMS), and nanotechnology, modern experimental satellite can be reduced to about 10-100 kilogram (a micro-satellite) and below to perform certain tasks a conventional large satellite used to achieve. Micro-satellite development practically integrate and transform the theory and knowledge of various science into valuable technology product such that it is considered as one of the most cost-effective project among other engineering projects. Development of microsatellites in the newly industrialized counties (NICs) became a pedagogic tool to incubate new space engineers and to boost industrial progress [2].

The high-precision and high-performance attitude determination and control system (ADCS) of the micro/nano satellite are the basic conditions for a satellite to run efficiently as the accomplishment of the mission of satellites relies on the performance of this instrument as well as being determined by the precision of the attitude control. From the international development trend, ADCS are approximately 40% of total development costs and this is the critical technology in the development of the micro/nano satellite [3].

Spacecraft attitude determination and control covers the entire range of techniques for determining the orientation of a spacecraft and then controlling it so that the spacecraft points in some desired direction. The attitude estimation and attitude control problems are coupled, but they can be considered separately to some extent [4].

The “Attitude Determination” algorithms used in the SSS-1P project are “TRAID” and “Extended Kalman Filter”. Harold Black had developed the algebraic method for the point-by-point determination of a spacecraft’s attitude from a set of two vector observations in 1964 [4]. Shuster later renamed this algorithm TRIAD [5]. The Kalman filter first was used for attitude determination by James Farrell in 1970 [6]. Even more, a constant-gain filter had been proposed earlier [7] and Kalman filters for attitude estimation had appeared previously in contractor reports [8].

To design and simulate a satellite, the main steps that must be followed can be introduced as [9]: derivation of the satellite motion, linearization around the orbit trajectory and study stability, observability, controllability.

In [10] the modeling of a microsatellite with the objective of SIL test is studied. The non-linear equations are drived for the designed ADCS based on [11, 12, 13]. The main characteristics of this research is implementing four truster as controllers, Unscented Kalman Filter and Nonlinear $H_\infty$ Robust Control Law. Even more, in this study the dominant sources of attitude disturbance torques introduced as the gravity-gradient torque, aerodynamic torque, and magnetic disturbance. The results of SIL show that the attitude estimation errors are within 3° and the UKF error can converge after 240 seconds even though the initial attitude error is large. The designed ADCS can fulfill the mission requirement by achieving attitude determination error of less than 5°.

In [14] the linearized model of a cubesat is drived and the block diagram introduced is highly applicable for most of the systems even in a microsatellite system.

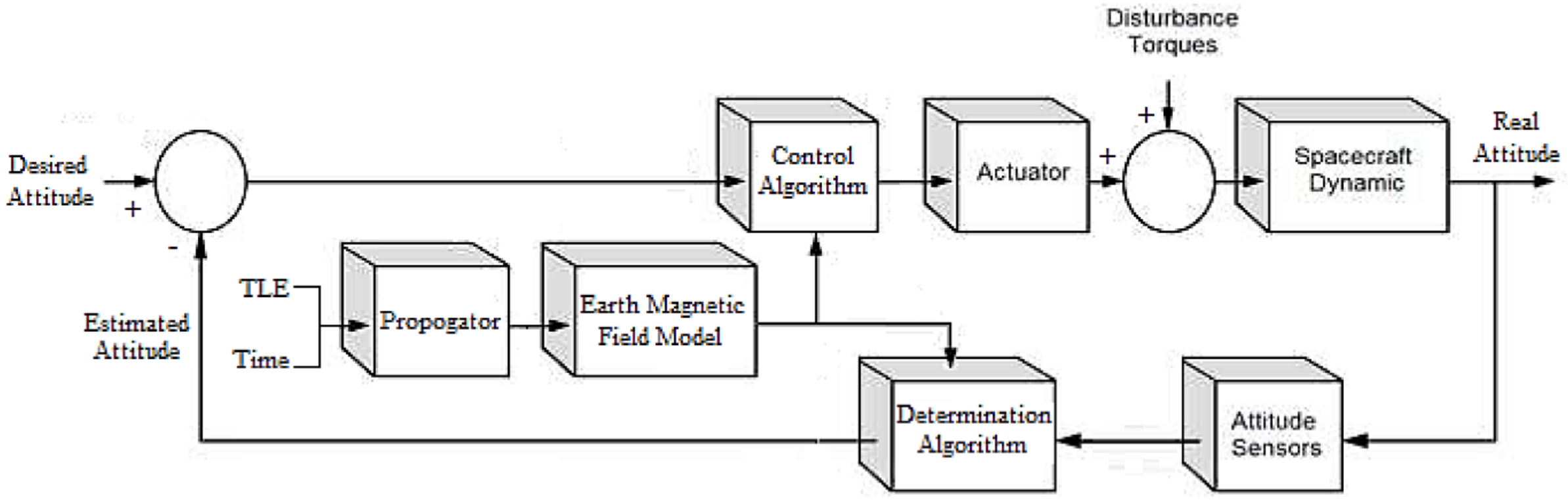

Figure 1-12 Block diagram of ADCS [14].

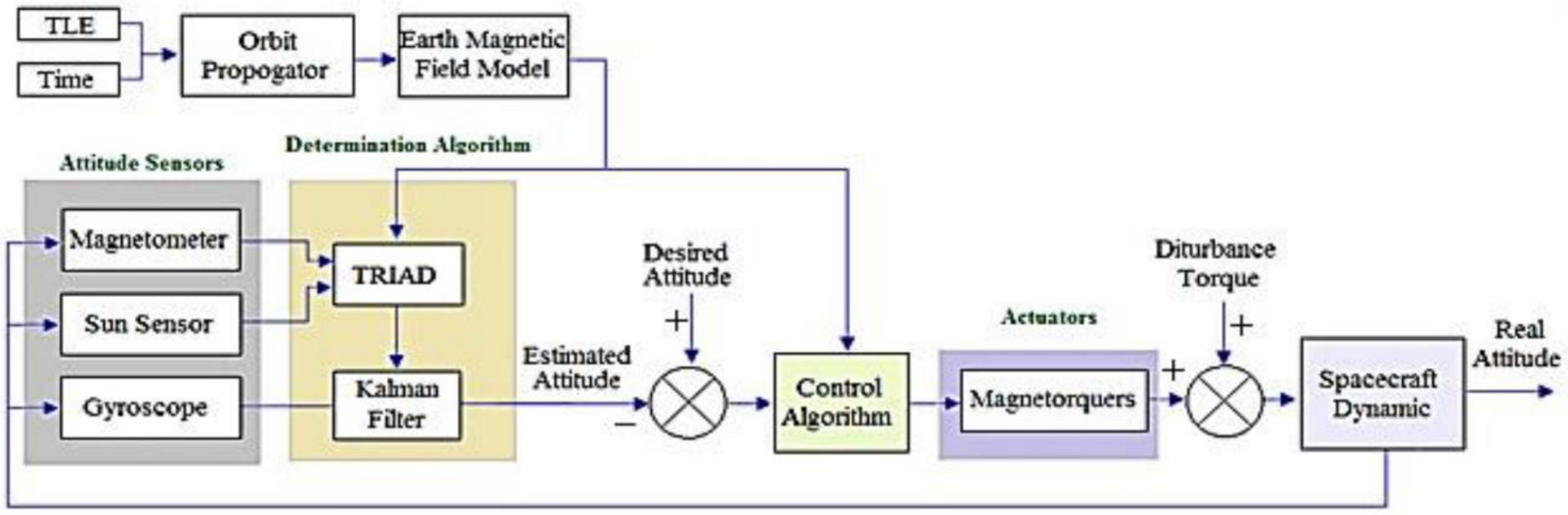

Figure 1-13 Overall view of the ADCS system *[14]*.

In addition, the extended kalman filter they have applied and the applied algorithm is also of a high quality material for the future studies and similar to the model of the SSS-1P micro-satellite.

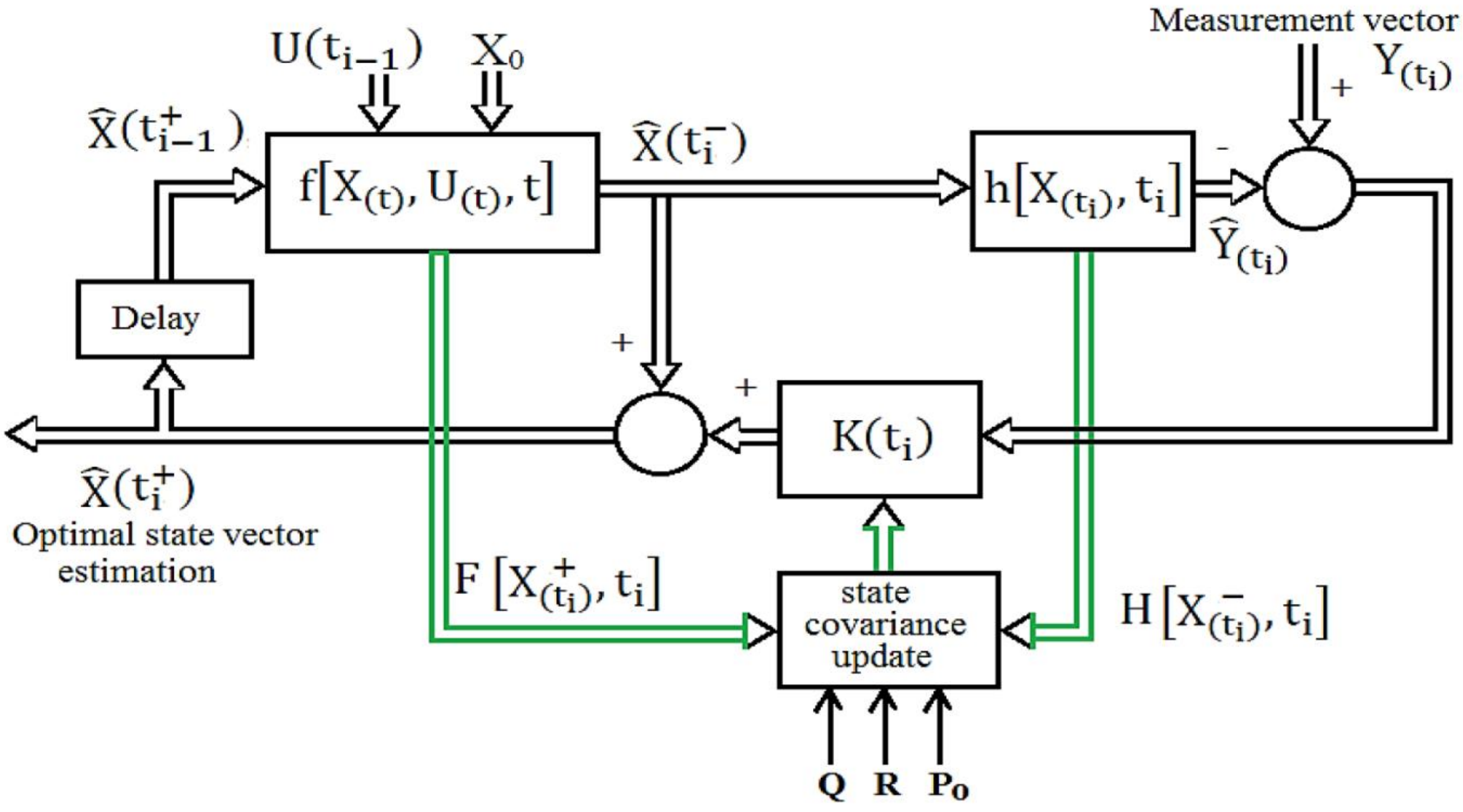

Figure 1-14 Block diagram of extended Kalman filter *[14, 38]*.

The objective of this paper was to establish the design of an effective attitude determination and control system for KufaSat based on purely magnetic actuation. The software selection comprised of TRIAD algorithm, extended Kalman filter as attitude determination software and B.dot algorithm, Quaternion feedback regulator algorithm as control software. Simulation results showed that the B-dot detumbling controller is capable of reducing the angular rate of the satellite to below the requirement [14]. The control algorithm used in SSS-1P for detumbling controller is also B-dot.

Even more, the TRIAD and QUEST methods are low cost attitude determination algorithms that they are introduced and compared with EKF in [15]. The table below presents the results of this comparassion.

Table 1-7 RMS for Attitude Determination Methods *[15]*.

| | Roll Motion | Pitch Motion | Yaw Motion |
|---|---|---|---|
| Extended Kalman Filter | 0.14 | 0.26 | 0.36 |
| TRIAD | 0.49 | 0.59 | 0.85 |
| QUEST | 0.53 | 0.57 | 1.01 |

Though not as accurate as an extended Kalman filter, TRIAD and QUEST algorithms produce estimates within three degrees of actual attitude. Because they are subject to measurement noise, TRlAD and QUEST have higher RMS values than the extended Kalman filter after 75 seconds. Despite these slight disadvantages, TRIAD and QUEST are computationally very simple, and can be used for a quick calculation attitude determination or to validate filter performance. In this scenario, TRIAD and QUEST RMS values were similar because there were only two measurements. With more measurements, QUEST will theoretically outperform TRIAD because it will be able to produce an attitude estimate by using all measurements, whereas TRIAD can only process two [15]. Consequently, in this thesis I am not going to rely on TRIAD only and the algorithm designed will input data to the TRIAD algorithm, then the TRIAD algorithm will calculate the attitude matrice and the singular value decomposition of that matrice will be calculated then SVD will be inserted to the EKF so it will provide better outcome. More information about this will be introduced in the future sections.

The contents above would give the reader a good underestanding about general design of the ADCS. The additional concepts and cosiderations will be discussed now. A magnetic field model is essential for emulation of magnetometers and to be able to calculate the disturbance torque generated by the satellite magnetic residual. It also makes it possible to simulate how much torque the magnetorquers produce. Lastly, a magnetic field model makes it possible to predict magnetometer measurements on board the satellite and predict how the magnetic field varies in the future.

A combination of sources creates the magnetic field surrounding the Earth. More than 90% of this field is generated by the Earth’s outer core and this is referred to as the Main Field, which varies slowly [16]. The Main Field can be described as that of a bar magnet with north and south poles residing within the Earth and field lines extending out into space [16].

Predictive models are the WMM, EMM, HDGM, and HDGM-RT. These differ in what they predict; the WMM predicts only the main magnetic field generated by the Earth's internal dynamo, while the EMM, HDGM and HDGM-RT include contributions from the Earth's crust. The HDGM also includes a basic model of the external field. The HDGM-RT includes a real-time model of the Earth's external field. They also differ in how often they are updated; the WMM and EMM are updated once every five years, while the HDGM is updated every year.

Historic models that we use are the IGRF, the gUFM, the USHistoric model, as well as the IGRF+. The IGRF is the accepted international scientific model of the Earth's field going back to the year 1900. The IGRF+ is a combination of these models that uses an interpolation from 1890 to 1900 to ensure a smooth transition. Finally, the USHistoric is a model based on a polynomial interpolation of early magnetic data in the continental United States [17]. This thesis cosidered and implemented IGRF and WMM as the basic models. Both of them can be used for the calculation. Other disturbances and models will be introduced in chapter 2.

Even more, the Simplified General Perturbations (SGP) model is one of the important part in this design. The SGP model series began development in the 1960s [18] and became operational in the early 1970s [19]. The development culminated in Simplified General Perturbations-4 (SGP4), and although the name is similar, the mathematical technique is very different from the original SGP technique [20]. The first release of the refined SGP4 propagator source code was Spacetrack Report Number 3 [21]. Spacetrack Report Number 3 officially introduced five orbital propagation models to the user community—SGP, SGP4, SDP4, SGP8 and SDP8—all "generally" compatible with the Two-Line Elements (TLE) data [20].

In [22] satellite designed and build by the Norwegian University of Science and Technology test satellite (NUTS) and lunched in 2016. The satellite is built and designed by students at NTNU as a part of the Norwegian Student Satellite Program, which is organized by Norwegian Centre for Space-related Education. The satellite can be classified as CubeSat based on the specification. The attitude determination section includes an Extended Quaternion Estimator (EQUEST) and an Extended Kalman Filter (EKF) [23, 24, 25, 22]. There are four controller designed for this satellite: the B-dot controller, the Detumbling controller, the Pointing controller and the Tumbling controller [22, 26, 27, 28, 29].

In [22] the software design explained clearly and in a compromised way. Different packgaes and their appliactions are also explained from a software engineering point of view. The running platform for this project was FreeRTOS. The software including but not limited to the implementation of EKF, drivers, controller and so on. In the system manager level packages such as logging, voting algorithm, commanding system and several other additional codes are developed. The final package also missing several parts also it is not tested on the hardware at all.

In addition, the [22] approach is creating the code packages all by themselves and from the scratch. Several other litratures have been studied carefully and methods such as designing the system based on UML diagrams, converting the simulink model and etc. have found. The best approach based on my opinion is a combination of many of these methods for the design and implementatin of the source code.

Development of C code directly from designed algorithm has the advantage of implementation of the functions or methods that are not convertible by matlab. Development of c code using MEX toolbox of matlab has the advantage of time efficiency but it will result in multiple additional files and it will increase the complexity of the code in underestanding and furthur development. Finally, using open source packages for several parts such as communication protocols, drivers, predictive packages such as IGRF, WMM, sun vector prediction, SGP4 model and so on, requires a good underestaning of the packages and testing all possible situations so the security and reliability of the package could be guaranteed.

The design of a controller requires a mathematic modeling followed by the adjusting of some model parameters. However to overcome the controller to a single piece of hardware, involves the codification of this mathematical based model into an appropriate firmware description suited to work correctly in a specific platform. Recently a model driven development approach, firstly defined by system engineers, has been frequently used as way to reduce the time of development of embedded systems, producing rapid and reliable product in a short time development cycle. This model driven approach is basically used for test and is known as X-In-The-Loop. These tests provide four levels of testing configurations: MIL (Model-In-The-Loop), SIL (Software-In-The-Loop), PIL (Processor-In-The-Loop) and HIL (Hardware-In-The-Loop). Each of the configuration levels provides some advances and reduces the gap in the development process that initiate with the mathematical model and ends at the firmware running in a stand-alone microprocessor platform [30, 31].

In [32, 31], simulations SIL and PIL were performed to obtain an attitude determination and control system (ADCS) for the microsatellite CKUTEX from Cheng Kung University. The SIL simulation is made using the MATLAB software and after, the PIL simulation is implemented using a PIC microcontroller to the implantation of the ADCS algorithm while the satellite dynamics is implemented in NI-PXI platform and coded by Labview software. According to the authors, the attitude determination and control system that was obtained and tested, provided good results once the plant dynamic response obeyed the project specifications.

In [31, 33], the MATLAB software is used with the purpose of generate the code of an entire control system and the dynamics of two satellites (represented by two robots). In this work a physical simulator using two industrial robots is assembled for a simulation of proximity operations between satellites as on-orbit servicing (OOS) activities. A model of the satellites dynamics, its control, actuators and sensors (constituting the named Application Control System - ACS) is made in MATLAB/Simulink by the tool named Real-Time

Workshop (RTW). It generates a code supported by the operating system VxWorks that on the other hand, operate according this code, the monitoring and control system of the facility where the movement commands desired are sent to the robots. It is a HIL simulation where controllers, actuators, state observers, among other modeled modules in the MATLAB/Simulink can be removed of the ACS and included in their own hardware if necessary.

In [31, 34], a true co-simulation approach is studied for control application running on a Field Programmable Gate Array processor (FPGA). The proposed approach adopts the LABVIEW Real-Time environment (from National Instruments©) to perform the simulation of an artificial satellite' dynamic model. This study simulates a reaction wheel and its control in Simulink exclusively for the controller design. Subsequently, the generated code is implemented on the FPGA. The entire system model attitude control is previously built and simulated in Simulink. The RTW generates the code that represents the satellite dynamics' mathematical model inside LABVIEW environment. In the case of the reaction wheel and its control, RTW generates the C code to simulate the dynamic model that must be adapted to the LABVIEW rules for FPGA programming.

In [31, 35], the Simulink is used together with another MATLAB toolbox called xPC Target. This toolbox allows to perform prototyping, testing and development of real-time systems for running in general-purpose computers. The MATLAB/Simulink tools were used to generate executable code for the target computer from the model built in Simulink. The goal is to achieve simulation and real-time implementation of an algorithm integrating inertial navigation and GPS, using the validation and testing of the proposed algorithm in the FlightGear flight simulator (inside MATLAB), running on the host computer.

## 1.4 Dissertation's Structure

In chapter one and two, I will introduce the fundamental dynamic and kinematic non-linear equations. The linearization for a specific prerational condition and the assumptions to achieve the linearization will be introduced. The model will be linearized for a specific operating condition for the design and development of the controllers. The disturbances including atmospheric drag, solar radiation and so on are also introduced with regarding to the operating Altitude (500 ~ 600 km). The brief introduction of different magnetic field models and their differences will be introduced as well. In addition, the SGP4 predictive model to estimate the position and the velocity of the satellite discussed.

In chapter three, the thesis describes the software development phase, some additional algorithms that I implemented but they are not a part of mathematical models, such as system manager, time manager and so on with pseudo code. The architecture of the software is also described in this section.

Finally, the numerical simulations are analyzed and the simple compression between tests results from PIL, SIL, and MATLAB simulation are presented. This section will presents the validation and the trustworthy of the final package. Logical reading sequence among the thesis chapter and the dissertation's structure presented in the figure below.

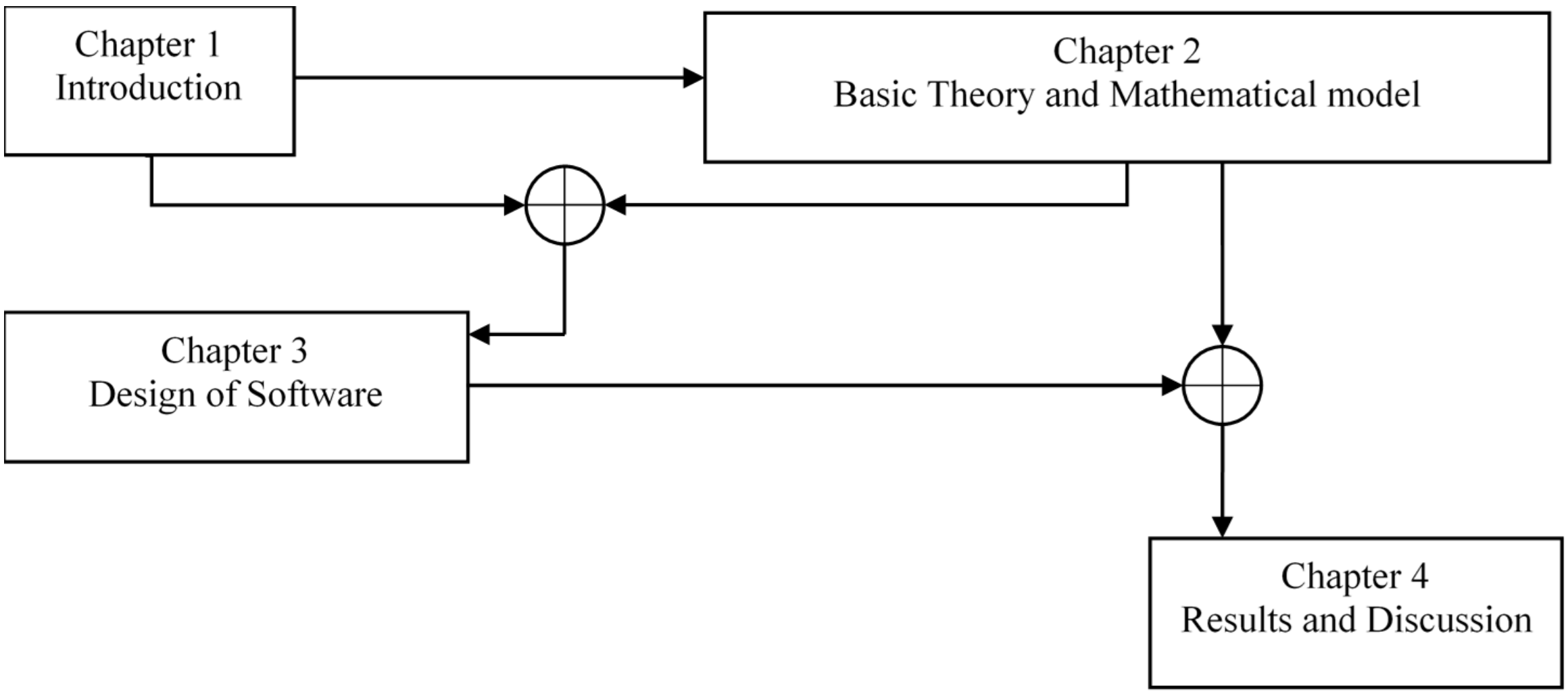

Figure 1-15 Logical reading sequence of the chapters.

# 2. Basic Theory & Mathematical model

## 2.1 Reference frames

Three reference frames are most important and commonly used for attitude determination.

**Earth Centred Inertial Frame**

The origin of the frame is located at the centre of the Earth. The z axis shows the geographic North Pole while the x axis is directed toward the Vernal Equinox. The y axis completes the coordinate system as the cross product of z and x axes.

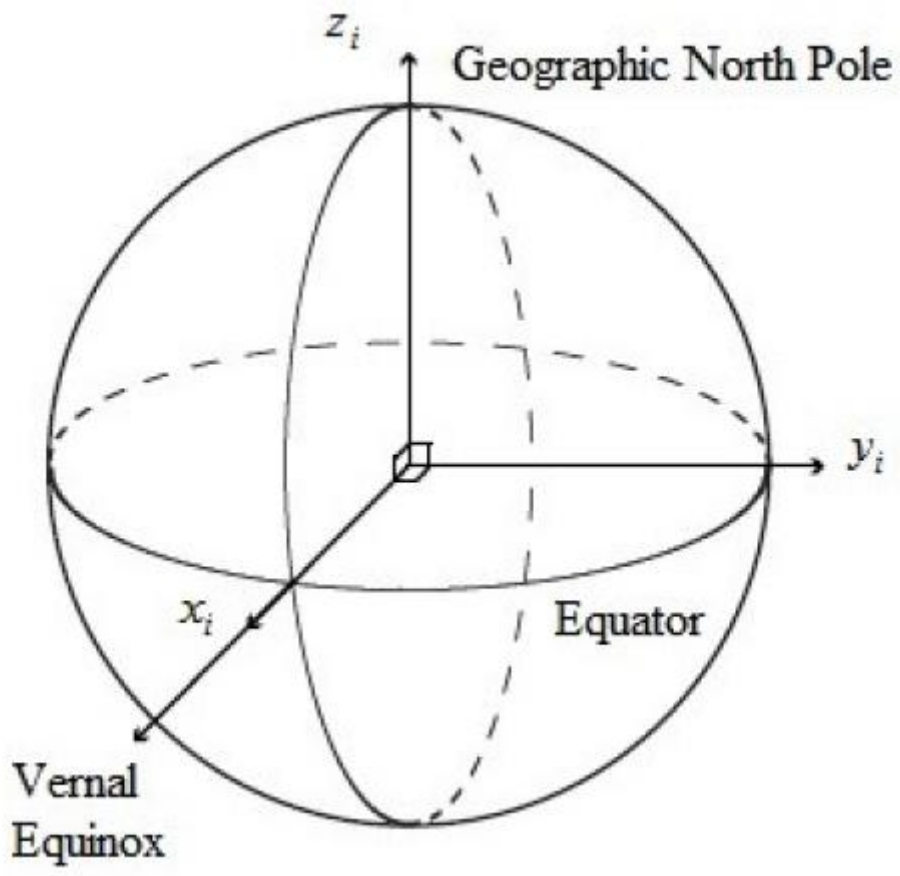

Figure 2-1 Representation of ECI frame *[51]*.

**Orbit Reference Frame**

The origin of the frame is at the mass centre of the spacecraft. The z axis is in nadir direction (towards the centre of the Earth) and the y axis is tangential to the orbit. The x axis completes to the orthogonal right-hand system.

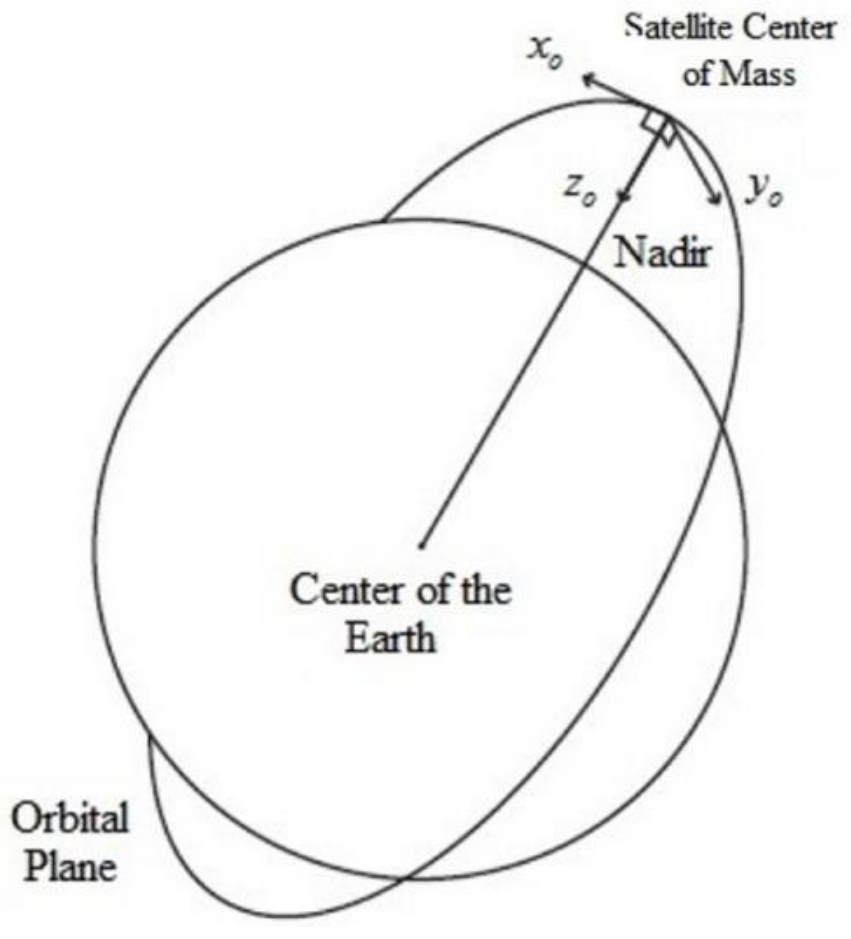

*Figure 2-2 Representation of Orbital frame . [51]*

**Satellite Body Frame**

The origin of the frame is located at a specified point in the spacecraft body. Three parameters named as Euler angles set the condition of the body frame related to the reference coordinate system.

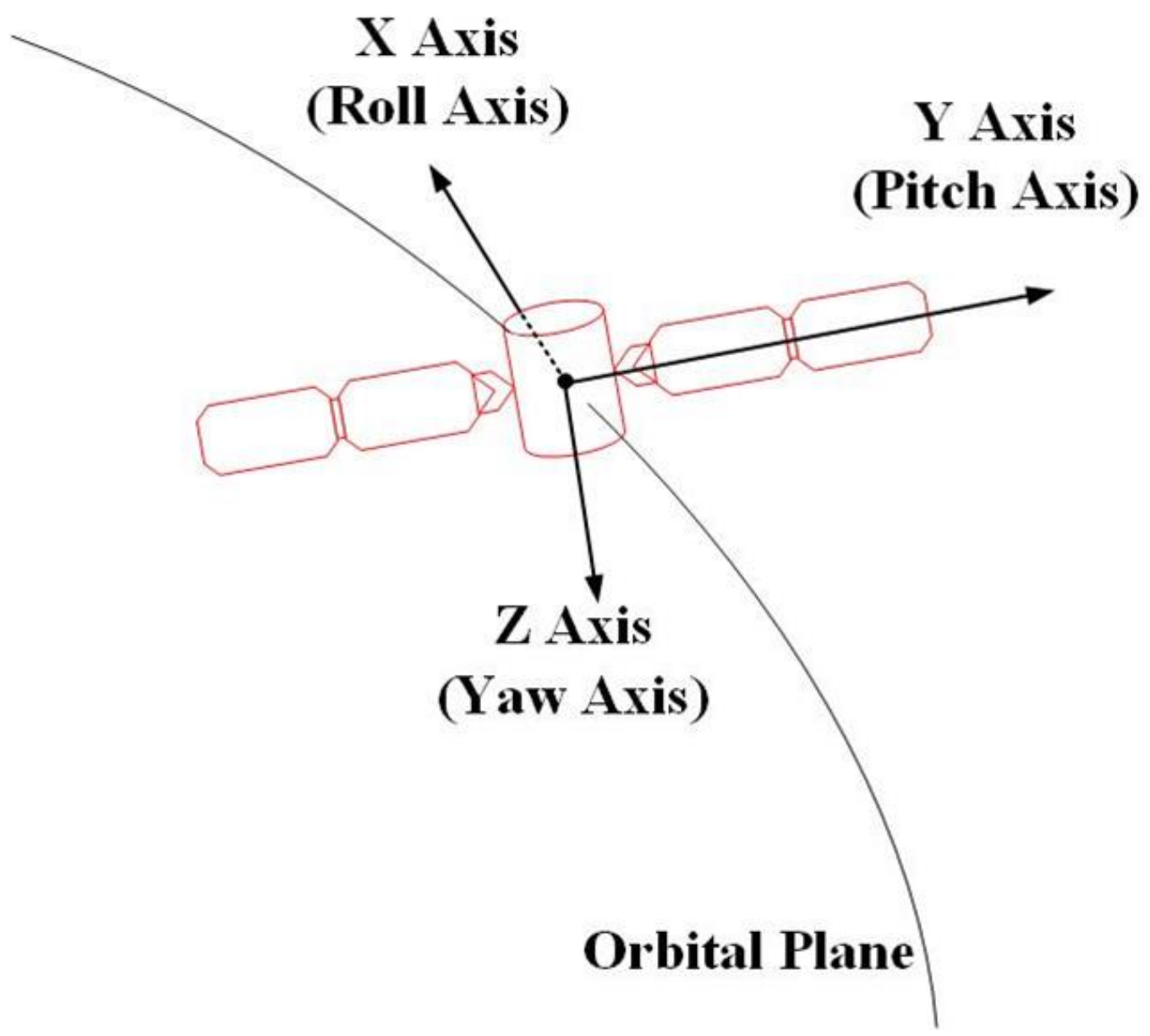

Figure 2-3 Representation of Satellite Body frame *[51]* .

## 2.2 Mathematical model

The Dynamic Model of the Satellite starting with Euler's moment equation [36] and we will have from [37]

$$\dot{H} = T_{dist} - \omega \times H$$

And

$$H = I\omega$$

And the Kinematic model of the satellite can be described by

$$\dot{q} = \frac{1}{2}\Omega q \equiv \frac{1}{2}\begin{bmatrix} 0 & \omega_3 & -\omega_2 & \omega_1 \\ -\omega_3 & 0 & \omega_1 & \omega_2 \\ \omega_2 & -\omega_1 & 0 & -\omega_3 \\ -\omega_1 & -\omega_2 & -\omega_3 & 0 \end{bmatrix} q$$

Satellite dynamics represented in standard form, by having $I_s$ as the satellite inertia matrix, $\omega$ as the angular velocity of the satellite and $M_w$ as the total angular momentum of the wheels, all in satellite's coordinate system and $T_{dist}$ is the total disturbance torque. Therefor, the dynamic equation will be as below

$$\frac{d}{dt}(I_s\omega) + \dot{M}_w = T_{dist} - \omega \times I_s\omega - \omega \times M_w$$

Which is

$$\dot{\omega} = -I_s^{-1}(\omega \times I_s\omega) - I_s^{-1}\omega \times M_w - I_s^{-1}\dot{M}_w + I_s^{-1}T_{dist}$$

Using S () matrix operation for writing the cross product we will have

$$S(\omega) = \begin{bmatrix} 0 & -\omega_3 & \omega_2 \\ \omega_3 & 0 & -\omega_1 \\ -\omega_2 & \omega_1 & 0 \end{bmatrix}$$

We will have

$$\dot{\omega} = -I_s^{-1}S(\omega)I_s\omega - I_s^{-1}S(\omega)M_w - I_s^{-1}\dot{M}_w + I_s^{-1}T_{dist}$$

The complete nonlinear dynamic model of the satellite is finalized by noting that control torque in the body coordinate system is $T_c$ and resulting in the rate of change of the total angular momentum from wheels as

$$\dot{M}_w = -T_c$$

And the dynamic of the satellite will be actuated by the reaction wheels will be as below

$$\dot{\omega} = -I_s^{-1}S(\omega)I_s\omega - I_s^{-1}S(\omega)M_w + I_s^{-1}T_c + I_s^{-1}T_{dist}$$

The kinematics of the satellite cab be described by using the attitude quaternion below to represent a rotation

$$q = \begin{bmatrix} q1 \\ q2 \\ q3 \\ q4 \end{bmatrix}$$

The orientation of the satellite is calculated rotating from the inertial frame to the satellite frame by Directional Cosine Matrix $DCM_{SI}$. The rotation matrix is parametrized by the quaternion $DCM_{SI}(q)$. Finally, a vector will be measured in inertial frame $V_I$ can be represented in the satellite body coordinate as

$$V_S = DCM_{SI}(q)V_I$$

And the kinematics of the satellite is

$$\frac{d}{dt}\begin{bmatrix} q1 \\ q2 \\ q3 \\ q4 \end{bmatrix} = \begin{bmatrix} 0 & \omega_3 & -\omega_2 & \omega_1 \\ -\omega_3 & 0 & \omega_1 & \omega_2 \\ \omega_2 & -\omega_1 & 0 & -\omega_3 \\ -\omega_1 & -\omega_2 & -\omega_3 & 0 \end{bmatrix} q$$

We will compact the notation and the compact notation is subsequently referred to in block diagrams.

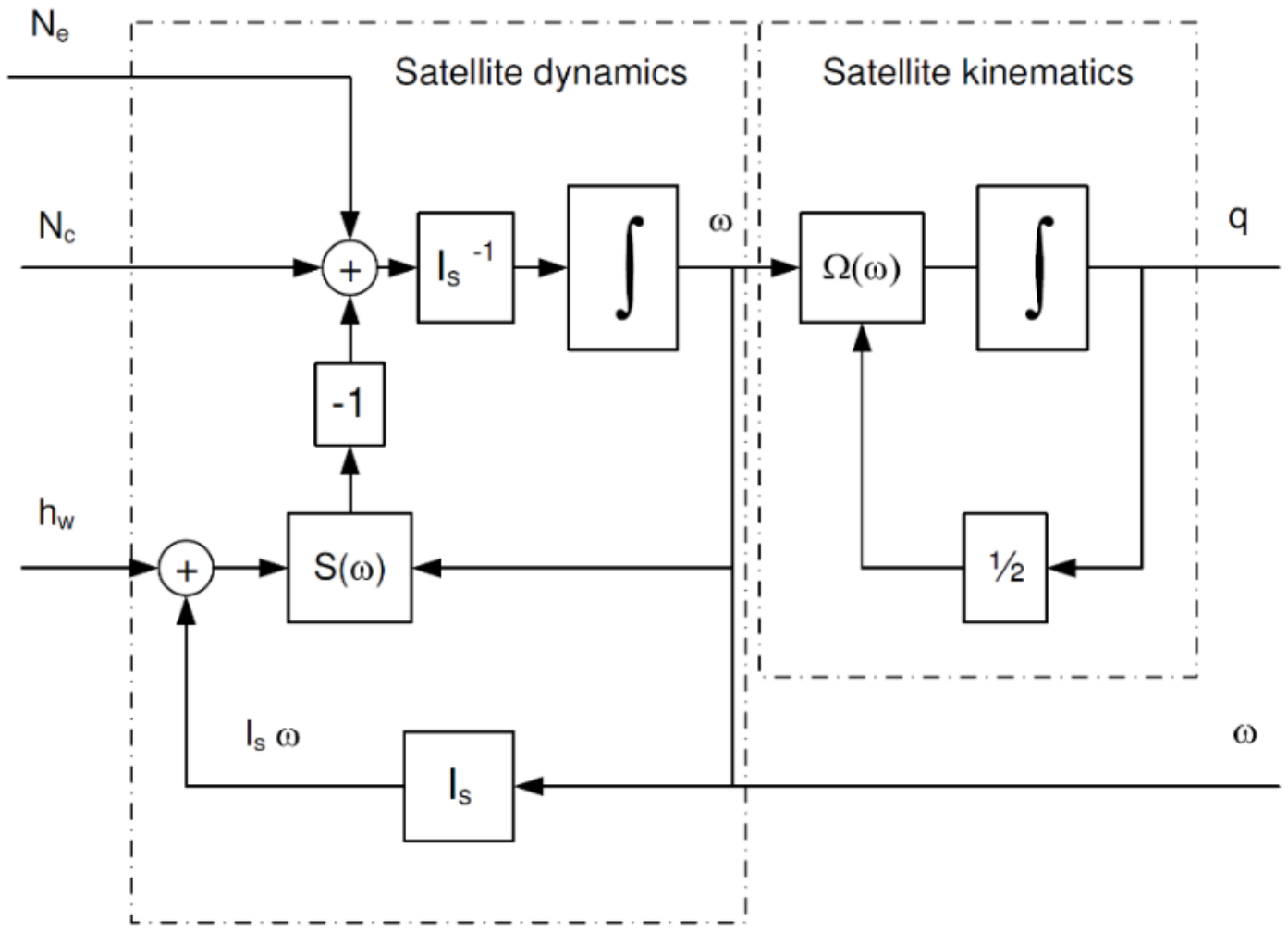

*Figure 2-4 Satellite block diagram [37].*

$$\dot{q} = \frac{1}{2}\Omega(\omega)q$$

Defining a vector part of the first three components of the $q$ and denote this by $g$ which is called Gibbs vector we will have

$$g = \begin{bmatrix} q1 \\ q2 \\ q3 \end{bmatrix}$$

The we can have

$$\dot{g} = -\frac{1}{2}\omega \times g + \frac{1}{2}q_4\omega$$

$$\dot{q} = -\frac{1}{2}\omega^T g$$

or written out in component form

$$\frac{d}{dt}\begin{bmatrix} q1 \\ q2 \\ q3 \\ q4 \end{bmatrix} = -\frac{1}{2}\begin{bmatrix} 0 & -\omega_3 & \omega_2 \\ \omega_3 & 0 & -\omega_1 \\ -\omega_2 & \omega_1 & 0 \\ \omega_1 & \omega_2 & \omega_3 \end{bmatrix}\begin{bmatrix} q1 \\ q2 \\ q3 \end{bmatrix} + \frac{1}{2}\begin{bmatrix} q_4 & 0 & 0 \\ 0 & q_4 & 0 \\ 0 & 0 & q_4 \\ 0 & 0 & 0 \end{bmatrix}\begin{bmatrix} \omega_1 \\ \omega_2 \\ \omega_3 \end{bmatrix}$$

Or

$$\frac{d}{dt}\begin{bmatrix} g \\ q_4 \end{bmatrix} = \frac{1}{2}\begin{bmatrix} -S(\omega) \\ -\omega^T \end{bmatrix} g + \frac{1}{2}q_4\begin{bmatrix} I_{3\times3} \\ 0 \end{bmatrix}\omega$$

This operation preserves unit length of the quaternion. The satellite model with nonlinear dynamic and kinematic equations will be like it should be noted that the dynamic of the satellite has a structure that is invariant with the choice of coordinate system. As long as all quantities are based on the same coordinate system, the structure of the equation is unchanged. The values of elements in the inertia tensor do change. The effects caused by a rotation of both measured values and reference coordinates for the inertia tensor to a satellite geometry defined coordinate system is achieved by rotating the various quantities in the dynamic equation

$$\frac{d}{dt}\left(I_s\omega_p\right) = -\omega_p \times I_s\omega_p - \omega_p \times M_{w_p} - T_{c_p} + T_{dist_p}$$

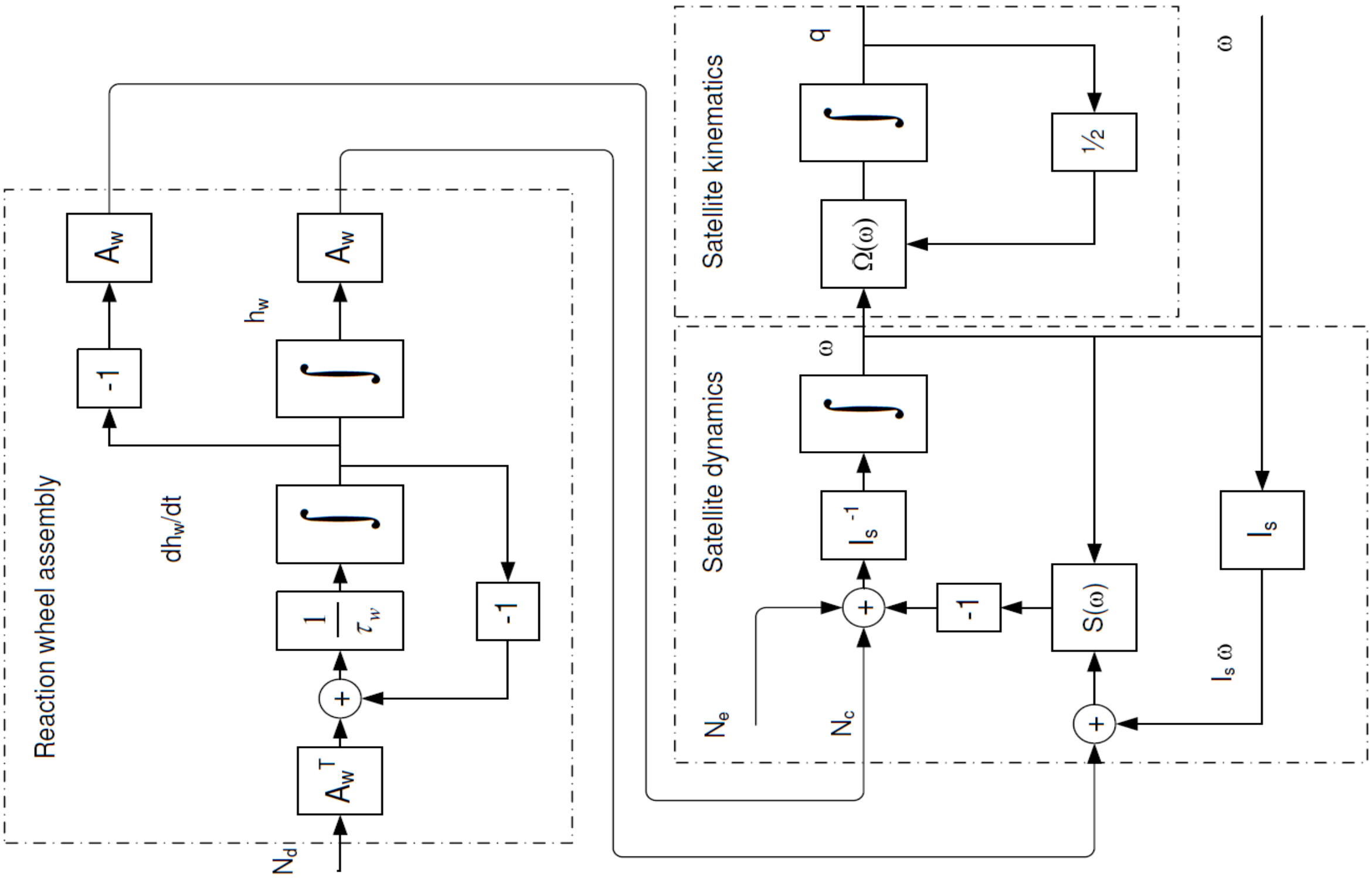

*Figure 2-5 Satellite dynamics and kinematics model with reaction wheel assembly as actuator [37].*

Then rotating to the body frame using

$$\omega_p = DCM_{pb}\omega_b$$

$$M_{w_p} = DCM_{pb}M_{w_b}$$

$$T_p = DCM_{pb}T_b$$

$$I_s = DCM_{pb}I_{sb}A_{pb}^T$$

By substitution and simplification, we will have

$$\frac{d}{dt}(I_{sb}\omega_b) = -\omega_b \times I_{sb}\omega_b - \omega_b \times M_{w_b} - T_b + T_{dist\,b}$$

This non-linear equation of motion can be linearized in an arbitrary point of operation $\left(\bar{\omega}, \bar{g}, \overline{q_4}, \bar{h}\right)$ in order to drive the set of linear state space equations. This is needed to enable a stringent stability and performance analysis for the linear control systems. The deviation from steady state (point of linearization) is denoted by a tilde above the variables, $\omega = \bar{\omega} + \tilde{\omega}$ but the quaternion representation of attitude poses a specific problem. With $dg$ denoting the orientation at time $t + dt$ relative to the attitude at time $t$, then, since

$$\begin{bmatrix} \tilde{g} \\ \widetilde{q_4} \end{bmatrix} = \begin{bmatrix} dg_1 \\ dg_2 \\ dg_3 \\ dq_4 \end{bmatrix} = \begin{bmatrix} e_1 \sin\left(\frac{1}{2}\omega dt\right) \\ e_2 \sin\left(\frac{1}{2}\omega dt\right) \\ e_3 \sin\left(\frac{1}{2}\omega dt\right) \\ \cos\left(\frac{1}{2}\omega dt\right) \end{bmatrix} \cong \begin{bmatrix} \frac{1}{2}\omega_1 dt \\ \frac{1}{2}\omega_2 dt \\ \frac{1}{2}\omega_3 dt \\ 1 \end{bmatrix}$$

Then $\frac{d}{dt}q_4 = 0$ and $S(\omega)dg = 0$ hence

$$\frac{d}{dt}\tilde{g} = \frac{1}{2}S(\omega)\tilde{g} + \frac{1}{2}\widetilde{q_4}I_{3\times3}\omega = \frac{1}{2}I_{3\times3}\omega$$

$$h = \bar{h} + \tilde{h}; \frac{d}{dt}h = \frac{d}{dt}\tilde{h}$$

The state vector of linear equation of motion is

$$x = (\tilde{\omega}_1, \tilde{\omega}_2\,, \tilde{\omega}_3\,, \tilde{g}_1, \tilde{g}_2, \tilde{g}_3, h_1, h_2, h_3)^T$$

Which has the control input of $u = T_c$. Assuming external and disturbance torques have zero mean. The state space equation is

$$\dot{x}(t) = A(t)x(t) + B_u(t)T_c(t) + B_d(t)T_{dist}(t)$$

Where

$$A_{ij} = \frac{\delta f_i}{\delta x_j}; \qquad B_{ij} = \frac{\delta f_i}{\delta u_j}$$

And

$$f = \begin{bmatrix} -I_s^{-1}S(\omega)I_s\omega - I_s^{-1}S(\omega)M_w + I_s^{-1}(T_c + T_{dist}) \\ -\frac{1}{2}S(\omega)g + \frac{1}{2}q_4 I_{3\times3}\omega \\ -\frac{1}{2}\omega^T g \\ -T_c \end{bmatrix}$$

Using symbolic manipulation to calculate the Jacobians we will have

$$A = \begin{bmatrix} I_s^{-1}A_{\omega,\omega} & 0 & I_s^{-1}A_{\omega,h} \\ \frac{1}{2}I_{3\times3} & 0 & 0 \\ 0 & 0 & 0 \end{bmatrix}$$

$$B_u = \begin{bmatrix} I_s^{-1} \\ 0 \\ -I_{3\times3} \end{bmatrix}$$

$$B_d = \begin{bmatrix} I_s^{-1} \\ 0 \\ 0 \end{bmatrix}$$

And

$$A_{\omega,\omega} = \left[A_{\omega,1}, A_{\omega,2}, A_{\omega,3}\right]$$

Also

$$A_{\omega,1} = \begin{bmatrix} \omega_2 I_{31} - \omega_3 I_{21} \\ -2I_{31}\omega_1 - \omega_2 I_{32} - \omega_3 I_{11} + h_3 \\ I_{21}\omega_1 + \omega_2 I_{22} + I_{23}\omega_3 - \omega_2 I_{11} - h_2 \end{bmatrix}$$

$$A_{\omega,2} = \begin{bmatrix} I_{31}\omega_1 + 2I_{32}\omega_2 + \omega_3 I_{33} - \omega_3 I_{22} - h_3 \\ \omega_3 I_{21} - \omega_1 I_{32} \\ -\omega_1 I_{11} - 2I_{12}\omega_2 - I_{13}\omega_3 + \omega_1 I_{22} + h_1 \end{bmatrix}$$

$$A_{\omega,3} = \begin{bmatrix} -I_{21}\omega_1 - \omega_2 I_{22} - 2I_{23}\omega_3 + \omega_2 I_{33} + h_2 \\ \omega_1 I_{11} + I_{12}\omega_2 + 2I_{13}\omega_3 - \omega_1 I_{33} - h_1 \\ \omega_1 I_{23} - \omega_2 I_{13} \end{bmatrix}$$

$$A_{\omega,h} = \begin{bmatrix} 0 & \omega_3 & -\omega_2 \\ -\omega_3 & 0 & \omega_1 \\ \omega_2 & -\omega_1 & 0 \end{bmatrix}$$

The nominal condition is expressed through the parameters $\omega = \bar{\omega}$, the average angular rate of the satellite, and $h = \bar{h}$, the average stored angular momentum expressed in satellite body coordinates. This linear model accepts an arbitrary moment of inertia tensor, which enables subsequent use for both controller design and analysis of sensitivity (robustness) properties. Uncertainties include magnitude and rotation of the inertia tensor and alignment of wheels. The basic dynamic properties change with the resulting angular momentum of the wheels. The changes in linear dynamics could be analyzed should the satellite be demanded to rotate along one of its axes, e.g. during maneuvers.

Assuming an earth pointing satellite in a circular orbit has orbit period $T_0$. The orbit rate is $\omega_0 =$

$\frac{2\pi}{T_0}$. The nominal angular rate of the satellite is then $\omega = (0, -\omega_0, 0)\ \left[\frac{rad}{s}\right]$. The negative sign arrives from the definition of the orbit coordinate system with its x-axis along the satellite's velocity vector and the z-axis pointing towards Nadir (Center of the Earth). Investigation of properties of the satellite should be done by linearizing the system at $\omega = (0, -\omega_0, 0)$.

## 2.3 Contollers

### B-dot controller

The magnetic damping uses the **Bdot** control law to damp the satellite's rotating rate. The control process of the magnetic damping mode is as follows.

(1) Calculate $\boldsymbol{B}_{dot} = (\boldsymbol{B}_{b,k} - \boldsymbol{B}_{b,k-2})/(2T)$, where $T$ is the control cycle period and is set to be 0.25s, and $k$ is the control step.

(2) Calculate expected normal magnetic moment $\boldsymbol{M}_n = -\boldsymbol{B}_{dot} / |\boldsymbol{B}_{dot}|$;

(3) Generate maximum magnetic moment $\boldsymbol{M}_{\max}$ along $\boldsymbol{M}_n$: let $k_{UM} = 9/12(\mathrm{V/Am^2})$ be the control voltage to magnetic moment ratio, $U_{\max} = 9$ be the maximum control voltage, $M_{\max} = 12$ be the maximum magnetic moment. The control voltage of the magnetorquers is

$$\boldsymbol{U}_m = k_m \frac{U_{\max}\boldsymbol{U}_0}{\max(\boldsymbol{U}_0)} \mathrm{sign}(\boldsymbol{M}_n)$$

Where $\boldsymbol{U}_0 = k_{UM}\boldsymbol{M}_n M_{\max}$ , $k_m$ is the time-division work coefficient designed for the situation when the mast has not been deployed: $k_m = \begin{cases} k \ mod \ 2, & Mflag = 0 \\ 1, & Mflag \neq 0 \end{cases}$

### Rate damping

The rate damping uses the angular velocity feedback to damp the satellite's rotating rate. The control process of the rate damping mode is as follows.

(1) Calculate the expected torque $\boldsymbol{T}_{E_RD}$: $\boldsymbol{T}_{E_RD} = -\boldsymbol{K}_{d_RD}\boldsymbol{w}_{bo}$, where

$$\boldsymbol{K}_{d_RD} = \begin{cases} \mathrm{diag}(34.1221,\ 34.0776,\ 4.7492), & Mflag \neq 0 \\ \mathrm{diag}(5.1101\ \ 5.0611\ \ 9.4984), & Mflag = 0 \end{cases}$$

(2) Calculate the $\boldsymbol{T}_{E_RD}$'s vertical component $\boldsymbol{T}_{EV_RD}$ with respect to the magnetic field vector:

$$\boldsymbol{T}_{EV_RD} = \boldsymbol{T}_{E_RD} - \left(\boldsymbol{T}_{E_RD} \cdot \frac{\boldsymbol{B}_b}{|\boldsymbol{B}_b|}\right)\frac{\boldsymbol{B}_b}{|\boldsymbol{B}_b|}$$

(3) Calculate the expected magnetic moment $\boldsymbol{M}_c$:

$$\boldsymbol{M}_c = -\frac{\boldsymbol{T}_{EV_RD} \times \boldsymbol{B}_b}{|\boldsymbol{B}_b|^2}$$

(4) Calculate the control voltage: $\boldsymbol{U}_m = \text{Saturation}(k_{UM}\boldsymbol{M}_c, U_{max})$.

**Attitude Maneuver**

The attitude maneuver uses the maneuver law to draw the satellite back to the Earth pointing status from an arbitrary attitude. The control process of the attitude maneuver mode is as follows.

(1) Calculate the Earth pointing error angle

$$A_{zn} = \text{acos}(\boldsymbol{C}'_{bo}[0,0,1]^T \; g \; [0,0,1]^T);$$

(2) Calculate the target attitude, which makes the nearest way to pull the satellite from current attitude back to Earth-pointing. Let the current attitude quaternion with respect to the target attitude be $\boldsymbol{Q}_{br}$, there is

$$\boldsymbol{Q}_{br} = \boldsymbol{Q}'_{EU}(\text{Normal}([0,0,1]^T \times (\boldsymbol{C}_{bo}[0,0,1]^T)), A_{zn})$$

where Normal is the normalizing a vector, $\boldsymbol{Q}_{EU}$ means to calculate the quaternion from the Euler axis and the Euler angle.

(3) Convert $\boldsymbol{Q}_{br}$ to the Euler axis $[lx, ly, lz]^T$ and the Euler angle $Sita$.

(4) Calculate the expected maneuver angular speed $W$ as the relationship shown in Fig 9, where $W_{\max} = 0.15$ deg/s, $SitaD = 2$ deg;

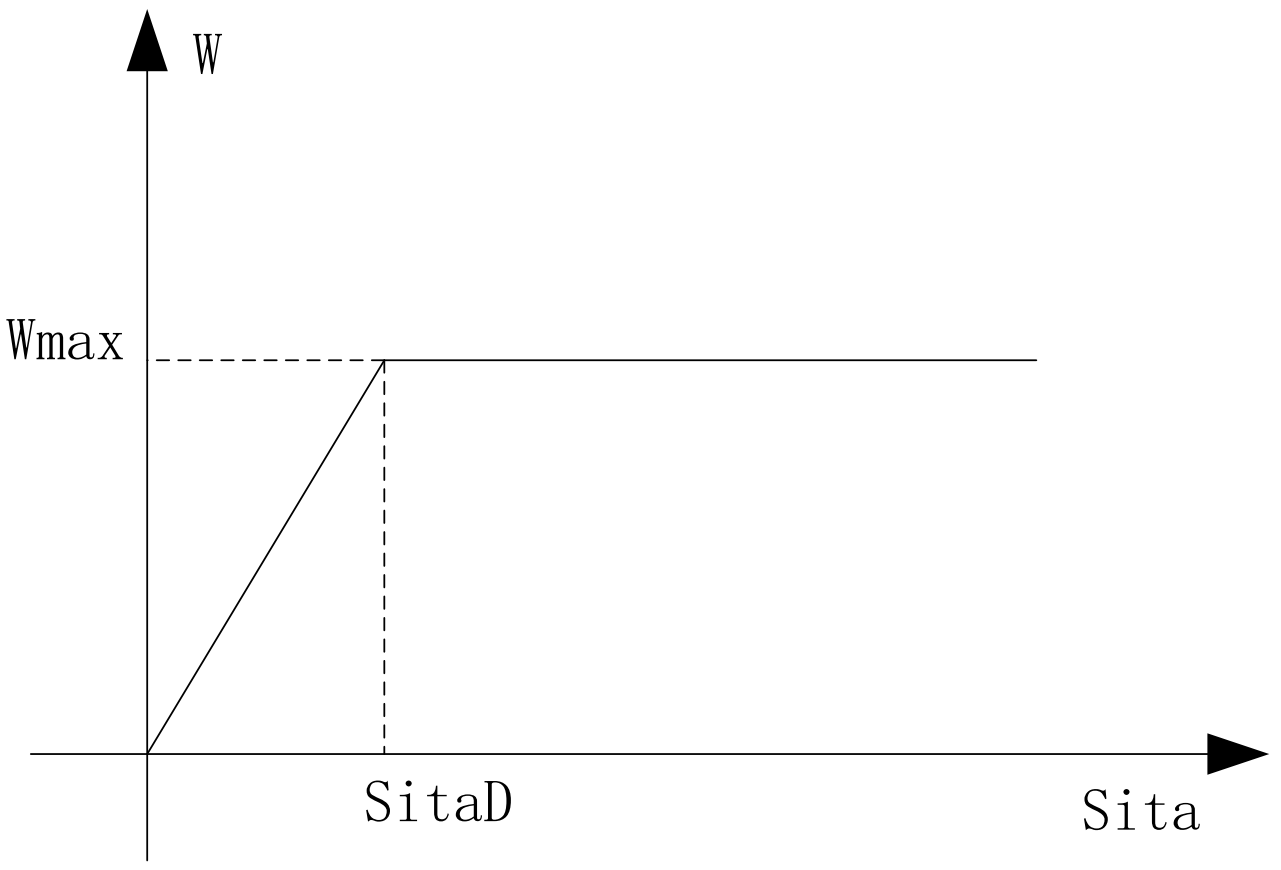

Figure 2-6 the maneuver angular speed

(5) Calculate the expected maneuver angular velocity $\boldsymbol{W} = W[lx, ly, lz]^T$;

(6) Calculate the expected torque $\boldsymbol{T}_{E_MV} = -\boldsymbol{K}_{d_MV}(\boldsymbol{W} - \boldsymbol{w}_{bo})$, where

$$\boldsymbol{K}_{d_MV} = \begin{cases} \text{diag}(29.5725,\ \ 29.5339,\ \ 4.1160), & Mflag \neq 0 \\ \text{diag}(4.4287,\ \ 4.3863,\ \ 8.2320), & Mflag = 0 \end{cases}$$

(7) Calculate the $\boldsymbol{T}_{E_MV}$'s vertical component $\boldsymbol{T}_{EV_MV}$ with respect to the magnetic field vector:

$$\boldsymbol{T}_{EV_MV} = \boldsymbol{T}_{E_MV} - \left(\boldsymbol{T}_{E_MV} \cdot \frac{\boldsymbol{B}_b}{|\boldsymbol{B}_b|}\right) \frac{\boldsymbol{B}_b}{|\boldsymbol{B}_b|}$$

(8) Calculate the expected magnetic moment $\boldsymbol{M}_c$:

$$\boldsymbol{M}_c = -\frac{\boldsymbol{T}_{EV_MV} \times \boldsymbol{B}_b}{|\boldsymbol{B}_b|^2}$$

(9) Calculate the control voltage:

$$\boldsymbol{U}_m = \text{Saturation}(k_{UM}\boldsymbol{M}_c, U_{max})$$

**Earth pointing Stabilization**

The Earth pointing stabilization works when the Earth-pointing angle is small to further stabilize the satellite. The control process of the Earth pointing stabilization is as follows.

(1) Calculate the target attitude quaternion $\boldsymbol{Q}_{br} = [qr1, qr2, qr3, qr4]^T$ as in the attitude maneuver mode.

(2) Calculate the expected torque $\boldsymbol{T}_{E_EP}$:

$$\boldsymbol{T}_{E_EP} = -\boldsymbol{K}_{p_EP}[qr1, qr2, qr3]^T - \boldsymbol{K}_{d_EP}\boldsymbol{w}_{bo}$$

where

$$\boldsymbol{K}_{p_EP} = \begin{cases} \text{diag}(0.3780, \quad 0.3775, \quad 0.0263), & Mflag \neq 0 \\ \text{diag}(0.0566, \quad 0.0561, \quad 0.0263), & Mflag = 0 \end{cases}$$

$$\boldsymbol{K}_{d_EP} = \begin{cases} \text{diag}(22.7481, \quad 22.7184, \quad 3.1661), & Mflag \neq 0 \\ \text{diag}(3.4067, \quad 3.3740, \quad 6.332), & Mflag = 0 \end{cases}$$

(3) Calculate the $\boldsymbol{T}_{E_EP}$'s vertical component $\boldsymbol{T}_{EV_EP}$ with respect to the magnetic field vector:

$$\boldsymbol{T}_{EV_EP} = \boldsymbol{T}_{E_EP} - \left(\boldsymbol{T}_{E_EP} \cdot \frac{\boldsymbol{B}_b}{|\boldsymbol{B}_b|}\right) \frac{\boldsymbol{B}_b}{|\boldsymbol{B}_b|}$$

(4) Calculate the expected magnetic moment $\boldsymbol{M}_c$:

$$\boldsymbol{M}_c = -\frac{\boldsymbol{T}_{EV_EP} \times \boldsymbol{B}_b}{|\boldsymbol{B}_b|^2}$$

(5) Calculate the control voltage:

$$\boldsymbol{U}_m = \text{Saturation}(k_{UM}\boldsymbol{M}_c, U_{max})$$

## 2.4 Attitude Determination

The main attitude determination method is using the magnetometer-gyroscope filtering technique. The backup attitude determination method is using the star sensor and the gyroscope. This is for the reason that star sensor performance is limited by the satellite attitude when it is pointing to the Earth or the satellite's high angular velocity.

### Magnetometer-gyroscope Filtering

The magnetometer-gyroscope filtering technique can provide attitude information at any attitude and at any rate within the gyroscope measurement limit. It is explained as follows.

Let $\boldsymbol{Q}_{bi} = [q_1, q_2, q_3, q_4]^T$ be the inertial quaternion. Let $\boldsymbol{w}_{bi} = [w_x, w_y, w_z]^T$ be the inertial angular velocity. Let $[c_x, c_y, c_z]^T$ be the gyroscope drift of each axis. Let $\boldsymbol{B}_i = [b_{i_x}, b_{i_y}, b_{i_z}]^T$ be the inertial magnetic field. Let $Ix = \boldsymbol{I}_b(1,1)$, $Iy = \boldsymbol{I}_b(2,2)$, $Iz = \boldsymbol{I}_b(3,3)$. Define the state vector to be

$$\boldsymbol{X} = [q_1, q_2, q_3, q_4, w_x, w_y, w_z, c_x, c_y, c_z]^T$$

The dynamics equation is

$$X = f(x) + u + \varepsilon$$

where $\boldsymbol{\varepsilon}$ is the noise vector,

$$\boldsymbol{f} = \begin{bmatrix} 0.5w_z q_2 - 0.5w_y q_3 + 0.5w_x q_4 \\ -0.5w_z q_1 + 0.5w_x q_3 + 0.5w_y q_4 \\ 0.5q_1 w_y - 0.5q_2 w_x + 0.5q_4 w_z \\ -0.5q_1 w_x - 0.5q_2 w_y - 0.5q_3 w_z \\ -\dfrac{I_z - I_y}{I_x w_y w_z} \\ -\dfrac{I_x - I_z}{I_x w_z w_x} \\ -\dfrac{I_y - I_x}{I_z w_x w_y} \\ 0 \\ 0 \\ 0 \end{bmatrix}$$

$\boldsymbol{u}$ is the control vector

$$\boldsymbol{u} = [0,0,0,0,\frac{T_x}{I_x},\frac{T_y}{I_y},T_z,I_z]^T$$

The linearization matrix of $\boldsymbol{f}$ is

$$\boldsymbol{F} = \frac{\partial \boldsymbol{f}}{\partial \boldsymbol{x}} = \begin{bmatrix} 0 & 0.5w_z & -0.5w_y & 0.5w_x & 0.5q_4 & -0.5q_3 & 0.5q_2 & 0 & 0 & 0 \\ -0.5w_z & 0 & 0.5w_x & 0.5w_y & 0.5q_3 & 0.5q_4 & -0.5q_1 & 0 & 0 & 0 \\ 0.5w_y & -0.5w_x & 0 & 0.5w_z & -0.5q_2 & 0.5q_1 & 0.5q_4 & 0 & 0 & 0 \\ -0.5w_x & -0.5w_y & -0.5w_z & 0 & -0.5q_1 & -0.5q_2 & -0.5q_3 & 0 & 0 & 0 \\ 0 & 0 & 0 & 0 & 0 & -\frac{I_z - I_y}{I_x w_z} & -\frac{I_z - I_y}{I_x w_y} & 0 & 0 & 0 \\ 0 & 0 & 0 & 0 & -\frac{I_x - I_z}{I_y w_z} & 0 & -\frac{I_x - I_z}{I_y w_x} & 0 & 0 & 0 \\ 0 & 0 & 0 & 0 & -\frac{I_y - I_x}{I_z w_y} & -\frac{I_y - I_x}{I_z w_x} & 0 & 0 & 0 & 0 \\ 0 & 0 & 0 & 0 & 0 & 0 & 0 & 0 & 0 & 0 \\ 0 & 0 & 0 & 0 & 0 & 0 & 0 & 0 & 0 & 0 \\ 0 & 0 & 0 & 0 & 0 & 0 & 0 & 0 & 0 & 0 \end{bmatrix}$$

Let $\boldsymbol{B}_b = [b_x, b_y, b_z]^T$ be the magnetometer outputs, $[gx, gy, gz]^T$ be the gyroscope outputs. The measurement equation is

$$\boldsymbol{Z} = \boldsymbol{h}(\boldsymbol{X}) + \boldsymbol{v}$$

where $\boldsymbol{Z} = [bx, by, bz, gx, gy, gz]^T$, $\boldsymbol{v}$ is the noise vector, $\boldsymbol{h}$ is

$$\boldsymbol{h} = \begin{bmatrix} (q_1^2 - q_2^2 - q_3^2 + q_4^2)b_{i_x} + 2(q_1q_2 + q_3q_4)b_{i_y} + 2(q_1q_3 - q_2q_4)b_{i_z} \\ 2(q1q2 - q3q4)b_{i_x} + (-q1^2 + q2^2 - q3^2 + q4^2)b_{i_y} + 2(q1q4 + q2q3)b_{i_z} \\ 2(q1q3 + q2q4)b_{i_x} + 2(q2q3 - q1q4)b_{i_y} + (-q1^2 - q2^2 + q3^2 + q4^2)b_{i_z} \\ w_x + c_x \\ w_y + cy \\ wz + cz \end{bmatrix}$$

The linearization matrix of $\boldsymbol{h}$ is

$$\boldsymbol{H} = \frac{\partial \boldsymbol{h}}{\partial \boldsymbol{x}}$$
$$= \begin{bmatrix} 2(q_1b_{i_x} + q_2b_{i_y} + q_3b_{i_z}) & -2(q_2b_{i_x} - q_1b_{i_y} + q_4b_{i_z}) & -2(q_3b_{i_x} - q_4b_{i_y} - q_1b_{i_z}) & 2(q_4b_{i_x} + q_3b_{i_y} - q_2b_{i_z}) & 0 & 0 & 0 & 0 & 0 & 0 \\ 2(q_2b_{i_x} - q_1b_{i_y} + q_4b_{i_z}) & 2(q_1b_{i_x} + q_2b_{i_y} + q_3b_{i_z}) & -2(q_4b_{i_x} + q_3b_{i_y} - q_2b_{i_z}) & -2(q_3b_{i_x} - q_4b_{i_y} - q_1b_{i_z}) & 0 & 0 & 0 & 0 & 0 & 0 \\ 2(q_3b_{i_x} - q_4b_{i_y} - q_1b_{i_z}) & 2(q_4b_{i_x} + q_3b_{i_y} - q_2b_{i_z}) & 2(q_1b_{i_x} + q_2b_{i_y} + q_3b_{i_z}) & 2(q_2b_{i_x} - q_1b_{i_y} + q_4b_{i_z}) & 0 & 0 & 0 & 0 & 0 & 0 \\ 0 & 0 & 0 & 0 & 1 & 0 & 0 & 1 & 0 & 0 \\ 0 & 0 & 0 & 0 & 0 & 1 & 0 & 0 & 1 & 0 \\ 0 & 0 & 0 & 0 & 0 & 0 & 1 & 0 & 0 & 1 \end{bmatrix}$$

where $b_{i_x}, b_{i_y}, b_{i_z}$ is calculated from the *IGRF* model and the BD/GPS orbit determination results. The filtering equation is

$$\widehat{\boldsymbol{X}}_{k+1,k} = \widehat{\boldsymbol{X}}_k + \boldsymbol{f}(\widehat{\boldsymbol{X}}_k)T$$
$$\boldsymbol{P}_{k+1,k} = \boldsymbol{\Phi}_{k+1,k}\boldsymbol{P}_k\boldsymbol{\Phi}_{k+1,k}^T + \boldsymbol{Q}_k$$

$$\boldsymbol{K}_{k+1} = \boldsymbol{P}_{k+1,k}\boldsymbol{H}_{k+1}^T(\boldsymbol{H}_{k+1}\boldsymbol{P}_{k+1,k}\boldsymbol{H}_{k+1}^T + \boldsymbol{R}_{k+1})^{-1}$$
$$\boldsymbol{P}_{k+1} = (\boldsymbol{I} - \boldsymbol{K}_{k+1}\boldsymbol{H}_{k+1})\boldsymbol{P}_{k+1,k}(\boldsymbol{I} - \boldsymbol{K}_{k+1}\boldsymbol{H}_{k+1})^T + \boldsymbol{K}_{k+1}\boldsymbol{R}_{k+1}\boldsymbol{K}_{k+1}^T$$
$$\widehat{\boldsymbol{X}}_{k+1} = \widehat{\boldsymbol{X}}_{k+1,k} + \boldsymbol{K}_{k+1}[\boldsymbol{Z}_{k+1} - \boldsymbol{h}(\widehat{\boldsymbol{X}}_{k+1,k}, t_{k+1})]$$

where $T = 0.25\text{s}$ is the filtering period, $\boldsymbol{\Phi}_{k+1,k} = \boldsymbol{I} + T\boldsymbol{F}_k$, $\boldsymbol{Q}_k$ is the system noise set to be

$$\boldsymbol{Q}_k = diag(10^{-11}, 10^{-11}, 10^{-11}, 10^{-11}, 10^{-9}, 10^{-9}, 10^{-9}, 2 \times 10^{-12}, 2 \times 10^{-12}, 2 \times 10^{-12})$$

$\boldsymbol{R}_k$ is the measurement noise set to be

$$\boldsymbol{R}_k = diag(10^{-10}, 10^{-10}, 10^{-10}, 0,0,0)$$

The filtering yields the inertial attitude and the inertial angular velocity, that should be transformed to the orbital ones. BD/GPS orbit determination can generate the orbital frame quaternion relative to the inertial frame $\boldsymbol{Q}_{oi}$, and the orbital frame's angular velocity $\boldsymbol{w}_{oi}$. Then, the orbital attitude and angular velocity are calculated by

$$\boldsymbol{Q}_{bo} = \boldsymbol{Q}_{oi}' \otimes \boldsymbol{Q}_{bi}$$

$$\boldsymbol{w}_{bo} = \boldsymbol{w}_{bi} - \boldsymbol{C}_{bo}\boldsymbol{w}_{oi}$$

where $\boldsymbol{C}_{bo}$ is the correspondent transformation matrix of $\boldsymbol{Q}_{bo}$.

**Star Sensor and Gyroscope Attitude Determination**

The star sensor and the gyroscope can provide respectively the attitude information and the angular velocity information. Also, the star sensor and gyroscope filtering can be used to generate a higher precision. Now, the filtering technique is adopted and is explained as follows.

Define the state vector to be

$$\boldsymbol{X} = [q_1, q_2, q_3, q_4, c_x, c_y, c_z]^T$$

The dynamics equation is

$$X = f(x) + \varepsilon$$

where $\boldsymbol{\varepsilon}$ is the noise vector,

$$\boldsymbol{f} = \begin{bmatrix} 0.5q_2(g_z - c_z) - 0.5q_3(g_y - c_y) + 0.5q_4(g_x - c_x) \\ -0.5q_1(g_z - c_z) + 0.5q_3(g_x - c_x) + 0.5q_4(g_y - c_y) \\ 0.5q_1(g_y - c_y) - 0.5q_2(g_x - c_x) + 0.5q_4(g_z - c_z) \\ 0 \\ 0 \\ 0 \end{bmatrix}$$

The linearization matrix of $\boldsymbol{f}$ is

$$\boldsymbol{F} = \frac{\delta \boldsymbol{f}}{\delta \boldsymbol{x}} =$$

$$\begin{bmatrix} 0 & 0.5(g_z - c_z) & -0.5(g_y - c_y) & 0.5(g_x - c_x) & -0.5q_4 & 0.5q_3 & -0.5q_2 \\ -0.5(g_z - c_z) & 0 & 0.5(g_x - c_x) & 0.5(g_y - c_y) & -0.5q_3 & -0.5q_4 & 0.5q_1 \\ 0.5(g_y - c_y) & -0.5(g_x - c_x) & 0 & 0.5(g_z - c_z) & 0.5q_2 & -0.5q_1 & -0.5q_4 \\ -0.5(g_x - c_x) & -0.5(g_y - c_y) & -0.5(g_z - c_z) & 0 & 0.5q_1 & 0.5q_2 & 0.5q_3 \\ 0 & 0 & 0 & 0 & 0 & 0 & 0 \\ 0 & 0 & 0 & 0 & 0 & 0 & 0 \\ 0 & 0 & 0 & 0 & 0 & 0 & 0 \end{bmatrix}$$

Let the star sensor output be $Q_{bi} = [q_1, q_2, q_3, q_4]^T$. The measurement equation is

$$\boldsymbol{Z} = \boldsymbol{h}(\boldsymbol{X}) + \boldsymbol{v}$$

where $Z = [q_1, q_2, q_3, q_4]^T$, $\boldsymbol{v}$ is the noise vector, $\boldsymbol{h}$ is

$$h = [q_1, q_2, q_3, q_4, 0,0,0]^T$$

The linearization matrix of $\boldsymbol{h}$ is

$$\boldsymbol{H} = \frac{\partial \boldsymbol{h}}{\partial \boldsymbol{x}} = \begin{bmatrix} 1 & 0 & 0 & 0 & 0 & 0 & 0 \\ 0 & 1 & 0 & 0 & 0 & 0 & 0 \\ 0 & 0 & 1 & 0 & 0 & 0 & 0 \\ 0 & 0 & 0 & 1 & 0 & 0 & 0 \end{bmatrix}$$

The filtering equation is

$$\widehat{\boldsymbol{X}}_{k+1,k} = \widehat{\boldsymbol{X}}_k + \boldsymbol{f}(\widehat{\boldsymbol{X}}_k)T$$

$$\boldsymbol{P}_{k+1,k} = \boldsymbol{\Phi}_{k+1,k}\boldsymbol{P}_k\boldsymbol{\Phi}_{k+1,k}^T + \boldsymbol{Q}_k$$

$$\boldsymbol{K}_{k+1} = \boldsymbol{P}_{k+1,k}\boldsymbol{H}_{k+1}^T(\boldsymbol{H}_{k+1}\boldsymbol{P}_{k+1,k}\boldsymbol{H}_{k+1}^T + \boldsymbol{R}_{k+1})^{-1}$$

$$\boldsymbol{P}_{k+1} = (\boldsymbol{I} - \boldsymbol{K}_{k+1}\boldsymbol{H}_{k+1})\boldsymbol{P}_{k+1,k}(\boldsymbol{I} - \boldsymbol{K}_{k+1}\boldsymbol{H}_{k+1})^T + \boldsymbol{K}_{k+1}\boldsymbol{R}_{k+1}\boldsymbol{K}_{k+1}^T$$

$$\widehat{\boldsymbol{X}}_{k+1} = \widehat{\boldsymbol{X}}_{k+1,k} + \boldsymbol{K}_{k+1}[\boldsymbol{Z}_{k+1} - \boldsymbol{h}(\widehat{\boldsymbol{X}}_{k+1,k}, t_{k+1})]$$

where $T = 0.25s$ is the filtering period, $\boldsymbol{\Phi}_{k+1,k} = \boldsymbol{I} + T\boldsymbol{F}_k$, $\boldsymbol{Q}_k$ is the system noise set to

$$\boldsymbol{Q}_k = diag(10^{-13},10^{-13},10^{-13},10^{-13},10^{-13},10^{-13},10^{-13})$$

$\boldsymbol{R}_k$ is the measurement noise set to be

$$\boldsymbol{R}_k = diag(4 \times 10^{-8}, 4 \times 10^{-8}, 4 \times 10^{-8}, 4 \times 10^{-8})$$

Also, the inertial attitude and the inertial angular velocity should be transformed to the orbital ones using the orbit determination information.

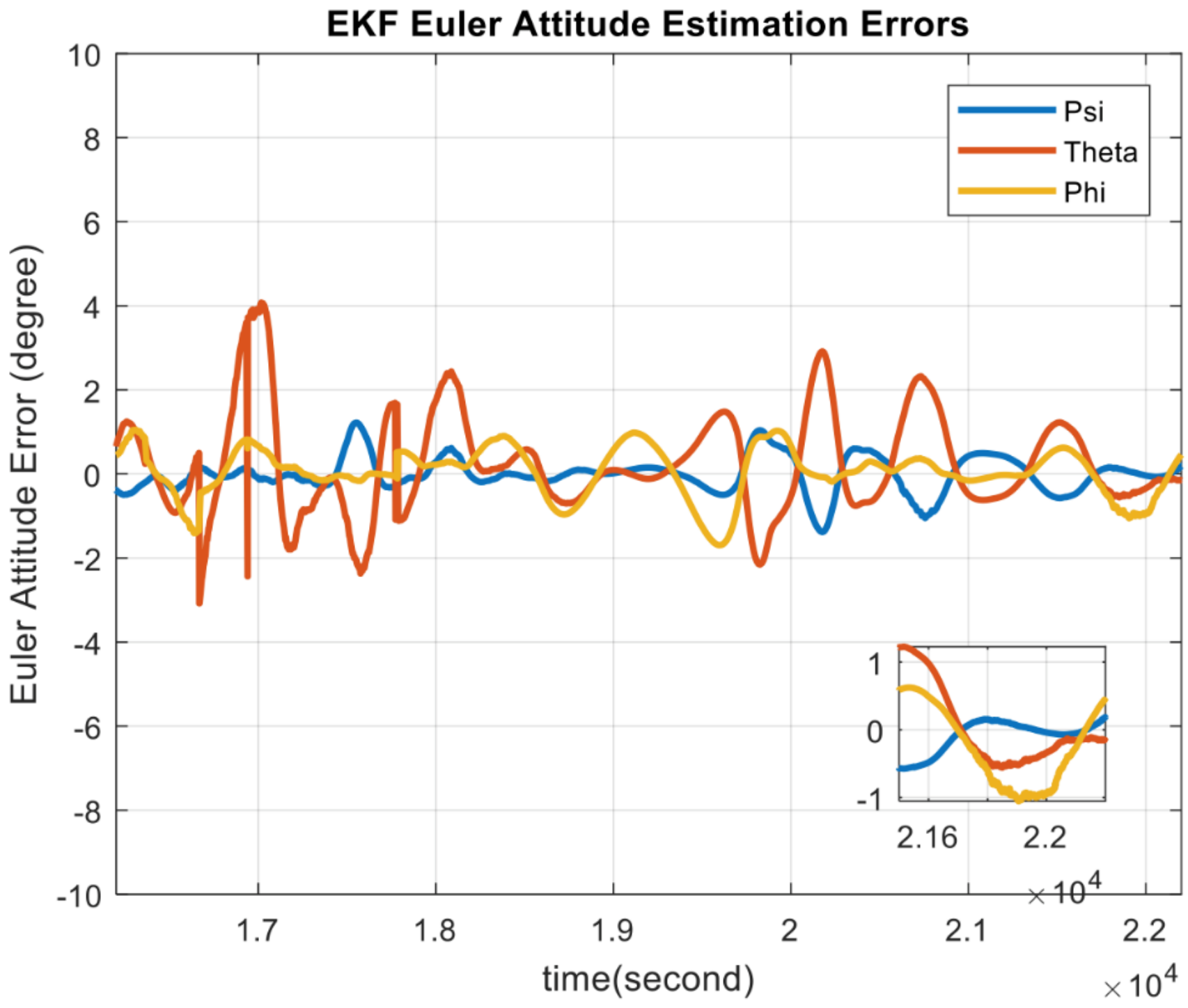

*Figure 2-7 The EKF Euler angles estimation errors during the orbit 3 up to orbit 4.*

**TRIAD**

The TRIAD estimator is a simple geometric solution for attitude determination problem. In order to estimate the orientation of one reference frame with respect to the other, the TRIAD method requires components of two vector pairs represented in the two reference frames. For the satellite attitude determination using the TRIAD method, two measured vectors in the BRC frame and their respective modelled vectors in the ORC frame are used to calculate a DCM. This DCM describes the orientation of the BRC frame relative to the ORC frame, or in other words the attitude of the satellite.

The first step to implement the TRIAD estimator is to construct two triads of orthogonal unit vectors. The first triad is constructed from two different measured vectors in BRC frame. Let $b_1$ and $b_2$ be the two vectors measured by two attitude sensors then the first triad of unit vectors $m_1$, $m_2$ and $m_3$ is determined as follows

$$m_1=\frac{b1}{||b1||}$$

$$m_2=\frac{b1 \times b2}{||b1 \times b2||}$$

$$m_3= m_1 \times m_2$$

The vector $b_1$ that is used to calculate the first unit vector $m_1$ in the triad is referred as the anchor vector. For better estimation accuracy, the output vector of the more accurate sensor should be chosen as the anchor. The second triad of unit vectors $r_1$, $r_2$ and $r_3$, is formed by two reference vectors $o_1$ and $o_2$ which are in fact the model vectors in the ORC corresponding to $b_1$ and $b_2$ respectively.

$$r_1=\frac{o1}{||o1||}$$

$$r_2=\frac{o1 \times o2}{||o1 \times o2||}$$

$$r_3= r_1 \times r_2$$

In the second step of the TRIAD implementation, the estimated DCM **C** is calculated from the two triads formed in the preceding step. Finally, the elements of the DCM **C** are then used to find the current estimated quaternion $\hat{q}$.

$$C = m_1 r_1^T + m_2 r_2^T + m_3 r_3^T$$

The Triad estimator is easy to implement and computationally less expensive, but its outputs are noisy for noisy sensor measurements.

For the satellite attitude determination using the TRIAD method, two measured vectors in the Body frame and their respective modelled vectors in the Orbital frame are used to calculate a DCM. This DCM describes the orientation of the Body frame relative to the Orbitak frame, or in other words the attitude of the satellite. The magnetometers and Sun sensors are used to calculate the DCM. Figure below indicates the attitude estimates from a TRIAD algorithm using the Sun Sensor and the magnetometer data. The TRIAD estimates are noisy and not available during the eclipse; therefore, these estimates are only used for EKF initializations in the mission ADCS. Singular Value Decomposition (SVD) technique is applied to calculate measurement noise covariance matrix as an input to the EKF. SVD estimations are applied to compute measurement noise covariance matrix from DCM calculated in Triad algorithm. So, the elements of “Rotation angle error covariance matrix” calculated for the SVD estimations are used in the EKF as the measurement noise covariance values. An EKF is designed to obtain the satellite’s angular motion parameters with the desired accuracy based on the data acquired from Triad.

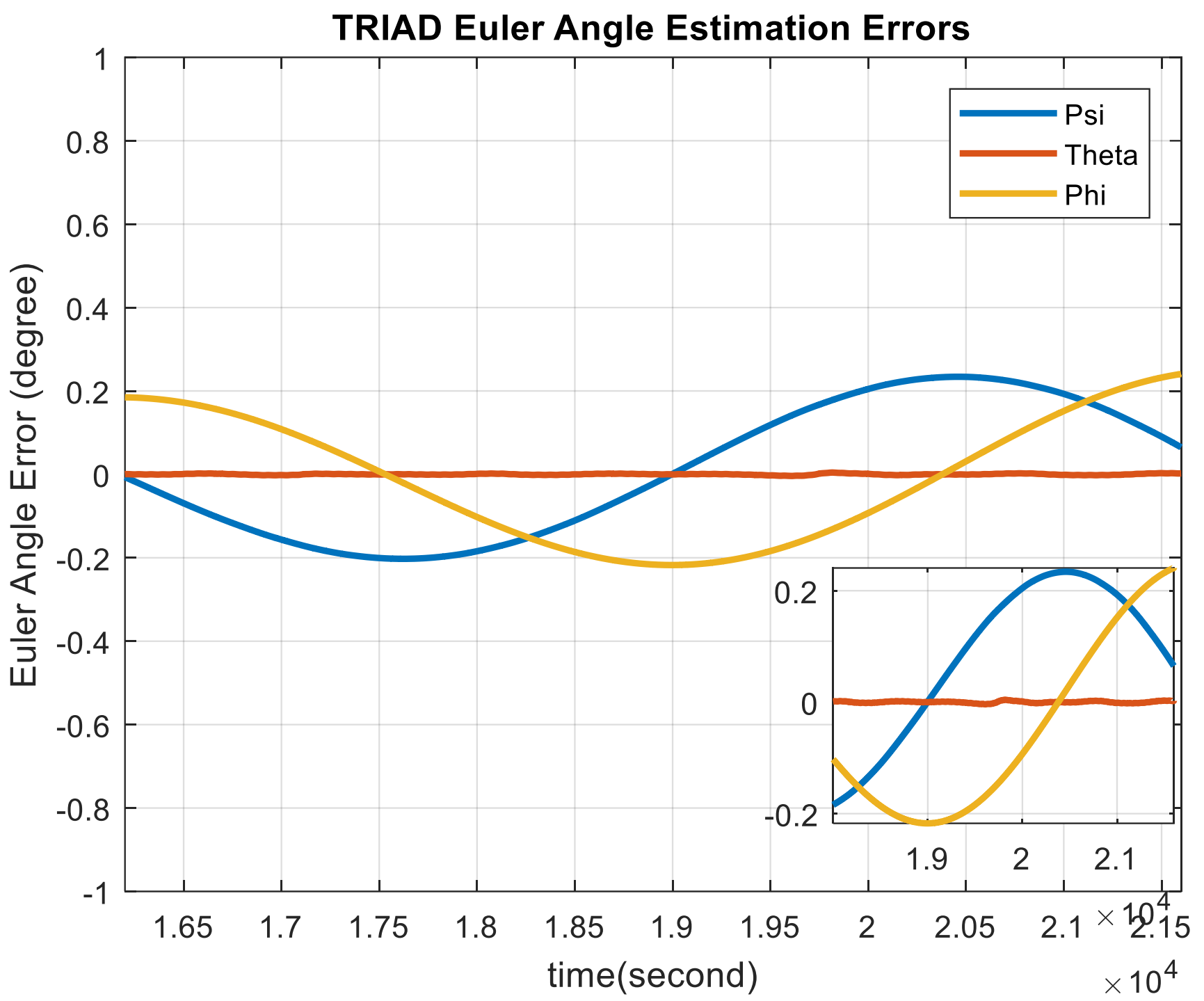

*Figure 2-9 The TRIAD Attitude estimates errors.*

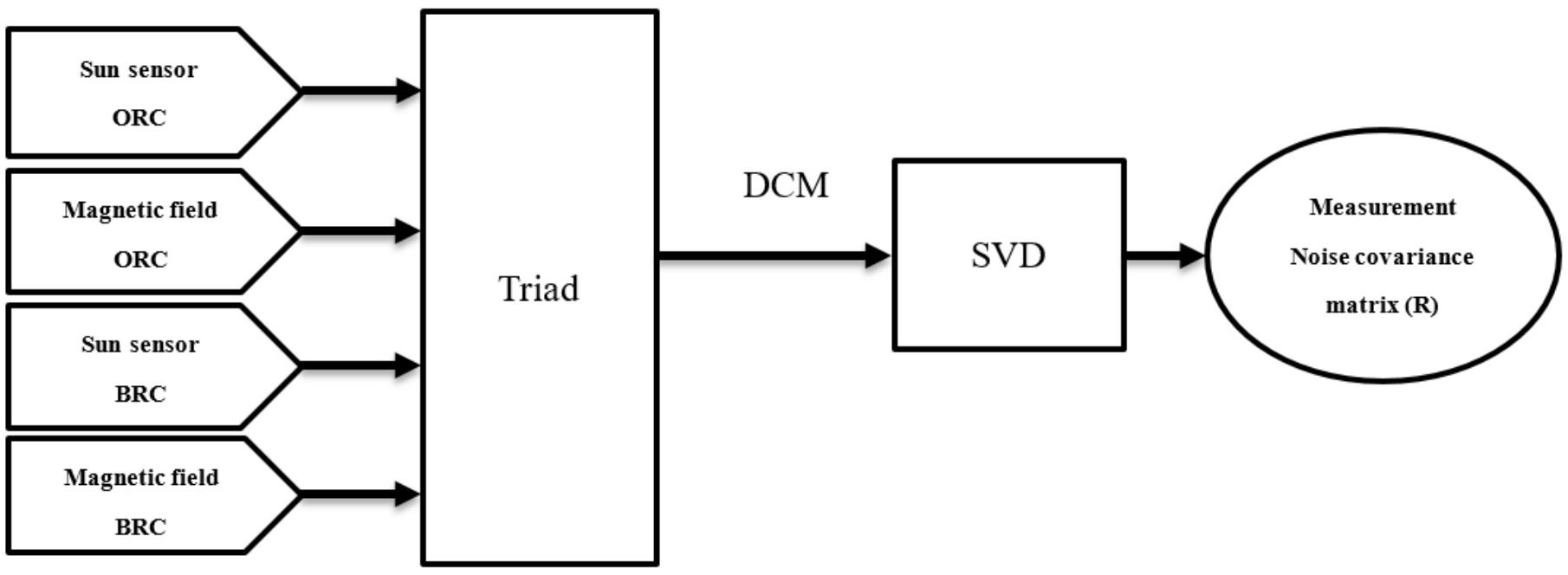

*Figure 2-8 Schematic of computing measurement noise covariance matrix from TRIAD.*

## 2.5 Diturbances and Eclipse

The following calculations in determination of the environment disturbances are presented for the worst circumstances. The obtained disturbances are used to estimate the initial characteristics of the actuators. Moreover, the performance of the actuators is simulated and evaluated in the presence of actual time-varying disturbance torques.

Table 2-1 Disturbance torques imposed on the SSS-1.

| Type | Torque | Considerations |
|---|---|---|
| **Aerodynamic Drag** | $T_a = 6.51 \times 10^{-6}\ \ Nm$ | • $T_a = \frac{1}{2}\rho C_d A_r V^2 (C_{p_a} - C_m)$<br>• $\rho = 3.76 \times 10^{-12}\ \frac{kg}{m^3}\ A_r = 0.23\ m^2$<br>• $V = 7612\ \frac{m}{s}\ C_{p_a} - C_m = 0.1\ \ C_d = 2.6$ |
| **Solar Radiation** | $T_s = 8.81 \times 10^{-7}\ \ Nm$ | • $T_s = \frac{\Phi}{c} A_s (1+q)\left(C_{p_s} - C_m\right) cos\varphi$<br>• $\Phi = 1367\ \frac{W}{m^2}\ \ c = 3 \times 10^8\ \frac{m}{s}$<br>• $A_s = 1.21\ m^2\ C_{p_s} - C_m = 0.1\ \ q = 0.6$ |
| **Gravity Gradient** | $T_g = 8.13 \times 10^{-6}\ \ Nm$ | • $T_g = \frac{3\mu}{2R^3}\ (I_{max} - I_{min}) \sin(2\theta)$<br>• $\mu = 3.986 \times 10^{14}\ \frac{m^3}{s^2}\ \ R = 6878137\ m$<br>• $\theta = 30°\ \ I_{max} - I_{min} = 5.108\ \frac{kg}{m^2}$ |
| **Magnetic Field** | $T_m = 2.40 \times 10^{-5}\ \ Nm$ | • $T_m = \frac{DM\lambda}{R^3}$<br>• $D = 0.5\ \ A.m^2\ \ R = 6878137\ m$<br>• $M = 7.8 \times 10^{15}\ \ Tesla.m^3\ \ \lambda = 2$ |
| **Total** | $T_{dis} = 3.952 \times 10^{-5}\ \ Nm$ | |

Being capable of determining when the satellite is in eclipse, gives the possibility of making more realistic sun sensor models and more accurate Sun vector measurements. Eclipse indication also helps the attitude estimation, because it is then possible to disregard sun sensor measurements during periods, where the satellite is in eclipse. The satellite will experience the light, umbra and penumbra periods when it is moving around the Earth. The minimum, maximum, mean and total duration of sunlight, penumbra and umbra periods can be calculated based on the orbital parameters. The satellite will be 63.3% of its time in sunlight and 36.4% in umbra periods. As expected, the penumbra periods are very short due to the low height of the orbit.

Table 2-2 eclipse condition analysis for SSS-1P.

| **Lighting conditions analysis result** | |
|---|---|
| **Period** | Mean Duration |
| **Sunlight** | 3595s |
| **Penumbra** | 18s |
| **Umbra** | 2067s |
| **Total** | 5680s |

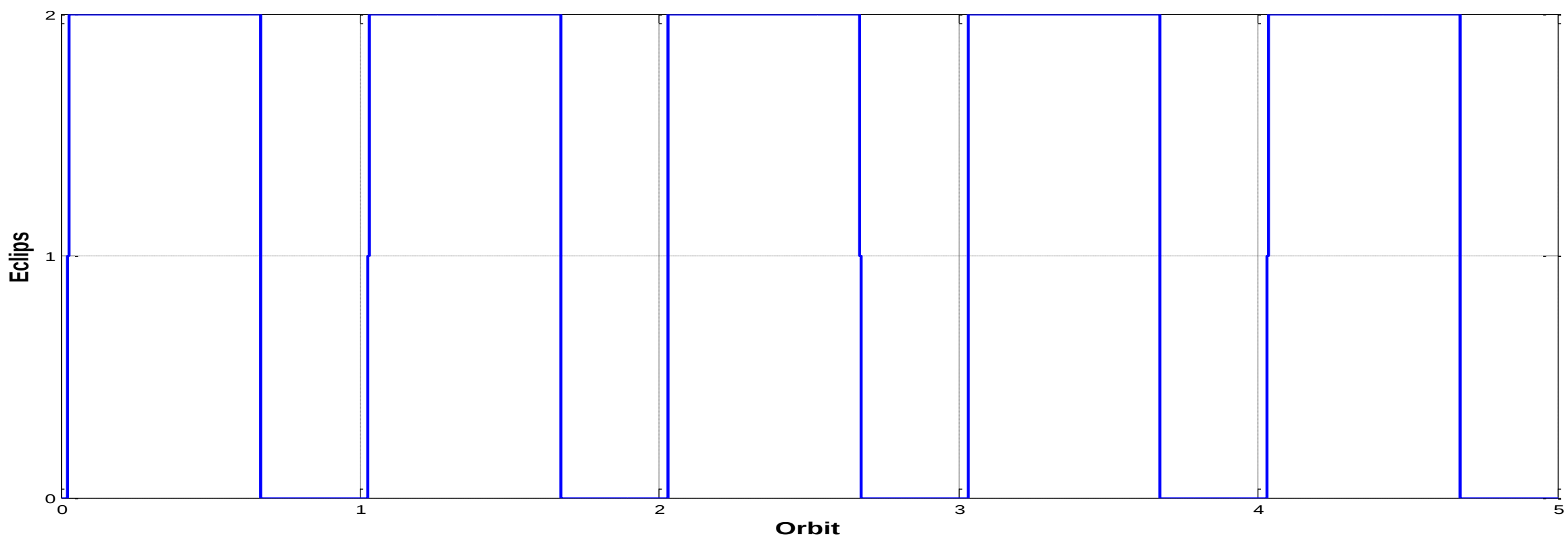

Figure 2-10 eclipse condition analysis for SSS-1P.

## 2.6 ADCS modes

### Idle Mode

The Idle Mode refers to the period, when there is no control activated. In the mission ADCS scenario, the satellite is in the Idle Mode after separation from the launch vehicle and tumbles with the rates $[10, 10, 10]$ *deg/s.* The RKF is initialized in the Idle Mode so that it may converge before activating the magnetic control.

### Safe Mode

The Safe Mode is also referred as the detumbling mode and is activated to damp the body rates, when the satellite is released from the launch vehicle. The magnetic B-dot controller is activated during the detumbling mode. As mentioned, the controller does not require any rate or attitude information and is fully based on the magnetometer readouts. However, for telemetry purposes, the rate information is taken from the RKF. Once the satellite body rates are less than *0.3°/s,* the satellite is taken into the stable spin mode by activating Magnetic Control Mode-2. The Safe Mode shall be activated at any instant in the mission orbit, if the satellite body rates exceed a threshold of *5°/s.*

### Nominal Mode

In this mode the satellite is toward nadir pointing and magnetometer and momentum wheel are working as actuators. This mode carries out till satellite start to imaging from target area. During this mode, EKF are performs as the Estimator method.

**Imaging Mode**

The RW control in Mode-2 is used with the Estimator mode 4 during the Imaging Mode. The sensor updates from all available sensors including the sun sensor and magnetometer are used during this mode. The Imaging Mode is only activated during the sunlit part of the orbit.

**Momentum Management**

The magnetic control in Mode 4 is used for unloading the RW angular momentum during the non-imaging part of orbit in the Nominal Mode. The attitude and the rate information are taken from the EKF in the Estimation Mode 4 with only the magnetometer updates available.

## 2.7 Initial Orbital Condition

The orbit is SSO with altitude of 500km (which can be adjusted with respect to piggy-back launch conditions). Here it is given 500km for related calculations. The local time of descending node is 10:30 AM. The launch time has not been defined at this stage.

Table 2-3 initial orbital parameters of SSS-1.

| **Type** | Sun Synchronous Orbit |
|---|---|
| **Altitude** | ~ 500 km |
| **Local time of descending node** | 10:30AM |
| **Semi-major axis** | 6878.14km |
| **Inclination** | 97.4065° |
| **Eccentricity** | 0 |
| **Perigee argument** | 0° |
| **RAAN** | 136.715° |
| **True anomaly** | 0° |
| **Launch site** | Taiyuan, China<br>(111°36'30.59"E, 38°50'56.71"N) |

## 2.8 Nadir Pointing

In the imaging process, it is required the satellite have a best performance. In this mode, the magnetorquer and momentum wheel will be used as actuators. For better performance, momentum wheel is switched to reaction wheel to apply required torque to the system. This mode can prepare the attitude accuracy of 0.05 degree according to statistical results. Low torques for precise fine pointing are applied by means of Reaction Wheel.

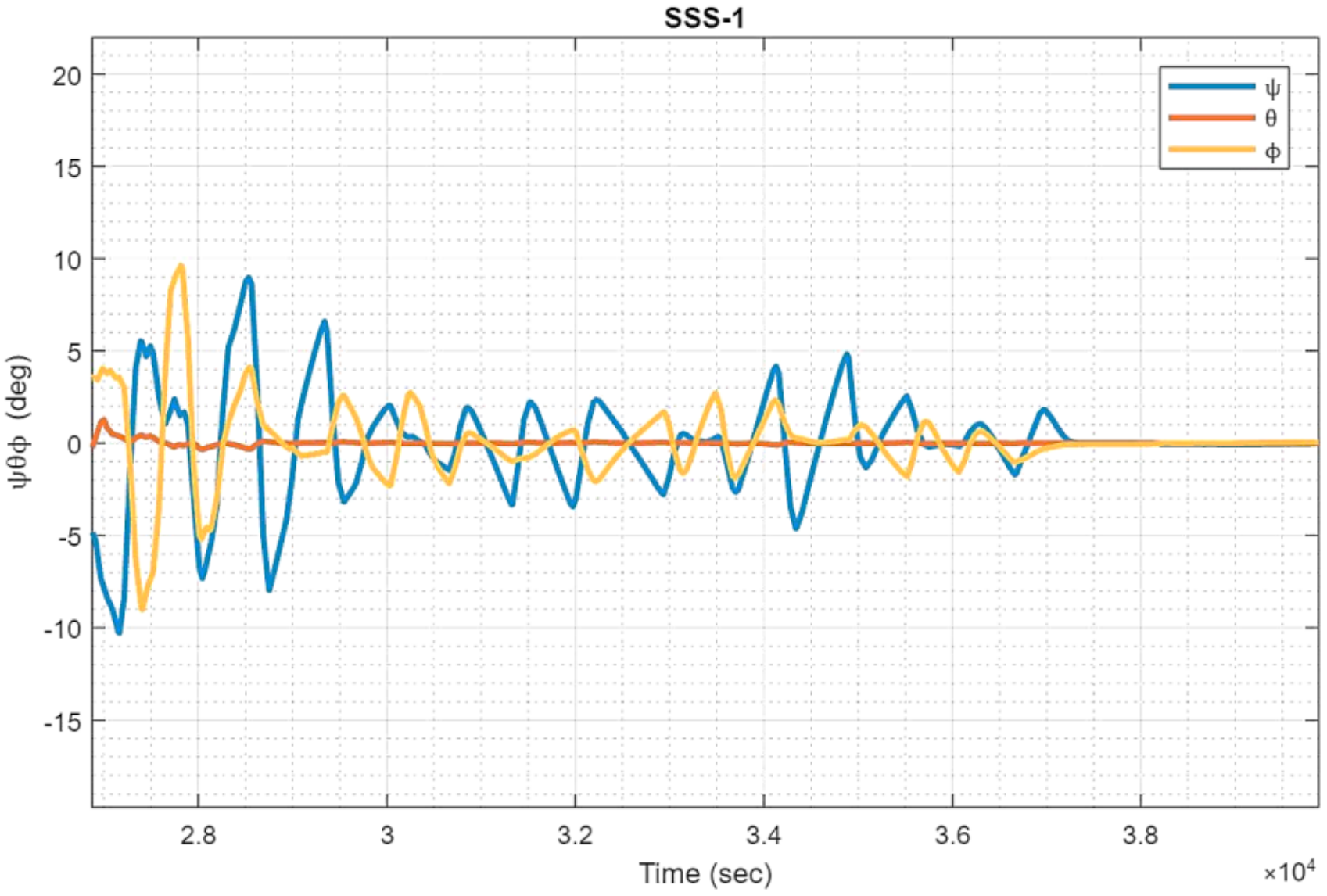

Figure 2-11 Time history of the Euler angles in nadir pointing mode.

## 2.9 Boom Deployment

In order to increase stability of the satellite in its mission, gravity gradient boom is deployed. When the nadir pointing accuracy reaches 5 degrees, the deployment mechanism becomes enable. In this process, the moment of inertia of the body in two lateral axes are increased. The disturbance torque generated during deployment will be cancelled out by the Magnetic torquers and momentum wheel.

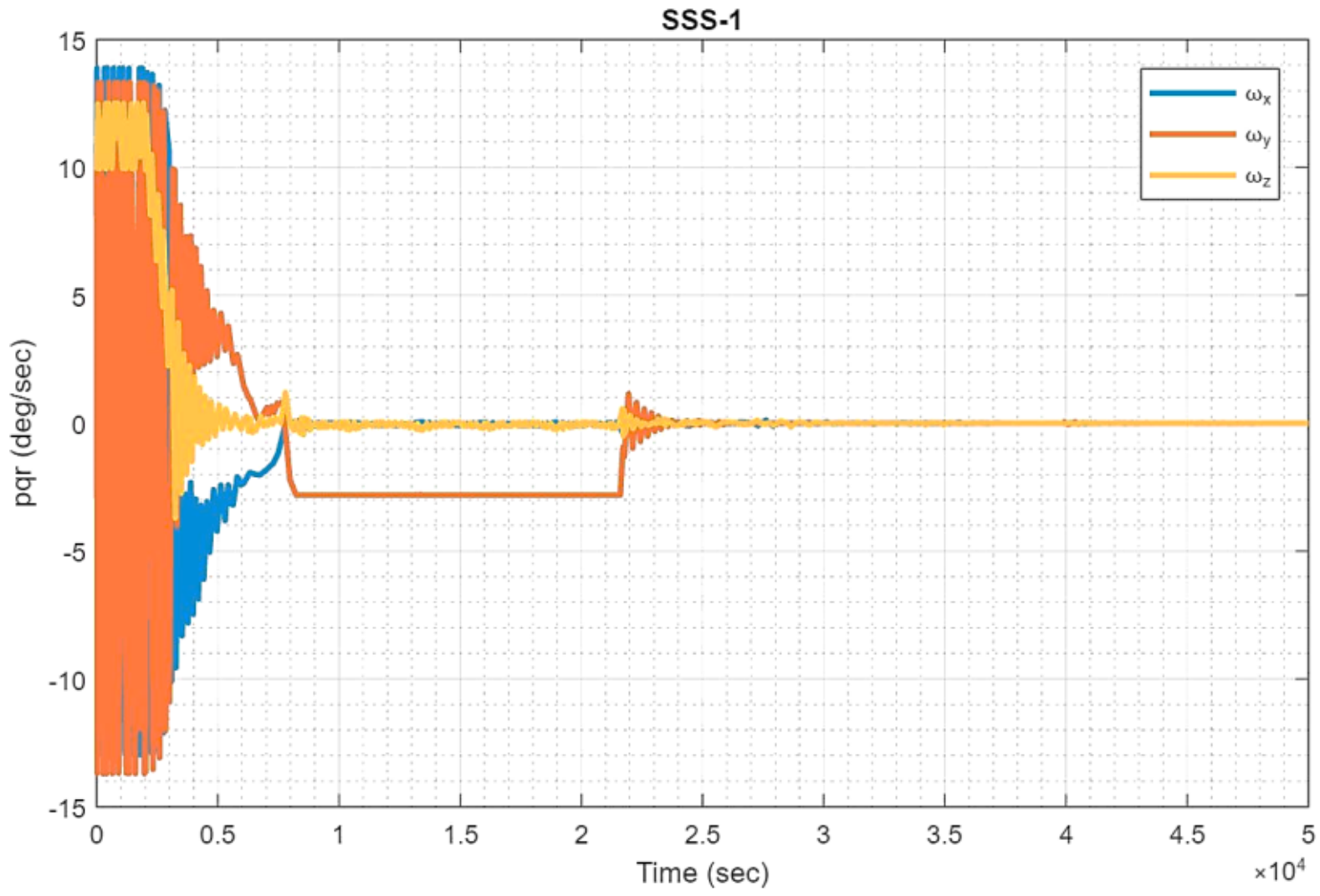

Figure 2-12 Time history of the angular velocity from safe mode to boom deployment.

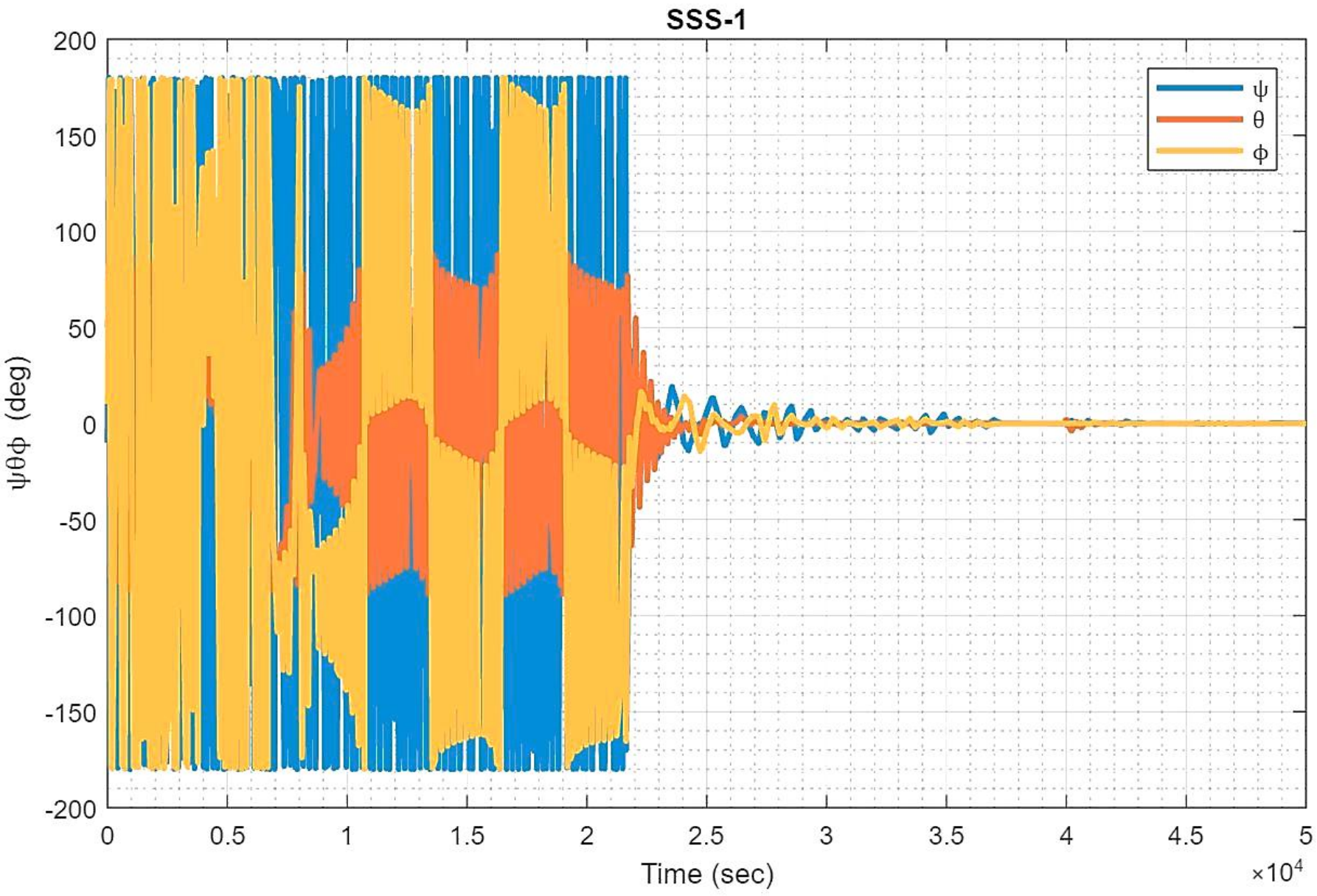

*Figure 2-13 Time history of the Euler angles from safe mode to boom deployment.*

## 2.10 Verification of the SIL test

In order to verify the simulation results with mission requirement, the below table has been provided and simulation results are compared to desired condition.

Table 2-4 statistical results of ADCS modes.

| **Item** | **System** | **ADCS design** |
|---|---|---|
| Accuracy (deg) | | |
| Normal Mode<br>Optional Mode | 5 deg<br>1 deg | 2<br>0.5 |
| Stability (deg/s) | | |
| Normal Mode<br>Optional Mode | 0.1<br>0.01 | 0.05<br>0.01 |

It's evident that, simulation results can address all the requirement successfully. Furthermore, these results demonstrate that the scenario that has been considered and designed for this mission has appropriate performance.

# 3. Design of Software

The majority of the software testing will occur after the satellite models has been integrated. This is due to the fact the software will deal with the onboard computer system and needs to be integrated into the rest of the satellites software before testing can being. The first test to occur will be testing the conversion of the raw data from the magnetometer and the rate gyros. These will be tested like mentioned in magnetorquer test plan but instead of using the raw values a read out from the onboard control unit will be used. To test the orbit propagator and the magnetic field model, hardware simulators will be used in place of the magnetometer and the rate gyros. The hardware simulators will generate measurements that SSS-1 is expected to sense in the orbit.

After the design, modeling and desktop simulation of attitude determination and control system, we need to plan a scenario to manufacture and test the final product step by step and systematically. The first step towards this goal is providing a real time simulation platform such that the simulation subsystems can be replaced with the corresponding real components. The next step is to convert the MATLAB/SIMULINK simulation codes to a lower level programing language (C/C++) applicable to the ADCS onboard computer.

Now the important test setup that can be developed are processor in the loop (PIL), software in the loop (SIL) and Hardware in the loop (HIL) platforms. The purpose of the HIL test is to check the performance of the provided package while working in the loop in simulation on the computer and all the packages are provided in C language. The purpose of the HIL test bed is to evaluate the performance of the attitude sensors and control actuates in the simulation loop one by one and all together as an integrated system. The purpose of the PIL test bed is to evaluate the performance of the systems software on the onboard computer in the simulation loop. In other words, we keep the satellite and space environment dynamic models in the real time desktop simulation while the attitude determination and control models are implemented on the onboard processor which is substituted in the simulation loop. Now, we will be able to apply all the ADCS components as well as the onboard processor in the HIL and SIL test platforms.

Simulating flight phase development and performance characteristics of space in a laboratory environment in order to test individual ADCS components and integrated system is another important but challenging step in verifying and validating the satellite's ADCS design.

The engineering or experimental model can be used to debug the functional test environment and the test procedures that will be used for the flight models. Using an engineering model also has the advantages of verifying the electrical ground support equipment interfaces and functionality as well as the satellite's compatibility with the ground segment. A possible software and hardware testing approach is outlined according to the flowchart shown in figure below.

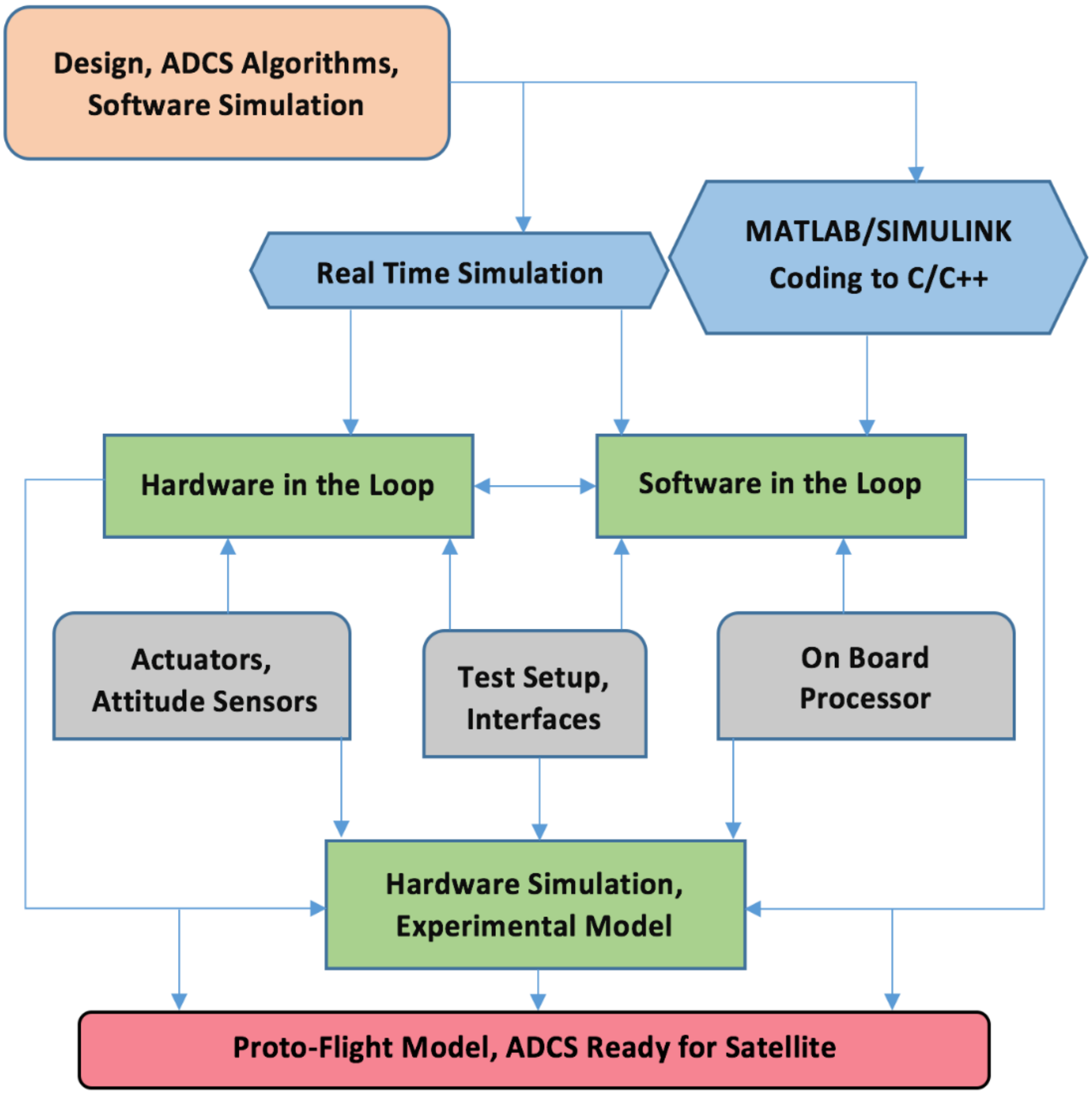

Figure 3-1 Five level architecture to test ADCS performance: 1- Software in the Loop, 2- Processor in the Loop, 3- Hardware in the Loop, 4- Experimental Model, 5- Satellite Flight Model

## 3.1 Pseudo code for Software-In-the-Loop Test

```
Algorithm ADCS Main

main () {
        Create RunTimeData and CurrentTime members for data structures.
        Function Initialize ADCS (RunTimeData, CurrentTime);
        If ADCS initialization → failed Then {
                Print "Error 0: The Initialization of ADCS failed ...!";
```

```
        }
        FOR i → 56761 do
                Function Orbit Calculation (RunTimeData, CurrentTime);
                If Orbit Calculation → failed Then {
                        Print "Error 1: The calculation of orbital parameters failed ...!";
                }
                CurrentTime Sec ← CurrentTime Sec + 1
                Function Time Manager (CurrentTime);
                Function Attitude Determination (RunTimeData, CurrentTime);
                If Attitude Determination → failed Then {
                        Print "Error 2: The calculation of Filters are failed ...!";
                }
                Function MANAGER (RunTimeData, CurrentTime);
                If MANAGER → failed Then {
                        Print "Error 3: The ADCS manager failed ...!";
                }
                Function Dynamic Model (RunTimeData, CurrentTime);
                If Dynamic Model → failed Then {
                        Print "Error 4: The calculation of dynamic model failed...!";
                }

        Return 0;
}
```

**Algorithm Time Manager**

```
Time Manager (CurrentTime) {
                int NOD = 0;

                        If CurrentTime sec > 60 Then {
                        CurrentTime sec ← 0
                        CurrentTime min ← CurrentTime min + 1
                                If CurrentTime min > 60 Then {
                                        CurrentTime min ← 0
                                        CurrentTime hr ← CurrentTime hr + 1
                                        If CurrentTime hr > 12 Then {
                                                CurrentTime hr ← 0
                                                CurrentTime day ← CurrentTime day + 1
                                                NOD ← Function Number of Days (CurrentTime)
                                                If CurrentTime day > NOD Then {
                                                        CurrentTime day ← 0
                                                        CurrentTime mon ← CurrentTime mon + 1
                                                        If CurrentTime mon > 12 Then {
                                                                CurrentTime mon ← 0
                                                                CurrentTime year ← CurrentTime
        year + 1
                                                        }
                                                }
                                        }
                                }
                        }
}
```

**Algorithm Attitude Determination**

```
Attitude Determination (RunTimeData, CurrentTime) {
        If RunTimeData sun_sensor_flag → 1 Then {
                Function TRIAD(RunTimeData)
        } ELSE If RunTimeData sun_sensor_flag → 0 Then {
                Function EKF(RunTimeData)
        }

        Return 0;
}
```

**Algorithm MANAGER**

```
MANAGER (RunTimeData, CurrentTime) {
        SWITCH RunTimeData BZ1 {
          case 2:
                {
                        Printf "\nI'm here ...";
                        Function Control_damp(RunTimeData);
                        FOR i → 3 do {
                                RunTimeData Tc[i] ← RunTimeData Tm[i];
                                tV[i] ← RunTimeData Omega_Body2Inertia[i];
                        }
                        C = Function MaxArray(tV, 3);
                        If C < 0.2*(pi / 180) Then {
                                RunTimeData BZ1 ← 3;
                        }
                }
                        break;
                case 3:
                {
                        Function Control_point(RunTimeData);
                        FOR i → 3 do {
                                RunTimeData Tc[i] ← RunTimeData Tm[i];
                                tV[i] ← RunTimeData Omega Body2Inertia[i];
                        }
                        C ← MaxArray(tV, 3);
                        If C > 0.5*(pi / 180) Then {
                                RunTimeData BZ1 ← 2;
                        }
                }
                        break;
                case 5:
                {
                        Function Control_point (RunTimeData);
                        Function Control_MWspin (RunTimeData, CurrentTime);
                        FOR i → 3 do {
                                RunTimeData Tc[i] ← RunTimeData Tm[i] + RunTimeData Tw[i];
                        }
                }
                        break;
                case 6:
                {
                        Function Control stable (RunTimeData);
```

```
                    Function Control MWdespin (RunTimeData, CurrentTime);
                    FOR i → 3 do {
                            RunTimeData Tc[i] ← RunTimeData Tm[i] + RunTimeData Tw[i]
                    }
                    If RunTimeData.BZ6 → 0 Then {
                            RunTimeData BZ1 ← 3
                            RunTimeData BZ6 ← 1
                    }
                break;
            }
        Return 0;
}
```

The pseudo code representing the algorithm for the sections that there will be no mathematical formulation in the previous sections. The algortihms for controllers, EKF and TRIAD filters are presented in the previous sections. The full code can be found in the appendices section of this thesis.

## 3.2 Data Structures

This sub-section states witch data structures exists in the system and which subsystems who should be able to access what data. The data structure Data is used for logging and to have a defined structure to send all current data in. The mode variable needs to be readable from all tasks, it is not used in the subsystems directly.

| Structure Name | Data | Data Type | | Structure Name | Data | Data Type |
|---|---|---|---|---|---|---|
| ADCS_DATA | | | | ADCS_DATA | | |
| | CounterLoop | int32_t | | | Tm[3] | double |
| | ECI_Pos[3] | double | | | Hw | int |
| | ECI_Pos_Prev[3] | double | | | Hw_set | int |
| | ECI_Vel[3] | double | | | BZ6 | int |
| | ECI_Vel_Prev[3] | double | | | BZ1 | int |
| | rsun_Inertia[3] | double | | | FailureFlag | int |
| | rsun_Body[3] | double | | | ZT1UpdateFlag | int |
| | rmag_Inertia[3] | double | | | ZT1 | int |
| | Delta_rmag_Body[3] | double | | | Kd_damp[3][3] | double |
| | rmag_Body_prev[3] | double | | | Kp_point[3][3] | double |
| | rmag_Body[3] | double | | | Kd_point[3][3] | double |
| | DCM_B2I[3][3] | double | | | Kp_stable[3][3] | double |
| | DCM_B2O[3][3] | double | | | Kd_stable[3][3] | double |
| | DCM_B2O_Prev[3][3] | double | | | SVD[4] | double |
| | Quat_Orbit2Intertia[4] | double | | | S_time | double |
| | Quat_B2O[4] | double | | | GyroDrift[3] | double |
| | Quat_B2O_prev[4] | double | | | PK1[7][7] | double |
| | Quat_B2I[4] | double | | | QK1[7][7] | double |
| | Quat_B2I_prev[4] | double | | | XK1[7] | double |
| | Omega_Orbit2Intertia[3] | double | | | RK[3][3] | double |
| | Omega_B2I[3] | double | | | XK[7] | double |
| | Omega_B2O[3] | double | | | PK[7][7] | double |
| | Omega_B2I_prev[3] | double | | Time | | |
| | Tg[3] | double | | | year | int16_t |
| | Tc[3] | double | | | mon | int16_t |
| | Tw[3] | double | | | day | int16_t |
| | GK[3] | double | | | hr | int16_t |
| | sun_sensor_flag | int | | | min | int16_t |
| | Mc[3] | double | | | sec | double |

Figure 3-2 List of Global data structures used in the ADCS to store data.

## 3.3 The SIL and PIL Test Algorithm

The algorithm below presents the sequence of the events and the general working flowchart of the program for both Software-In-the-Loop and Processor-In-the-Loop tests. The failure analysis in a basic level provided but needs to be improved.

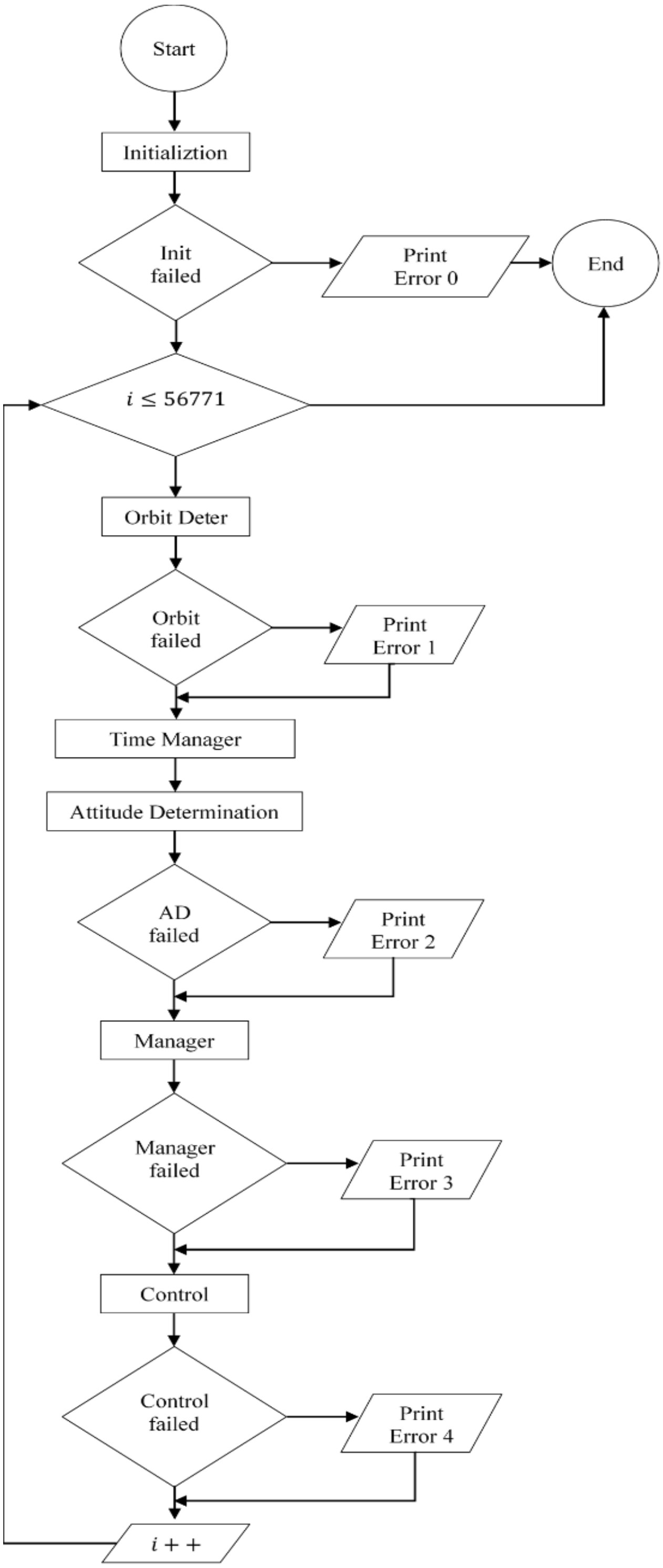

*Figure 3-3 flowchart of the program in SIL & PIL test.*

## 3.4 The APIs

| MANAGER () | | | |
|---|---|---|---|
| description | This function keeps track of time passed by breaking down the seconds passed to the new date and UTC time. | | |
| input | description | range / units | |
| TimeUTC | struct containing: | | |
| year | year | 1900 | 2100 |
| mon | month | 1 | 12 |
| day | day | 1 | 28,29,30,31 |
| hr | universal time hour | 0 | 23 |
| min | universal time min | 0 | 59 |
| sec | universal time sec | 0.000 | 59.999 |
| BZ1 | control mode | value based on the desined control mode. | |
| FailureFlag | | True = 1 = failure, False = 0 = No failure | |
| ZT1UpdateFlag | Ground stationupdate the flag. | True = 1 = updated, False = 0 = No update | |
| ZT1 | Ground stationupdate the flag. | desired value of the ground station of the control mode. | |
| coupling | Control_damp(), MaxArray(), Control_point(), Control_Mwspin(), Control_stable(), Control_Mwdespin() | | |
| dependencies | struct.h, adcs_manager.h, utils.h | | |

| TimeManager() | | | |
|---|---|---|---|
| description | This function keeps track of time passed by breaking down the seconds passed to the new date and UTC time. | | |
| input | description | range / units | |
| TimeUTC | struct containing: | | |
| sec | universal time sec | 0 | 60 |
| output | description | range / units | |
| TimeUTC | struct containing: | | |
| year | year | 1900 | 2100 |
| mon | month | 1 | 12 |
| day | day | 1 | 28,29,30,31 |
| hr | universal time hour | 0 | 23 |
| min | universal time min | 0 | 59 |
| sec | universal time sec | 0.000 | 59.999 |
| coupling | numberOfDays() | | |
| dependencies | struct.h, adcs_manager.h | | |

| printbuff() | | |
|---|---|---|
| description | This function prints the given array with the desired caption. | |
| input | description | range / units |
| *char | the desired caption t be printed before the array. | |
| *In | input array buffer. | |
| len | length of input buffer. | |
| output | | |
| prints the buffer data. | | |

| | |
|---|---|
| coupling | none. |
| dependencies | utils.h |

| Orbit () | | | |
|---|---|---|---|
| description | This function calculates and simulate the orbital parameters and situation. | | |
| input | description | range / units | |
| reci | position vector eci | km | |
| TimeUTC | struct containing: | | |
| year | year | 1900 | 2100 |
| mon | month | 1 | 12 |
| day | day | 1 | 28,29,30,31 |
| hr | universal time hour | 0 | 23 |
| min | universal time min | 0 | 59 |
| sec | universal time sec | 0.000 | 59.999 |
| ECI_Pos | position vector in eci frame | km | |
| ECI_Vel | velocity vector in eci frame | km/s | |
| ECI_Vel_Prev | previous position vector in eci frame | km/s | |
| ECI_Pos_Prev | previous position vector in eci frame | km | |
| output | | | |
| | description | range / units | |
| Quat_Orbit2Intertia | Quaternions | acceptable range for each mission phase | |
| DCM_B2I | Direct Cosine Matrix | acceptable range for each mission phase | |
| Omega_Orbit2Intertia | Angular Velocity | acceptable range for each mission phase | |
| ECI_Pos[3] | position vector in eci frame | acceptable range for each mission phase | |
| ECI_Vel[3] | velocity vector in eci frame | acceptable range for each mission phase | |
| coupling | sun(), norm(), magWmm() | | |
| dependencies | utils.h, sunPred.h, magPred.h, struct.h, orbit.h | | |

| MaxArray() | | |
|---|---|---|
| description | This function calculates the maximum value of the given array. | |
| input | description | range / units |
| *In | input array buffer. | |
| len | length of input buffer. | |
| output | | |
| MaxArray | maximum value of the given array. | |
| coupling | none. | |
| dependencies | utils.h | |

| MatProd() | | |
|---|---|---|
| description | This function calculates the result of matrices multiplication. | |
| input | description | range / units |
| *C1 | input buffer is the first matrix array (3x3). | |
| *C2 | input buffer is the second matrix array (3x3). | |
| output | | |
| C[3][3] | result of matrices multiplication. | |
| coupling | none. | |

| | |
|---|---|
| dependencies | utils.h |

| Unit() | | |
|---|---|---|
| description | This function calculates the unit vector of input vector. | |
| input | description | range / units |
| *In | input array buffer. | |
| output | | |
| *out | output buffer is the unit vector. | |
| coupling | none. | |
| dependencies | utils.h | |

| rotmtx2quat() | | |
|---|---|---|
| description | This function converts rotation matrix to quaternions. | |
| input | description | range / units |
| In[3][3] | rotation matrix. | |
| output | | |
| *q | output buffer are the Quaternions. | |
| coupling | none. | |
| dependencies | utils.h | |

| EKF() | | |
|---|---|---|
| description | This function implements EKF algorithm using noise covariance matrix and gyroscope measurements. | |
| input | description | range / units |
| SVD[4] | Singular Values Decomposition | |
| Quat_B2I[4] | Quaternions | |
| GyroDrift[3] | The gyroscope drift. | |
| rmag_Inertia[3] | magnetic field vector in inertial frame | |
| S_time | sampling time | |
| PK[7][7] | The EKF state matrices | |
| PK1[7][7] | The EKF state matrices | |
| QK1[7][7] | The EKF state matrices | |
| XK1[7] | The EKF state matrices | |
| RK[3][3] | The measurement noise Covariance Matrix RK (diagonal) | |
| output | | |
| XK1[7] | The new state matrix filtered value. | |
| coupling | TRIAD(), rotmtx2quat(), Unit(), norm() | |
| dependencies | utils.h, AttitudeDeter.h, struct.h | |

| norm() | | |
|---|---|---|
| description | This function calculates the norm of a vector. | |
| input | description | range / units |
| *In | input array buffer. | |
| len | length of input buffer. | |
| output | | |
| norm | norm of the input vector. | |
| coupling | none. | |

| | |
|---|---|
| dependencies | utils.h |

| AttitudeDetermination() | | |
|---|---|---|
| description | This Function decides on using TRIAD algorithm or EKF based on sun sensor flag. | |
| input | description | range / units |
| sun_sensor_flag | The sun sensor flag which has the value of True = 1 = Sun, False = 0 = eclipse. | |
| output | | |
| coupling | EKF(), TRIAD() | |
| dependencies | AttitudeDeter.h, struct.h | |

| TRIAD() | | |
|---|---|---|
| description | This Function implements TRIAD algorithm using measurements from sun sensor and magnetometer. | |
| input | description | range / units |
| rmag_Body[3] | magnetic field vector in body frame | |
| rmag_Inertia[3] | magnetic field vector in inertial frame | |
| rsun_Body[3] | sun vector in body frame | |
| rsun_Inertia[3] | sun vector in inertial frame | |
| output | | |
| Quat_B2I[4] | Quaternions | |
| SVD[4] | Singular Values Decomposition | |
| coupling | solver(), solver_initialize(), dsvd(), rotmtx2quat(), Unit(), norm() | |
| dependencies | svd.h, solver_initialize.h, solver.h, utils.h, AttitudeDeter.h, struct.h | |

| dsvd() | | |
|---|---|---|
| description | This function is for the SVD decomposition. | |
| input | description | range / units |
| a | mxn matrix to be decomposed, gets overwritten with u | |
| output | | |
| w | returns the vector of singular values of a | |
| v | returns the right orthogonal transformation matrix | |
| coupling | none. | |
| dependencies | defs_and_types.h | |

| sun() | | | |
|---|---|---|---|
| description | This function calculates the normed eci mean equator mean equinox (j2000) position vector of the sun given the Julian date. | | |
| input | description | range / units | |
| TimeUTC | struct containing: | | |
| year | year | 1900 | 2100 |
| mon | month | 1 | 12 |
| day | day | 1 | 28,29,30,31 |
| hr | universal time hour | 0 | 23 |
| min | universal time min | 0 | 59 |
| sec | universal time sec | 0.000 | 59.999 |
| output | | | |
| rsun | eci position vector of the sun normed to magnitude one | | |

| | |
|---|---|
| coupling | timeDatetime2jd(), timeUtc2tt(), timeJd2jc(), framePrecess() |
| dependenci es | frameTrans.h, timeConv.h, struct.h, sunPred.h |

| magWmm() | | | |
|---|---|---|---|
| description | This function Calculates the components of the Earth's magnetic field using the World Magnetic Model (WMM). | | |
| input | description | range / units | |
| reci | position vector eci | km | |
| TimeUTC | struct containing: | | |
| year | year | 1900 | 2100 |
| mon | month | 1 | 12 |
| day | day | 1 | 28,29,30,31 |
| hr | universal time hour | 0 | 23 |
| min | universal time min | 0 | 59 |
| sec | universal time sec | 0.000 | 59.999 |
| output | | | |
| Beci | geomagnetic field vector in nanotesla (nT) in eci frame | | |
| coupling | timeDatetime2years(), frameEci2ecef(), frameEcef2geod() | | |
| dependencie s | frameTrans.h, timeConv.h, struct.h, magPred.h | | |

| magIgrf() | | | |
|---|---|---|---|
| descriptio n | This function Calculates the components of the Earth's magnetic field using the International Geomagnetic Reference Field (IGRF) model. | | |
| input | description | range / units | |
| reci | position vector eci | km | |
| TimeUTC | struct containing: | | |
| year | year | 1900 | 2100 |
| mon | month | 1 | 12 |
| day | day | 1 | 28,29,30,31 |
| hr | universal time hour | 0 | 23 |
| min | universal time min | 0 | 59 |
| sec | universal time sec | 0.000 | 59.999 |
| output | | | |
| Beci | geomagnetic field vector in nanotesla (nT) in eci frame | | |
| coupling | timeDatetime2years(), frameEci2ecef(), frameEcef2geod() | | |
| dependen cies | frameTrans.h, timeConv.h, struct.h, magPred.h | | |

| timeGstime() | | | |
|---|---|---|---|
| description | This function finds the Greenwich sidereal time (iau-82). | | |
| input | description | range / units | |
| jdut1 | julian date of ut1 | days from 4713 bc. | |
| output | | | |
| gst | greenwich sidereal time | 0 rad | 2pi rad |
| coupling | timeJd2jc() | | |
| dependencies | timeConv.h, struct.h | | |

| timeTt2utc() | | | |
|---|---|---|---|
| description | This function converts terrestrial time to universal time, assuming the year is between 2005 and 2050. | | |
| input | description | range / units | |
| TimeTT | struct containing: | | |
| year | year | 1900 | 2100 |
| mon | month | 1 | 12 |
| day | day | 1 | 28,29,30,31 |
| hr | universal time hour | 0 | 23 |
| min | universal time min | 0 | 59 |
| sec | universal time sec | 0.000 | 59.999 |
| output | | | |
| TimeUTC | struct containing: | | |
| year | year | 1900 | 2100 |
| mon | month | 1 | 12 |
| day | day | 1 | 28,29,30,31 |
| hr | universal time hour | 0 | 23 |
| min | universal time min | 0 | 59 |
| sec | universal time sec | 0.000 | 59.999 |
| coupling | timeDays2datetime(), timeDatetime2days() | | |
| dependencies | timeConv.h, struct.h | | |

| timeUtc2tt() | | | |
|---|---|---|---|
| description | This function converts universal time to terrestrial time, assuming the year is between 2005 and 2050. | | |
| input | description | range / units | |
| TimeUTC | struct containing: | | |
| year | year | 1900 | 2100 |
| mon | month | 1 | 12 |
| day | day | 1 | 28,29,30,31 |
| hr | universal time hour | 0 | 23 |
| min | universal time min | 0 | 59 |
| sec | universal time sec | 0.000 | 59.999 |
| output | | | |
| TimeTT | struct containing: | | |
| year | year | 1900 | 2100 |
| mon | month | 1 | 12 |
| day | day | 1 | 28,29,30,31 |
| hr | universal time hour | 0 | 23 |
| min | universal time min | 0 | 59 |
| sec | universal time sec | 0.000 | 59.999 |
| coupling | timeDays2datetime(), timeDatetime2days() | | |
| dependencies | timeConv.h, struct.h | | |

| timeJc2jd() | | |
|---|---|---|
| description | This function finds the Julian day given the Julian century. | |
| input | description | range / units |
| | Julian date | days from 4713 bc. |
| output | | |
| Julian century of the modified julian date (jd - 2451545.0 )/ 36525. | | |
| coupling | none. | |
| dependencies | timeConv.h, struct.h | |

| timeJd2jc() | | |
|---|---|---|
| description | This function finds the julian century given the julian date. | |
| input | description | range / units |
| Julian century of the modified julian date (jd - 2451545.0 )/ 36525. | | |
| output | | |
| | Julian date | days from 4713 bc. |
| coupling | none. | |
| dependencies | timeConv.h, struct.h | |

| timeYears2datetime() | | | |
|---|---|---|---|
| description | This function converts the fractional year, to the equivalent month, day, hour, minute and second. | | |
| input | description | range / units | |
| years | fractional year | 1900.000 | 2100.999 |
| output | | | |
| Time | struct containing: | | |
| year | year | 1900 | 2100 |
| mon | month | 1 | 12 |
| day | day | 1 | 28,29,30,31 |
| hr | universal time hour | 0 | 23 |
| min | universal time min | 0 | 59 |
| sec | universal time sec | 0.000 | 59.999 |
| coupling | timeDays2datetime() | | |
| dependencies | timeConv.h, struct.h | | |

| timeDatetime2years() | | | |
|---|---|---|---|
| description | This function converts the fractional year, to the equivalent month, day, hour, minute and second. | | |
| input | description | range / units | |
| Time | struct containing: | | |
| year | year | 1900 | 2100 |
| mon | month | 1 | 12 |
| day | day | 1 | 28,29,30,31 |
| hr | universal time hour | 0 | 23 |
| min | universal time min | 0 | 59 |
| sec | universal time sec | 0 | 59.999 |
| output | fractional year | 1900.000 | 2100.999 |
| coupling | timeDatetime2days() | | |
| dependencies | timeConv.h, struct.h | | |

| timeDays2datetime() | |
|---|---|
| description | This function converts the day of the year, days, to the equivalent month, day, hour, minute and second. |

| input | description | range / units | |
|---|---|---|---|
| year | year | 1900 | 2100 |
| day | day of year plus fraction of a day | days | |
| output | | | |
| Time | struct containing: | | |
| year | year | 1900 | 2100 |
| mon | month | 1 | 12 |
| day | day | 1 | 28,29,30,31 |
| hr | universal time hour | 0 | 23 |
| min | universal time min | 0 | 59 |
| sec | universal time sec | 0.000 | 59.999 |
| coupling | none. | | |
| dependencies | timeConv.h, struct.h | | |

| timeDatetime2days() | | | |
|---|---|---|---|
| description | This function finds the fractional days through a year given the year, month, day, hour, minute and second. | | |
| input | description | range / units | |
| Time | struct containing: | | |
| year | year | 1900 | 2100 |
| mon | month | 1 | 12 |
| day | day | 1 | 28,29,30,31 |
| hr | universal time hour | 0 | 23 |
| min | universal time min | 0 | 59 |
| sec | universal time sec | 0 | 59.999 |
| output | day of year plus fraction of a day | days | |
| coupling | none | | |
| dependencies | timeConv.h, struct.h | | |

| timeJd2datetime() | | | |
|---|---|---|---|
| description | This function finds the year, month, day, hour, minute and second given the julian date. | | |
| input | description | range / units | |
| | julian date | days from 4713 bc | |
| output | | | |
| Time | struct containing: | | |
| year | year | 1900 | 2100 |
| mon | month | 1 | 12 |
| day | day | 1 | 28,29,30,31 |
| hr | universal time hour | 0 | 23 |
| min | universal time min | 0 | 59 |
| sec | universal time sec | 0 | 59.999 |
| coupling | timeDays2datetime | | |
| dependencies | timeConv.h, struct.h | | |

| timeDatetime2jd() | | |
|---|---|---|
| description | This function converts Universal Date and Time to Julian Date | |
| input | description | range / units |

| | | | |
|---|---|---|---|
| Time | struct containing: | | |
| year | year | 1900 | 2100 |
| mon | month | 1 | 12 |
| day | day | 1 | 28,29,30,31 |
| hr | universal time hour | 0 | 23 |
| min | universal time min | 0 | 59 |
| sec | universal time sec | 0 | 59.999 |
| output | | | |
| | julian date | days from 4713 bc | |
| coupling | none | | |
| dependencies | timeConv.h, struct.h | | |

| Solver() | | |
|---|---|---|
| description | This function solves four equations three unknown with "Least Square Method" | |
| input | description | range / units |
| q[0] | Quaternions | acceptable range for each mission phase |
| q[1] | Quaternions | acceptable range for each mission phase |
| q[2] | Quaternions | acceptable range for each mission phase |
| q[3] | Quaternions | acceptable range for each mission phase |
| dq[0] | Derivative of Quaternions | acceptable range for each mission phase |
| dq[1] | Derivative of Quaternions | acceptable range for each mission phase |
| dq[2] | Derivative of Quaternions | acceptable range for each mission phase |
| dq[3] | Derivative of Quaternions | acceptable range for each mission phase |
| output | | |
| | Angular velocity[3] | acceptable range for each mission phase |
| coupling | none | |
| dependencies | rtwtypes.h, solver_types.h, tmwtypes.h, solver_initialize.h, rt_nonfinite.h, defs_and_types.h, rtGetNaN.h, rtGetInf.h, solver.h | |

| Dynamic_Model() | | |
|---|---|---|
| description | This function implements the dynamic model of the satellite and calculate the result of applying control torque on the system. | |
| input | description | range / units |
| Omega_B2I_prev | Angular Velocity | acceptable range for each mission phase |
| DCM_B2I | Direct Cosine Matrix | acceptable range for each mission phase |
| DCM_B2O_Prev | Direct Cosine Matrix | acceptable range for each mission phase |
| Tc | magnetic moment | acceptable range for each mission phase |
| GRC | Gyro Draft | acceptable range for each mission phase |
| output | description | range / units |
| Tg | Gravity gradient moment | acceptable range for each mission phase |
| Omega_B2I | Angular Velocity | acceptable range for each mission phase |
| Quat_B2O | Quaternions | acceptable range for each mission phase |
| DCM_B2O | Direct Cosine Matrix | acceptable range for each mission phase |
| Quat_B2I | Quaternions | acceptable range for each mission phase |
| GK | Gyro Output | acceptable range for each mission phase |
| coupling | norm(), | |

| | |
|---|---|
| dependencies | struct.h, Dynamic.h, utils.h |

| Control_damp() | | |
|---|---|---|
| description | This function implements the B-dot controller. | |
| input | description | range / units |
| Kd_damp[3][3] | Kd gain of rate damping controller | |
| rmag_Body[3] | magnetic field vector in body frame | |
| output | description | range / units |
| Mc | Controller Output Control momentum | |
| Tm | Controller Output control torque | |
| coupling | norm(), MaxArray() | |
| dependencies | struct.h, control.h, utils.h | |

| Control_point() | | |
|---|---|---|
| description | This function implements the pointing controller. | |
| input | description | range / units |
| Quat_B2O[4] | Quaternions | acceptable range for each mission phase |
| Kp_point[3][3] | Kp gain of pointing controller | |
| Kd_point[3][3] | Kd gain of pointing controller | |
| rmag_Body[3] | magnetic field vector in body frame | |
| output | description | range / units |
| Mc | Controller Output Control momentum | |
| Tm | Controller Output control torque | |
| coupling | norm(), MaxArray() | |
| dependencies | struct.h, control.h, utils.h | |

| Control_stable() | | |
|---|---|---|
| description | This function implements the stabilization controller. | |
| input | description | range / units |
| Quat_B2O[4] | Quaternions | acceptable range for each mission phase |
| Kp_point[3][3] | Kp gain of stabilization controller. | |
| Kd_point[3][3] | Kd gain of stabilization controller. | |
| rmag_Body[3] | magnetic field vector in body frame | |
| output | description | range / units |
| Mc | Controller Output Control momentum | |
| Tm | Controller Output control torque | |
| coupling | norm(), MaxArray() | |
| dependencies | struct.h, control.h, utils.h | |

| Control_MWspin() | | |
|---|---|---|
| description | This function implements the momentum wheel controller. | |
| input | description | range / units |
| Hw | momentum wheel flag | |
| BZ6 | momentum wheel flag | |
| Quat_B2O[4] | Quaternions | acceptable range for each mission phase |
| Kp_point[3][3] | Kp gain of stabilization controller. | |
| Kd_point[3][3] | Kd gain of stabilization controller. | |
| output | description | range / units |
| Hw_set | Controller Output momentum wheel flag | |
| Tw | Controller Output Angular Velocity Moment | |
| coupling | none. | |
| dependencies | struct.h, control.h | |

| Control_MWdespin() | | |
|---|---|---|
| description | This function implements the momentum wheel controller reduces the spin. | |
| input | description | range / units |
| Hw | momentum wheel flag | |
| Omega_B2I[3] | Angular Velocity | |
| output | description | range / units |
| Tw | Controller Output Angular Velocity Moment | |
| coupling | none. | |
| dependencies | struct.h, control.h | |

| frameEcef2geod() | | |
|---|---|---|
| description | This function converts Earth-centered, Earth fixed (ECEF) coordinates to geodetic coordinates (World Geodetic System 1984 (WGS84)). | |
| input | description | range / units |
| recef | x,y,z coordinates of the point in kilometers. | |
| output | description | range / units |
| rgeod | Geodetic latitude and longitude in degrees, height above the Earth in kilometers. | |
| coupling | frameLatitudeRecur() | |
| dependencies | struct.h, frameTrans.h | |

| frameGeod2ecef() | | |
|---|---|---|
| description | This function converts geodetic coordinates (World Geodetic System 1984 (WGS84)) to Earth-centered, Earth fixed (ECEF) coordinates. | |
| input | description | range / units |
| rgeod | Geodetic latitude and longitude in degrees, height above the Earth in kilometers. | |
| output | description | range / units |
| recef | x,y,z coordinates of the point in kilometers. | |
| coupling | none. | |
| dependencies | struct.h, frameTrans.h | |

| frameEcef2eci() | | | |
|---|---|---|---|
| description | This function transforms a vector from the earth fixed (itrf) frame, to the eci mean equator mean equinox (j2000) frame. | | |
| input | description | range / units | |
| recef | position vector earth fixed | km | |
| TimeUTC | struct containing: | | |
| year | year | 1900 | 2100 |
| mon | month | 1 | 12 |
| day | day | 1 | 28,29,30,31 |
| hr | universal time hour | 0 | 23 |
| min | universal time min | 0 | 59 |
| sec | universal time sec | 0.000 | 59.999 |
| output | description | range / units | |
| reci | position vector in eci frame in kilometers. | | |
| coupling | framePrecess(), frameNutation(), frameSidereal(), framePolarm(), timeDatetime2jd(), timeUtc2tt(), timeJd2jc() | | |
| dependencies | struct.h, frameTrans.h, timeConv.h | | |

| frameEci2ecef() | | | |
|---|---|---|---|
| description | This function transforms a vector from the eci mean equator mean equinox (j2000) frame to the earth fixed (itrf) frame. | | |
| input | description | range / units | |
| reci | position vector in eci frame in kilometers. | | |
| TimeUTC | struct containing: | | |
| year | year | 1900 | 2100 |
| mon | month | 1 | 12 |
| day | day | 1 | 28,29,30,31 |
| hr | universal time hour | 0 | 23 |
| min | universal time min | 0 | 59 |
| sec | universal time sec | 0.000 | 59.999 |
| output | description | range / units | |
| recef | position vector earth fixed | km | |
| coupling | framePrecess(), frameNutation(), frameSidereal(), framePolarm(), timeDatetime2jd(), timeUtc2tt(), timeJd2jc() | | |
| dependencies | struct.h, frameTrans.h, timeConv.h | | |

| frameEci2teme() | | | |
|---|---|---|---|
| description | This function transforms a vector from the mean equator mean equinox frame (j2000) to the true equator mean equinox of date (teme) frame. | | |
| input | description | range / units | |
| reci | position vector in eci frame in kilometers. | | |
| TimeUTC | struct containing: | | |
| year | year | 1900 | 2100 |
| mon | month | 1 | 12 |
| day | day | 1 | 28,29,30,31 |
| hr | universal time hour | 0 | 23 |
| min | universal time min | 0 | 59 |
| sec | universal time sec | 0.000 | 59.999 |
| output | description | range / units | |
| rteme | position vector of date true equator, mean equinox km | | |
| coupling | framePrecess(), frameNutation(), frameSidereal(), framePolarm(), timeDatetime2jd(), timeUtc2tt(), timeJd2jc() | | |
| dependencies | struct.h, frameTrans.h, timeConv.h | | |

| frameTeme2eci() | | |
|---|---|---|
| description | This function transforms a vector from the true equator mean equinox of date, (teme) frame to the mean equator mean equinox (j2000) frame. | |
| input | description | range / units |
| rteme | position vector of date true equator, mean equinox km | |

| | | | |
|---|---|---|---|
| TimeUTC | struct containing: | | |
| year | year | 1900 | 2100 |
| mon | month | 1 | 12 |
| day | day | 1 | 28,29,30,31 |
| hr | universal time hour | 0 | 23 |
| min | universal time min | 0 | 59 |
| sec | universal time sec | 0.000 | 59.999 |
| output | description | range / units | |
| reci | position vector in eci frame in kilometers. | | |
| coupling | framePrecess(), frameNutation(), frameSidereal(), framePolarm(), timeDatetime2jd(), timeUtc2tt(), timeJd2jc() | | |
| dependencies | struct.h, frameTrans.h, timeConv.h | | |

## 3.5 Validation of Software

The World Magnetic Model Comparassion between Matlab Model and C Code. The full code is available in appendices.

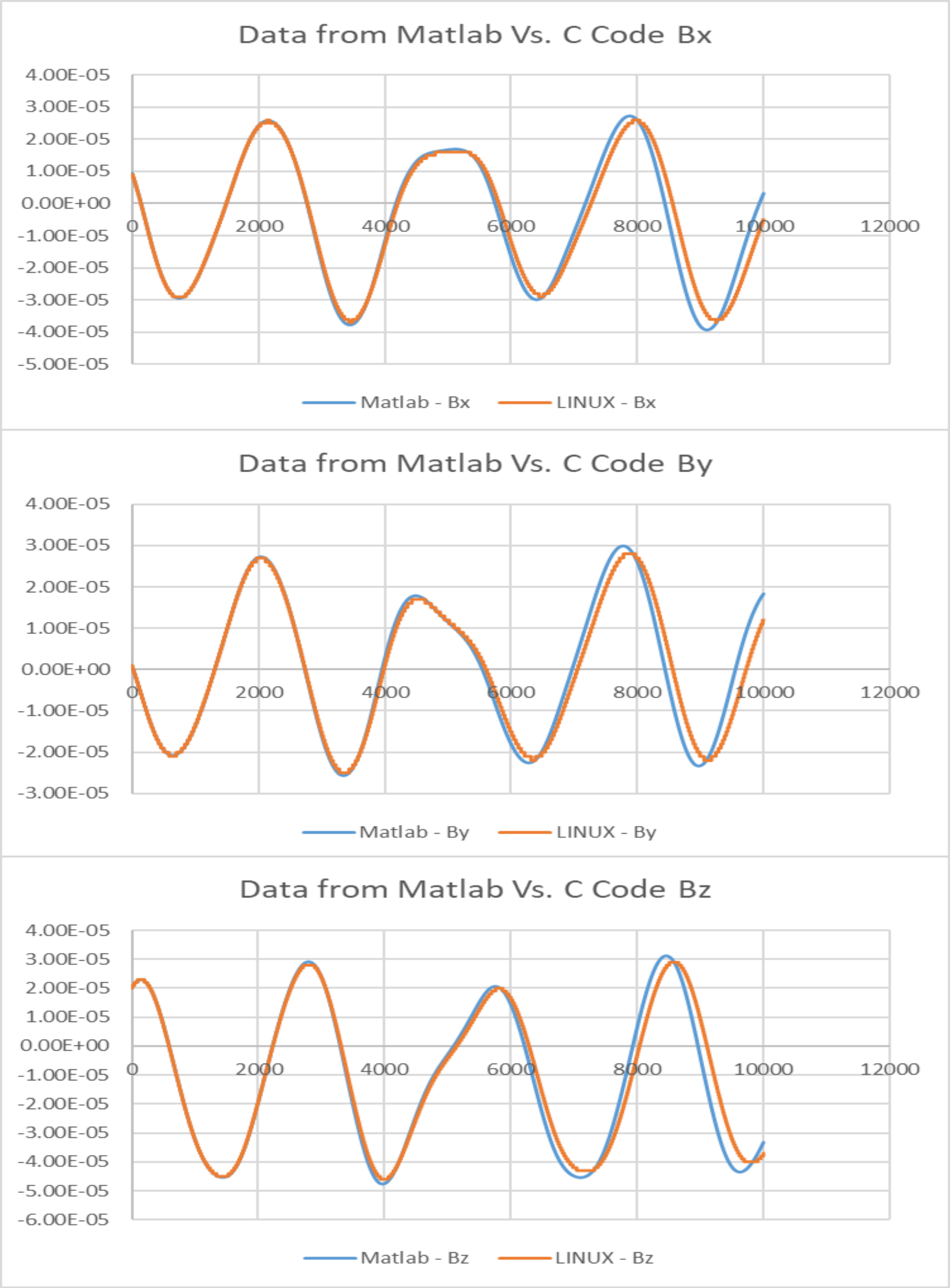

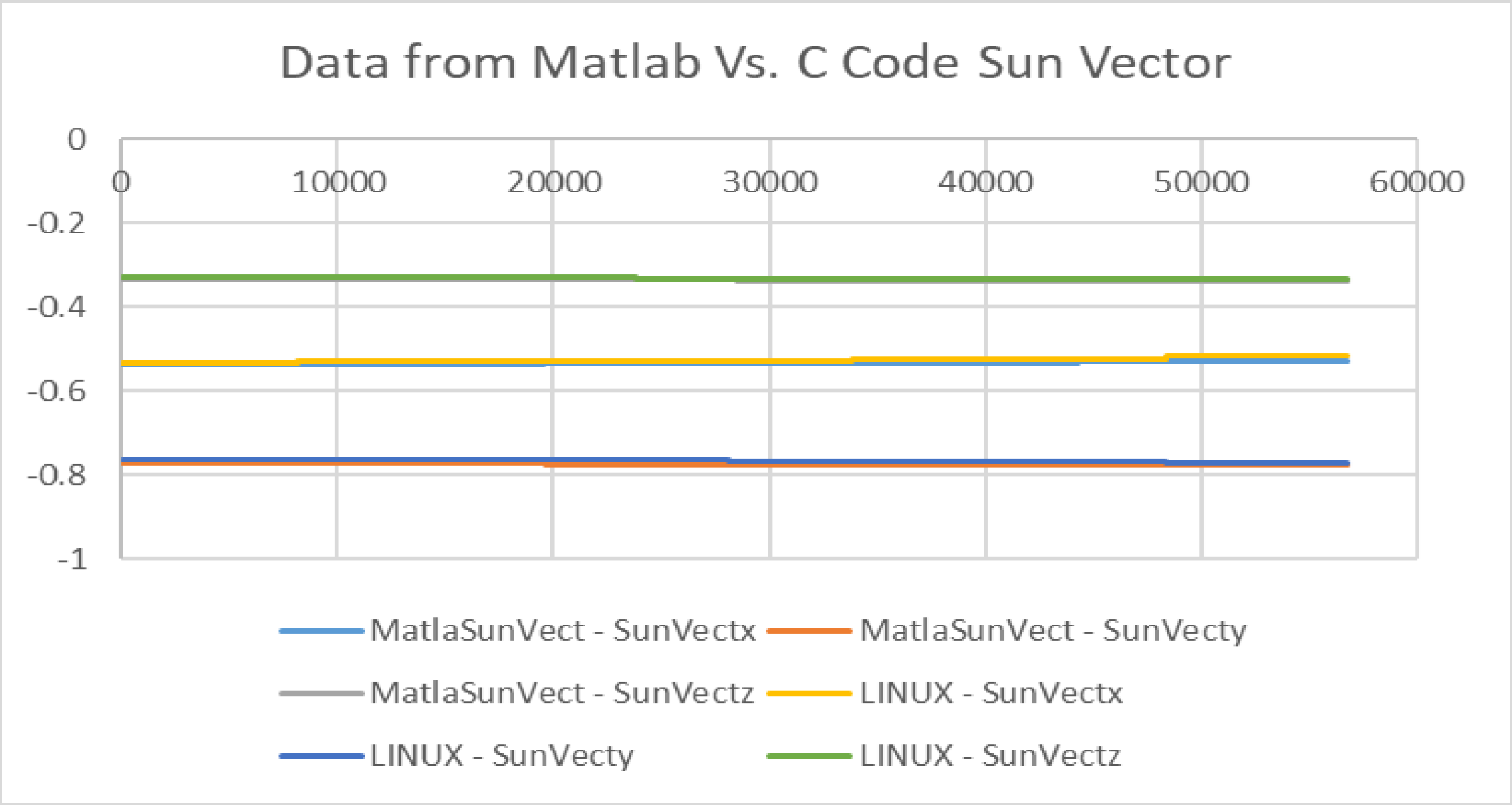
Data from Matlab Vs. C Code Sun Vector
0
-0.2
-0.4
-0.6
-0.8
-1
0
10000
20000
30000
40000
50000
60000
MatlaSunVect - SunVectx
MatlaSunVect - SunVecty
MatlaSunVect - SunVectz
LINUX - SunVectx
LINUX - SunVecty
LINUX - SunVectz

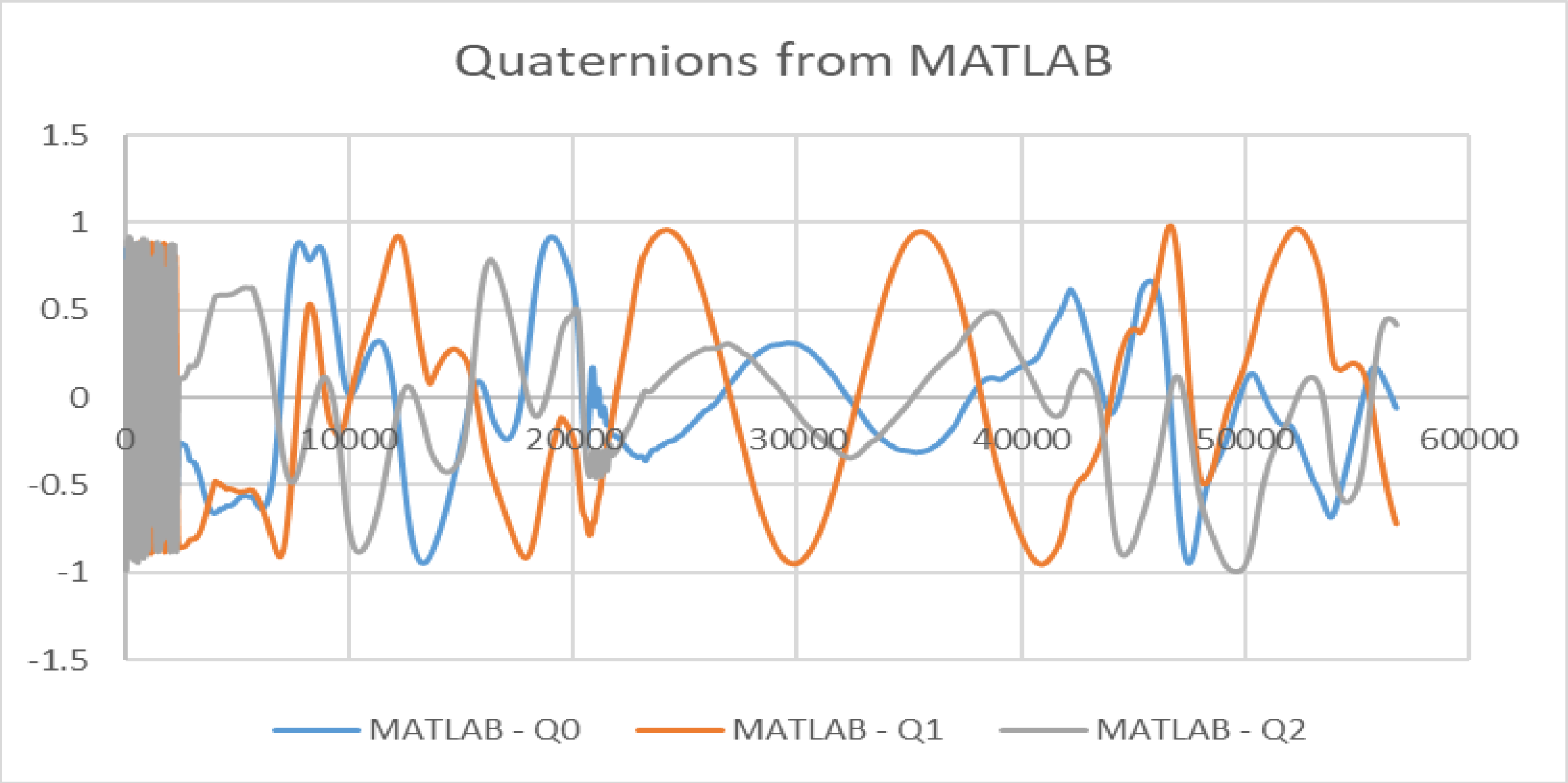

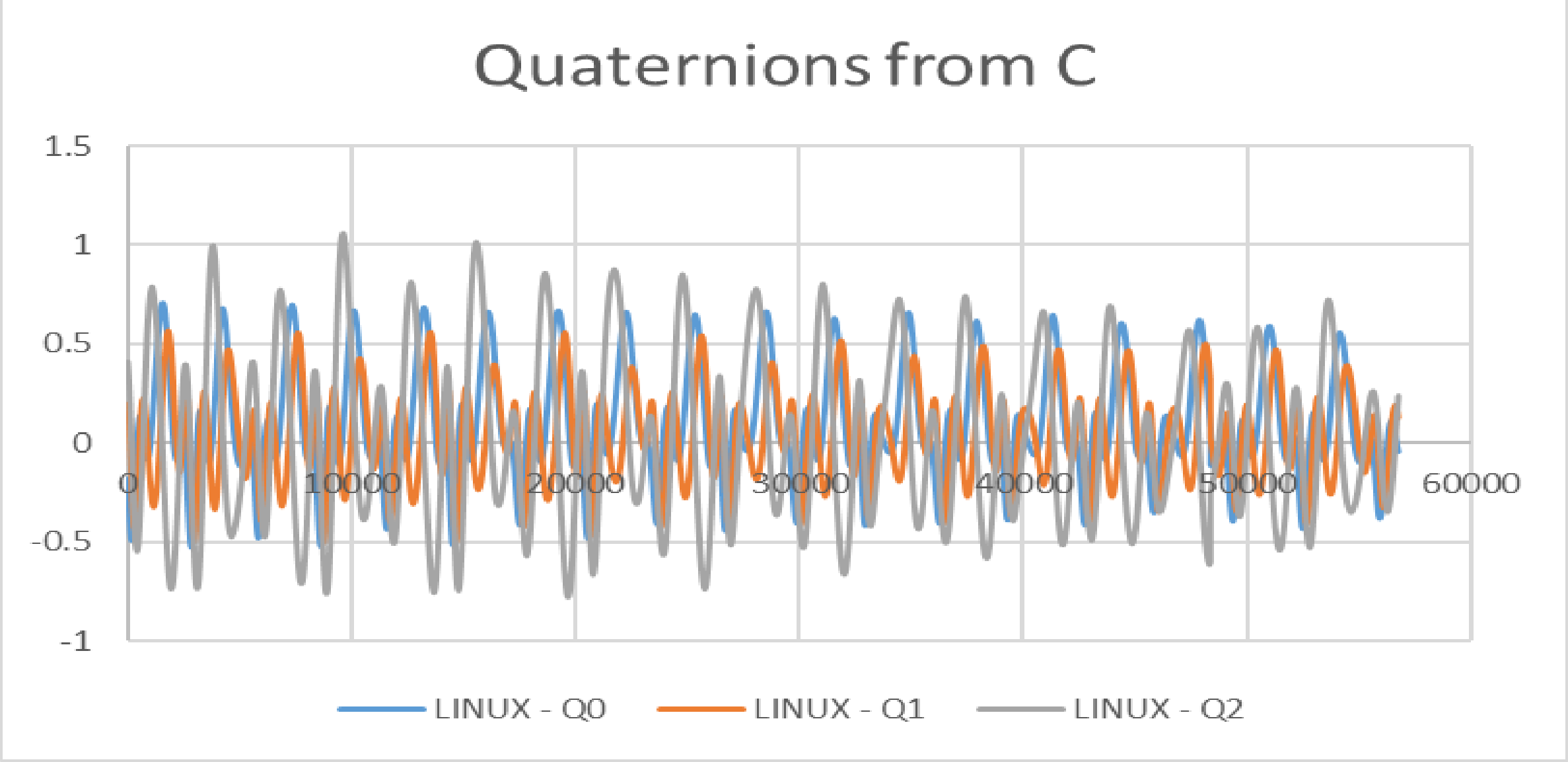

The Tracking and positioning of the satellite is compared from the output of C code and SGP4 model. The deviation in results is because of the fact that SGP4 model requires to be updated with GPS data regularly.

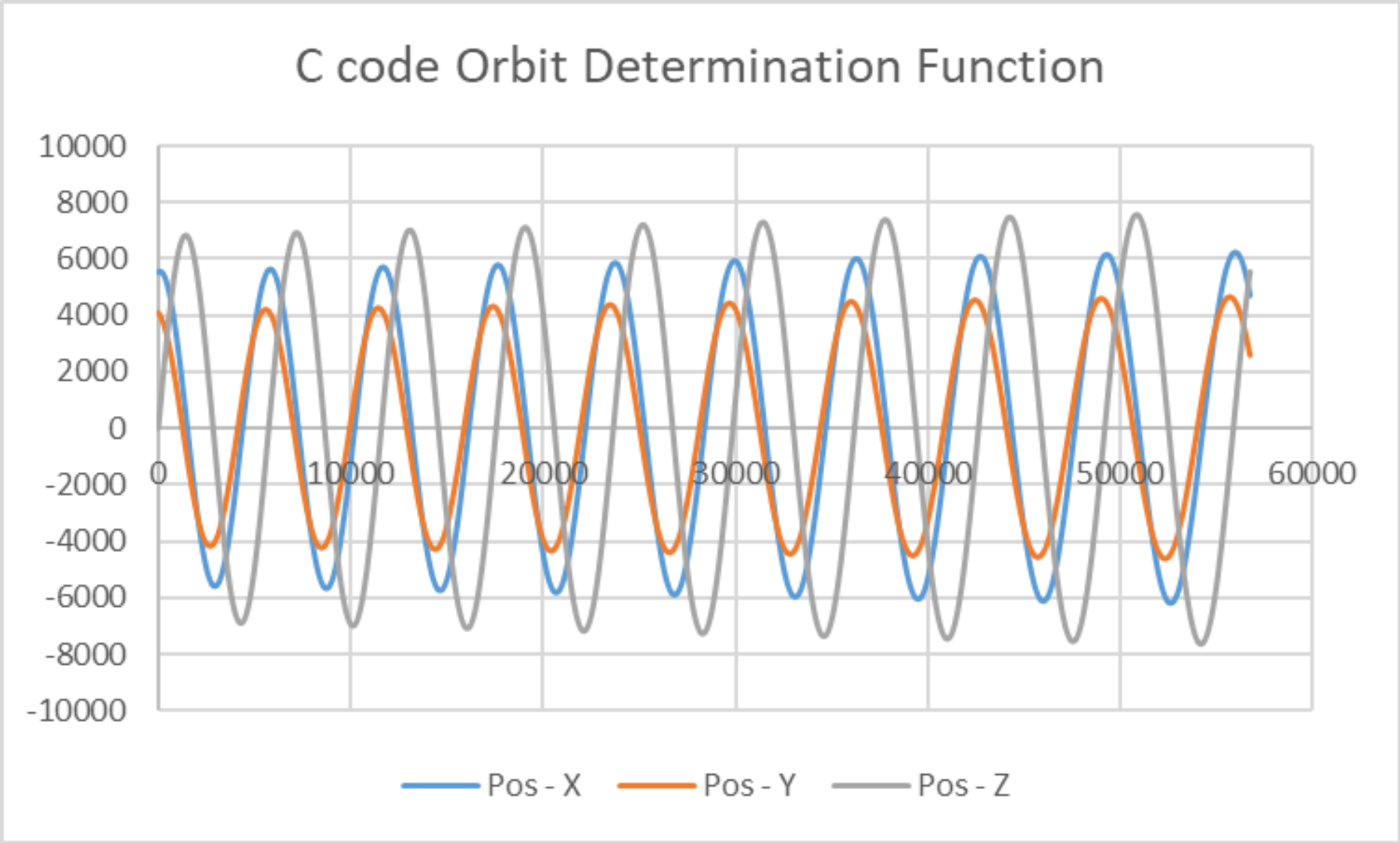

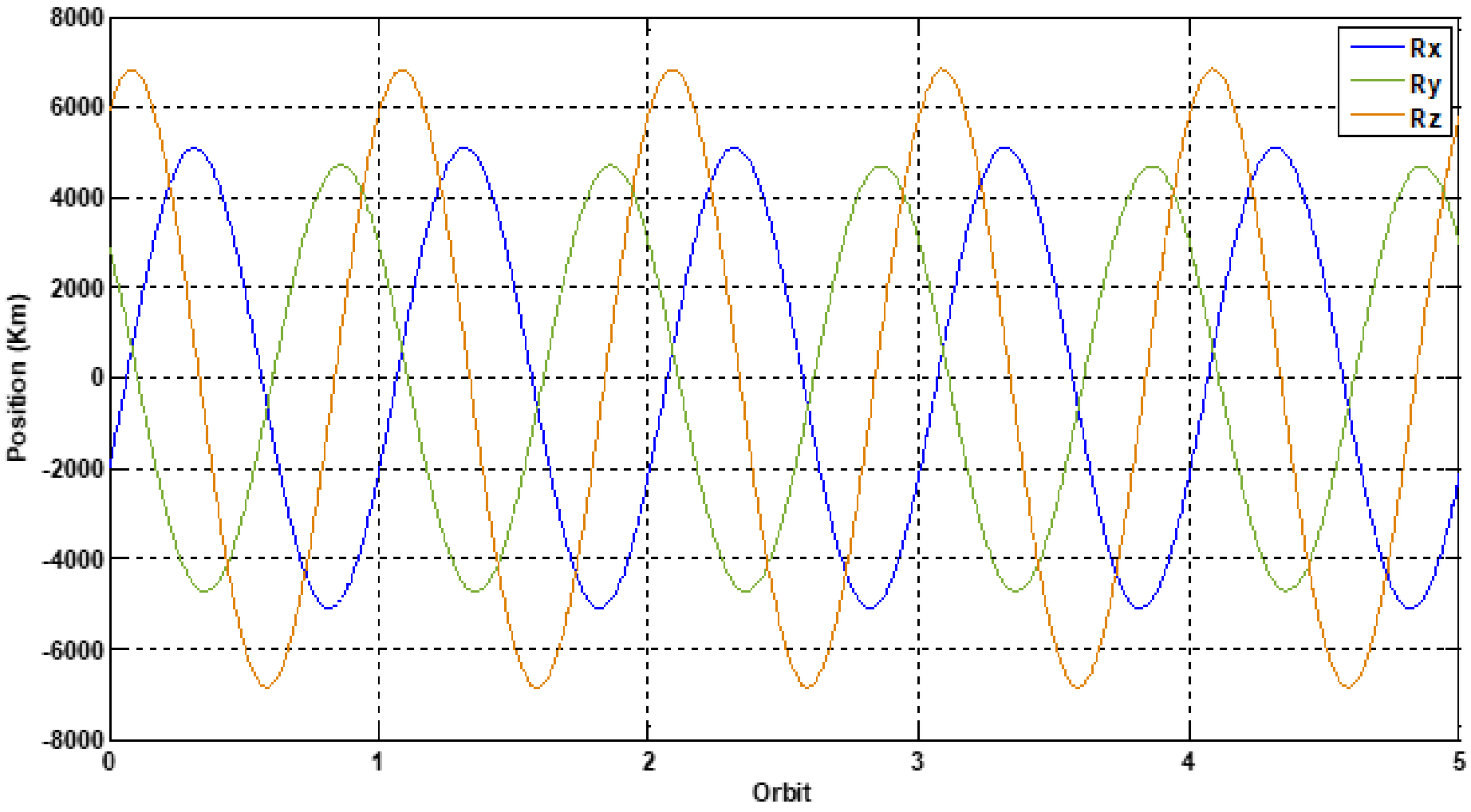

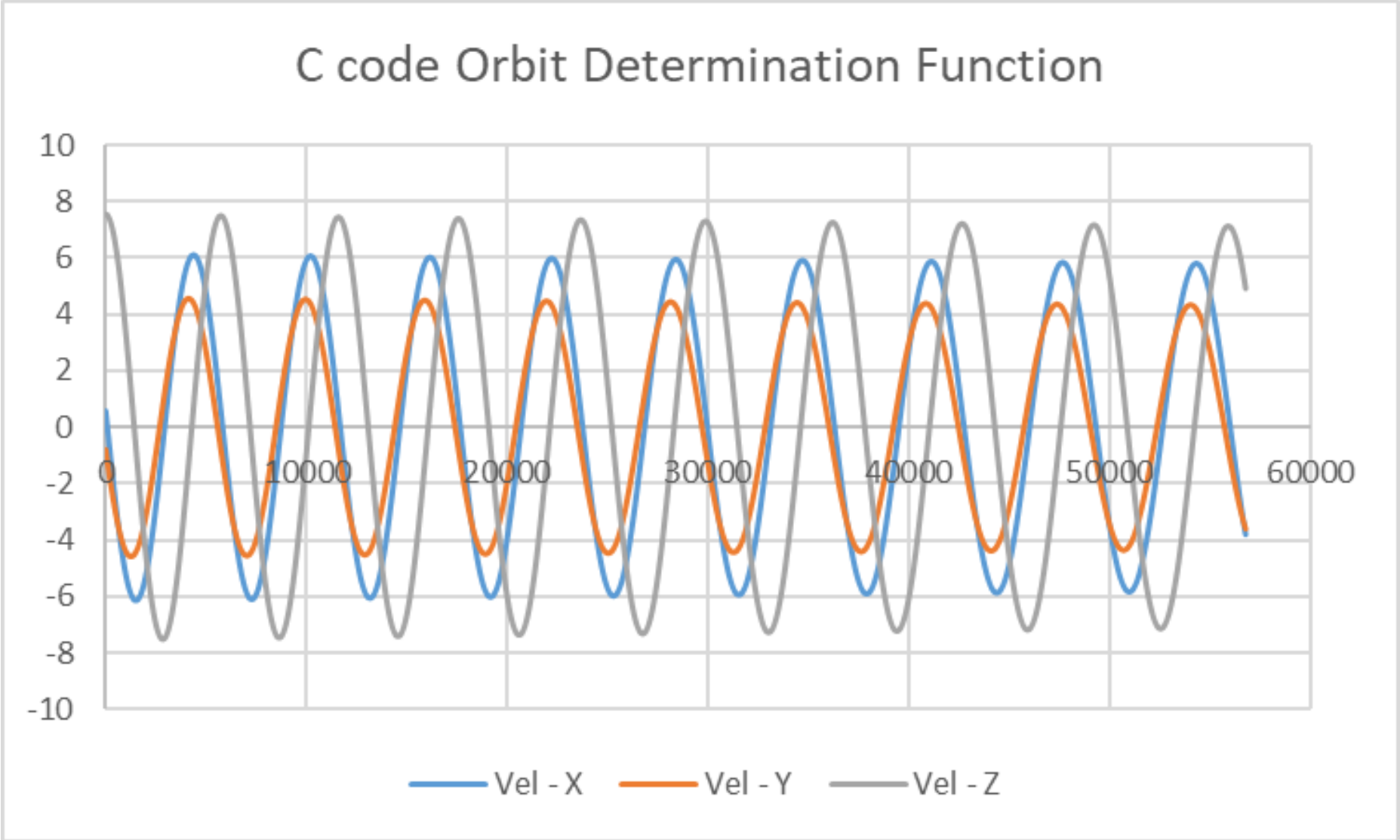

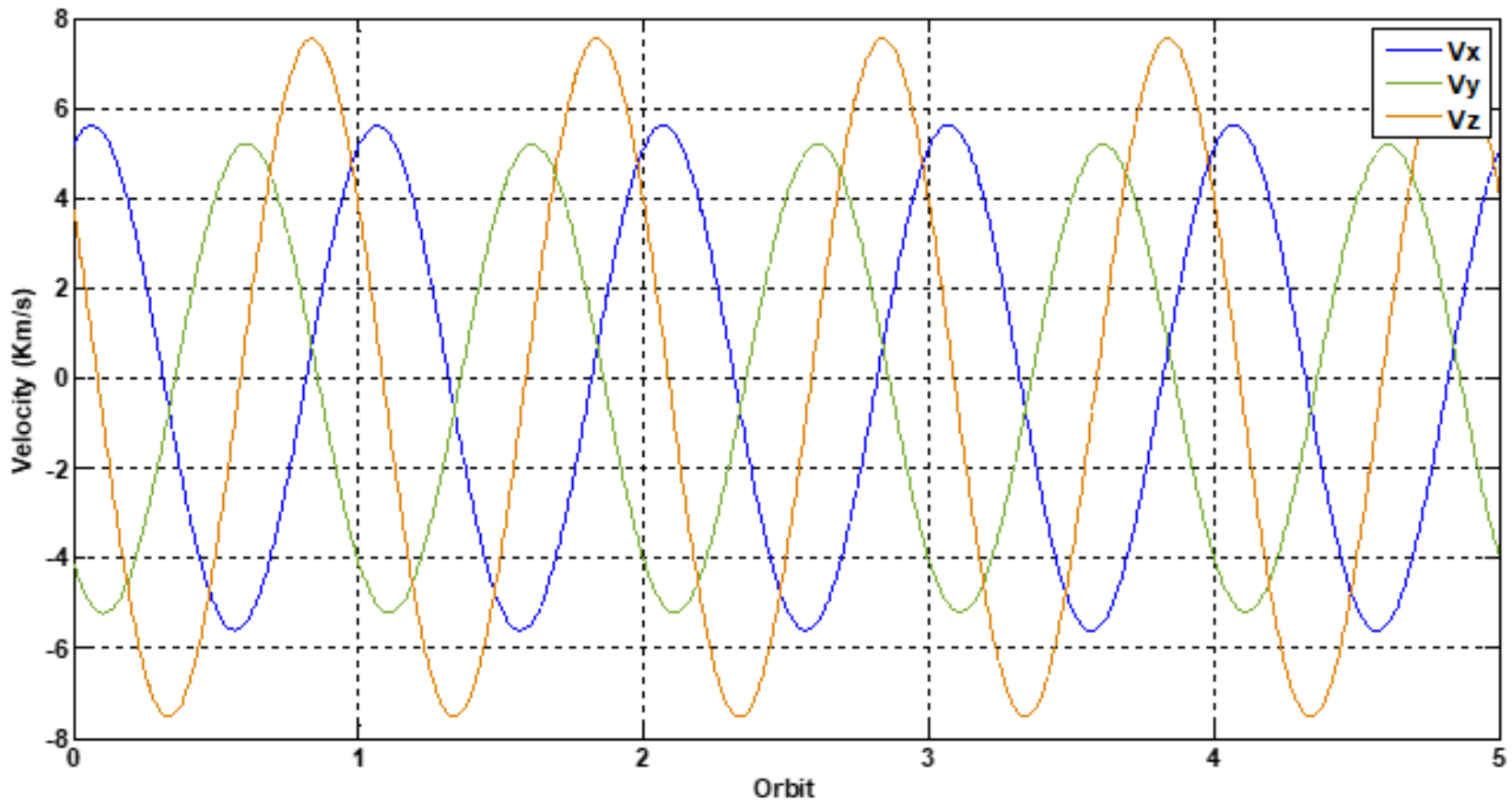

The Compatibility check test, SIL and PIL tests.

| OS Type | Compatibility Test Result |
|---|---|
| Linux - Ubuntu | ✓ |
| SylixOS | ✓ |
| Raspbian | ✓ |

In addition, the CPU can support x64 structures as well. The CPU usage for 5000 cycles with each 100 iterations of orbit calculation. The CPU processing requirement is passed. Tested on a single core CPU with a sequential algorithm. The processing power used in peak is 0.518 GHz, this includes the printing and other operations as well, the real loop time is about 2 seconds so the required processing power is less than the given number here for sure.

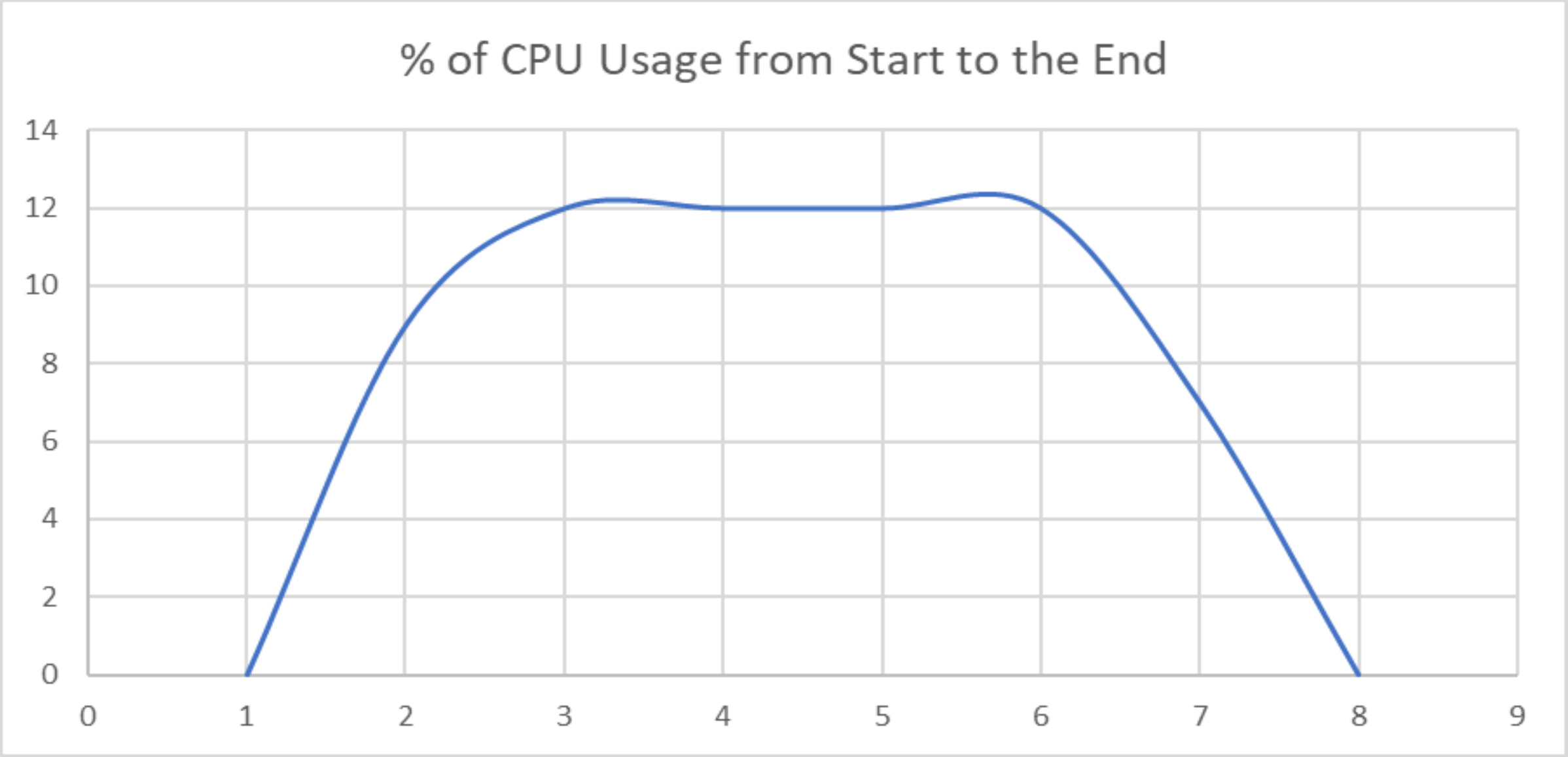

Inclusive samples as the total number of samples that are collected during the execution of the target function. The Exclusive samples are the number of samples that are collected during the direct execution of the instructions of the target function.

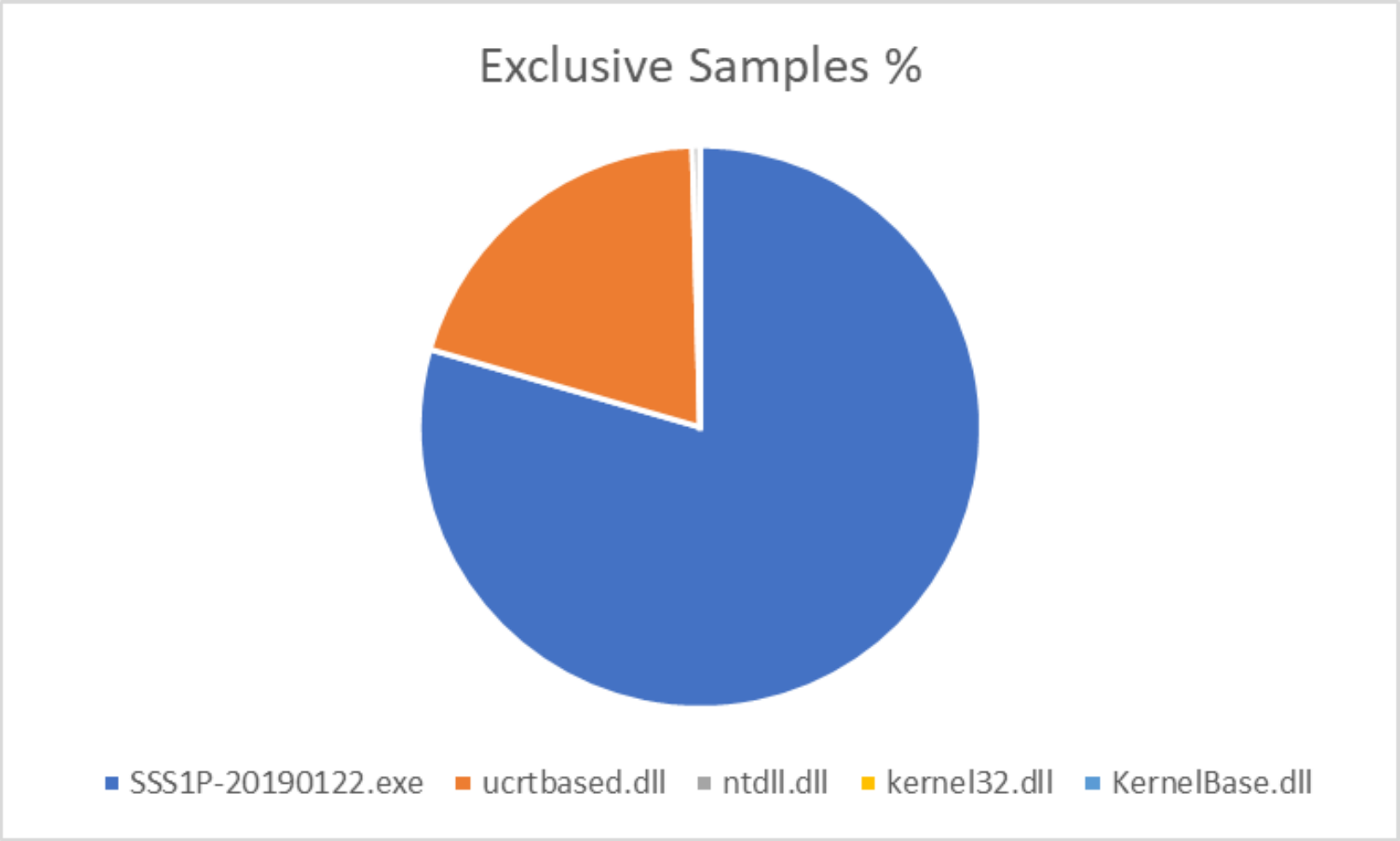

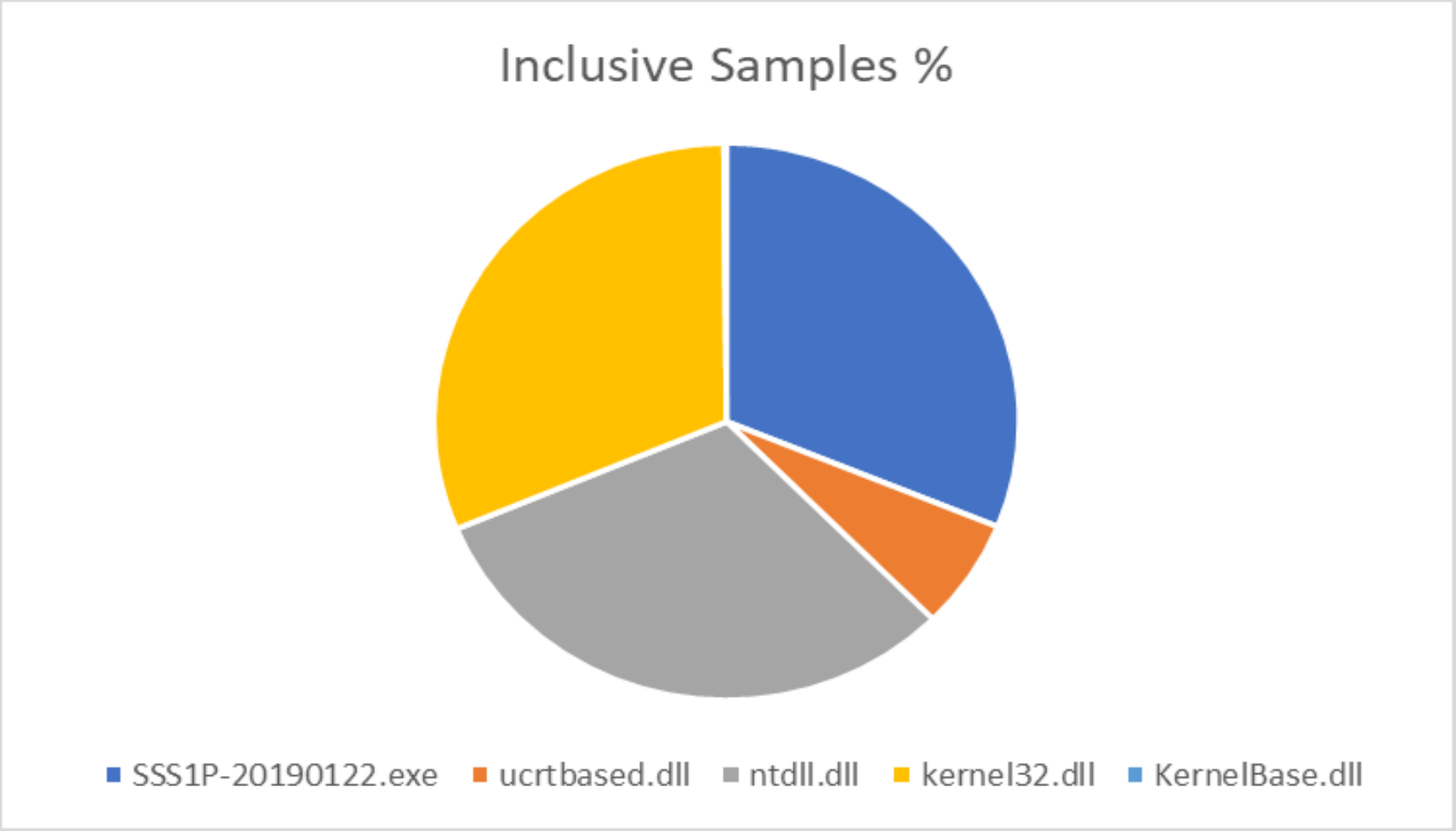

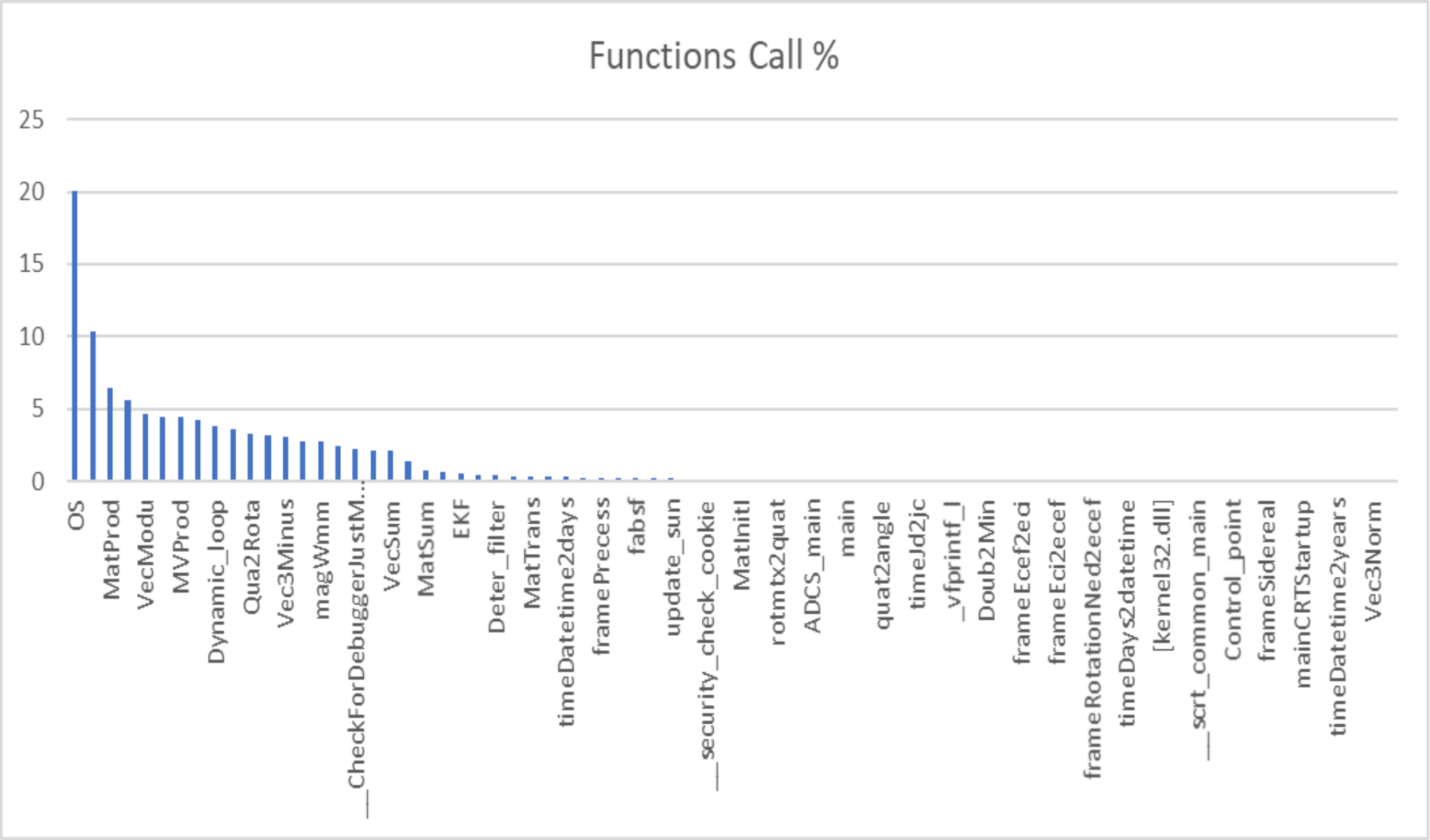

# 4. Final Tests, Results & Discussion

## 4.1 Introduction to tests

Verifies the correctness of the attitude control software and its control effect by running SSS-1P attitude control software (based on SylixOS operating system) in the on-board computer and establishing models of satellite attitude dynamics, sensors and actuators in the host computer SIMULINK software to form a closed loop in the form of serial communication. The test equipment can be listed as below:

- On-board computer
- Compilation environment and programming tool for on-board computer
- Host computer（Installing Matlab）
- Stabilized voltage supply
- MOXA card
- Multimeter
- Connecting cable

## 4.2 Attitude Control Software Transplantation Test

The original code of the attitude control software is based on Linux and the C language standard function library. The code conforms to the on-board operating system specification. The test can be demonstrated as below:

1) Establish models of satellite dynamics, sensors and actuators model (C codes) in the attitude control software to form a closed loop attitude control code;
2) Establish attitude control threads in the on-board software framework, move the closed-loop attitude control code into the attitude control threads, and compile and debug;
3) After compiling, establish a data output interface of the closed-loop attitude control code;
4) Run the on-board software to verify the logical correctness of the closed-loop attitude control code according to the output data.

### Results: Attitude Control Software Transplantation Test

Initially, no logic error occurred during the attitude control closed-loop code transplanting process.

## 4.3 Host Computer Simulink Modeling Test

In the host computer , the satellite attitude dynamics, sensors and actuators models are established in Simulink, and the serial communication model is established. The virtual serial port and serial port debugging software are used to verify that the serial port communication function of the host computer is normal.

1) Establish satellite attitude dynamics, kinematics model, and sensor models;
2) Establish a serial port model and write serial port sender and receiver data format conversion codes;
3) Establish a pair of virtual serial ports named port1and port2, port1 is opened from the SIMULINK model, port 2 is opened from Dasha serial port debugging software, set the responding and sending content, and run the SIMULINK model.
4) Analyze the corresponding sending and receiving content, and verify that the serial port communication function of the host computer is normal.

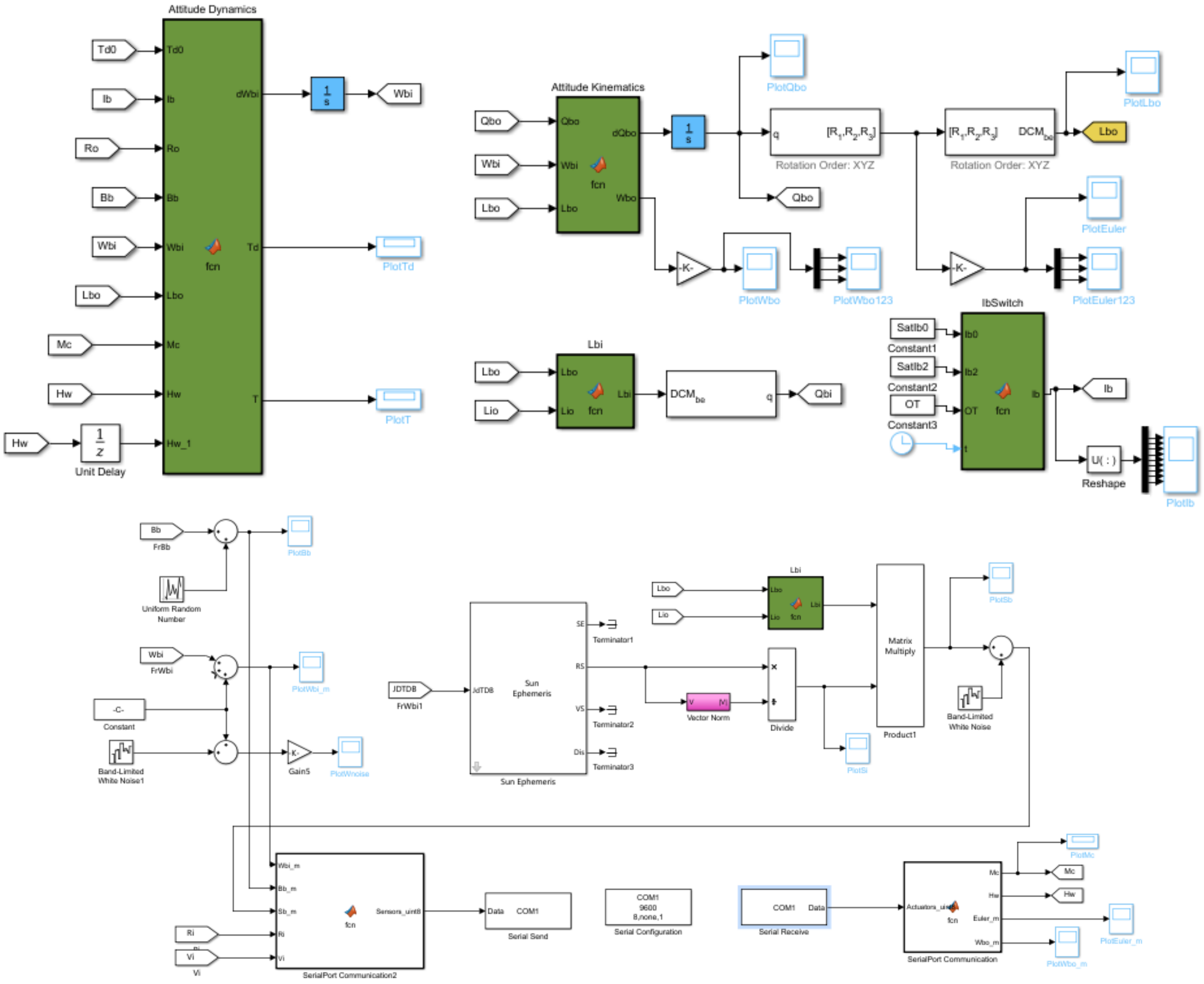

*Figure 4-2 Sensor and serial port model.*

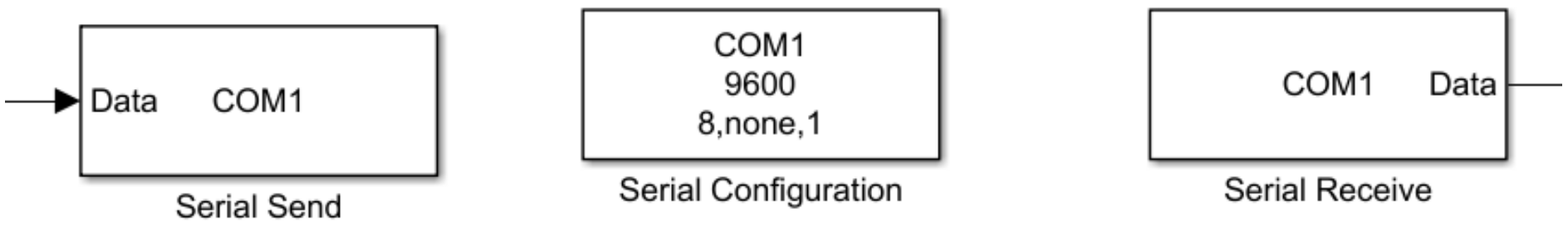

Figure 4-3 Serial port sender and receiver data format conversion code.

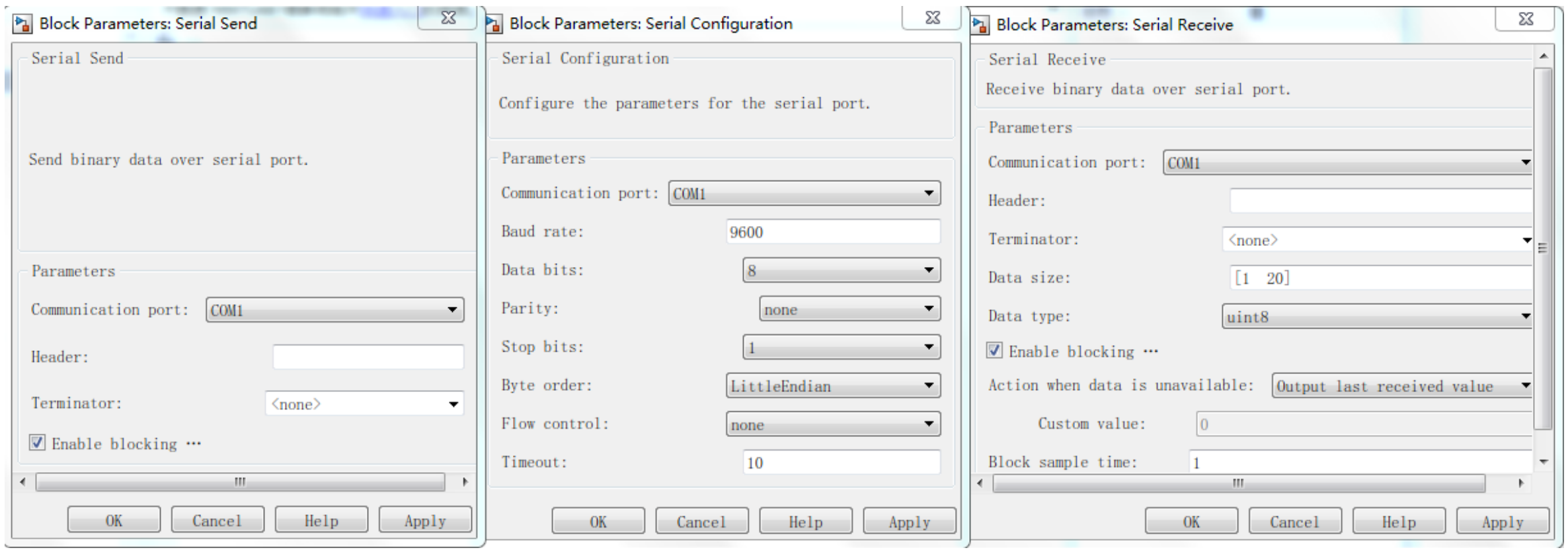

Figure 4-4 Serial Port Parameter Configuration.

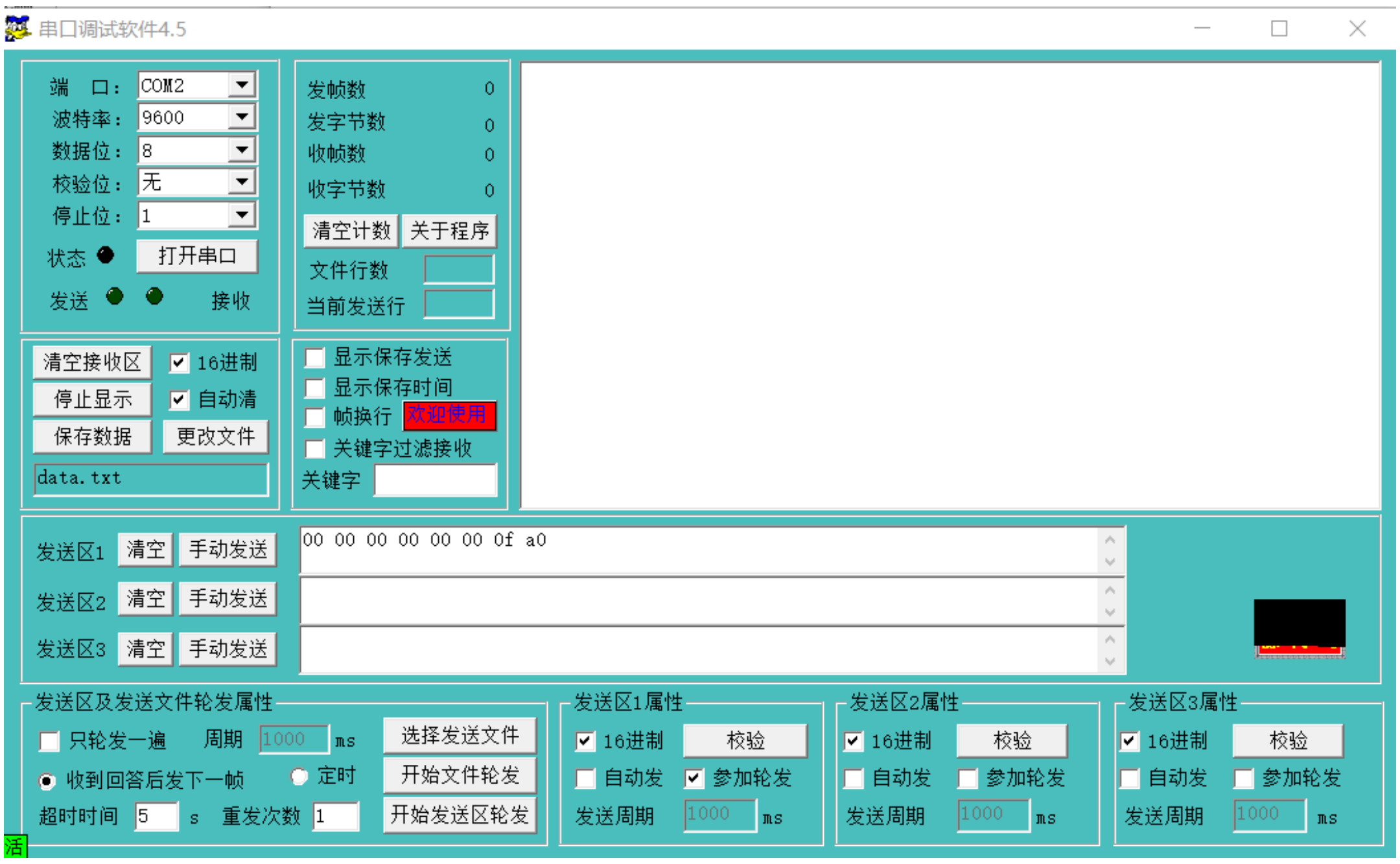

Figure 4-5 Setting the response content and method of the serial port debugging software.

## Results: Host Computer Simulink Modeling Test

When the SIMULINK model starts running, the data received from the serial port debugging software is consistent with the content sent by the SIMULINK model “Serial Send” module and after parsing, it conforms to the of the sensor data value ; The SIMULINK model “Serial Receive” module receives the same data as the serial port debugging software and after parsing, it conforms to the range of actuator command values. The serial communication function is normal and the data analysis is normal.

## 4.4 Serial communication test on on-board computer

Write the serial port receiving and sending function, transfer the two functions in the attitude control thread, search the free serial port in the on-board manual, verify the serial port communication function of the on-board computer by using the MOXA card and serial debugging software.

1) Write the function of the attitude control test dedicated serial port transceiver and data analysis (data format conversion) on the star software;
2) Search the manual of the on-board computer, select a free serial port, and test the wiring by using the multimeter to ensure that the wiring is consistent with the contact of the on-board computer;
3) Transfer the serial port transceiver function in the attitude control thread (receive the sensor data through the serial port receiving function, send the actuator command through the serial port sending function), the serial port address is RS232 address selected previously, and compile the on-board code;
4) Connect the selected serial port to the MOXA card, connect the other end to the PC, open it with Dasha serial debugging software, set the responding and sending content, start the on-board computer, program the compiled code, and run the attitude control thread;
5) Analyze the corresponding sending and receiving content, and verify that the serial port communication function of the satellite carrier is normal.

**Results: Serial communication test on on-board computer**

When the attitude control thread starts running, the data received by the serial debugging software is consistent with the content sent by "serial_send()" function,after parsing, it conforms to the value range of the actuator command; The data received by "serial_recv()" function is consistent with the content sent by the serial debugging software, after parsing, it conforms to the value range of sensors. The serial communication function is normal and the data analysis is normal.

## 4.5 Processor In the Loop Test

Use the MOXA card to connect the on-board computer to the host computer, run the SIMULINK model in the host computer, and start the attitude control thread on the star carrier to perform the closed-loop test of the attitude control software controller.

1) Connect the selected serial port of the on-board computer to MOXA card, connect the other end to the host computer, and select the corresponding serial port in the SIMULINK serial port model;

2) Turn on the power of the on-board computer, compile the star code, program and start the star carrier;

3) Run the SIMULINK model, after finishing compiling and the interface displays “T=0”, start the attitude control thread in the on-board computer;

4) After the end of the running, save the relevant data from the working space of Matlab in the host computer for analysis.

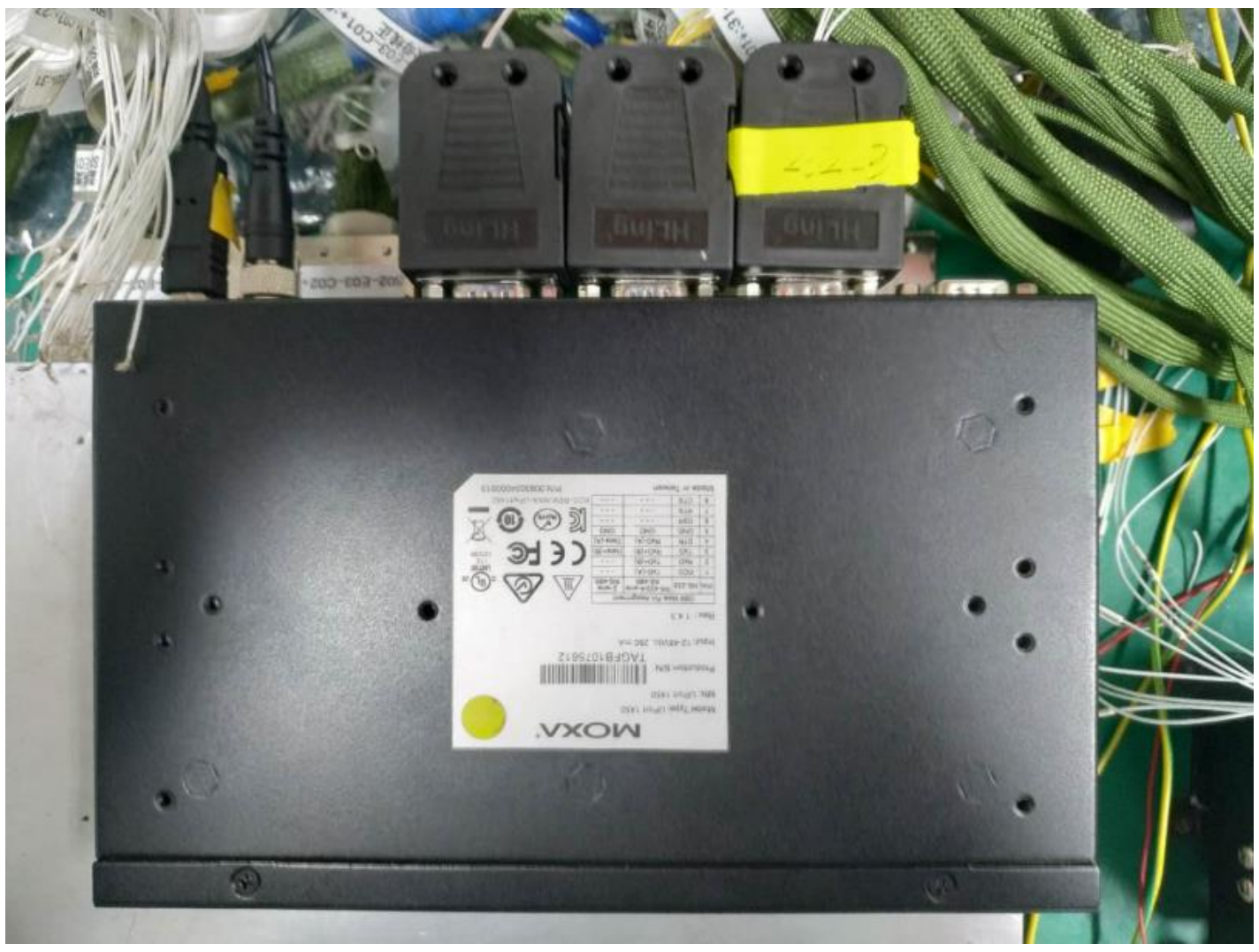

Figure 4-7 Connect selected serial port to MOXA card.

```
                         [[                      (R)
 [[[[         [[[[       [[            [[[[   [[[[
[[  [[          [[                   [[  [[  [[  [[
[[      [[  [[  [[    [[[[   [[  [[  [[  [[  [[
 [[     [[  [[  [[      [[   [[  [[  [[  [[   [[
  [[    [[  [[  [[      [[    [[[[   [[  [[    [[
   [[   [[  [[  [[      [[     [[    [[  [[     [[
    [[  [[  [[  [[      [[    [[[[   [[  [[      [[
[[  [[  [[  [[  [[      [[   [[  [[  [[  [[  [[  [[
 [[[[    [[[[  [[[[[[ [[[[[[ [[  [[   [[[[    [[[[
           [[
          [[    KERNEL: LongWing(C) 1.7.1
       [[[[   COPYRIGHT ACOINFO Co. Ltd. 2006 - 2018

SylixOS license: Commercial & GPL.
SylixOS kernel version: 1.7.1 Code name: Octopus

CPU      : TI TMS570LS3137 (Cortex-R4 160MHz VFPv3)
CACHE    : None Cache
PACKET   : SAST804 TMS570LS3137
ROM SIZE: 0x00200000 Bytes (0x00000000 - 0x001fffff)
RAM SIZE: 0x0003db00 Bytes (0x08002500 - 0x0803ffff)
BSP      : BSP version 1.0.0 for LongYuan
[root@sylixos:/root]# gnc
succeed to read
00 69 00 5e 00 64 08 3f fd fd 02 09 e1 df fa 1b 18 25 54 6f eb 3e 54 c6 00 58 77 08 96 83 f3 ba 4a 73 30 7b
succeed to send
14 05 2e e0 dd 37 00 00 03 47 0d 12 01 ca 08 3e 07 96 07 c8
欧拉角：8.391453 33.463791 4.581422   欧拉角速度：2.110098 1.942127 1.992280
succeed to read
00 6b 00 5b 00 62 08 19 fd cb 02 69 e0 ee fc 0d 17 56 54 72 1b 3e 51 9f 00 75 f4 08 7c 08 f3 a6 be 73 30 5a
succeed to send
17 d9 2e e0 da cd 00 00 04 36 0d e5 02 09 08 68 07 59 07 9e
欧拉角：10.783812 35.572594 5.210667   欧拉角速度：2.152531 1.881410 1.950349
succeed to read
00 6c 00 59 00 61 07 ee fd a0 02 c7 e0 1b fd fe 16 75 54 74 43 3e 4e 73 00 93 71 08 61 8d f3 93 34 73 30 31
succeed to send
1b ec 2e df d8 5c 00 00 05 2d 0e b4 02 38 08 7e 07 30 07 88
欧拉角：13.258311 37.640018 5.687293   欧拉角速度：2.174955 1.840540 1.928483
succeed to read
00 6e 00 56 00 60 07 c0 fd 7b 03 24 df 66 ff ec 15 83 54 76 65 3e 4b 43 00 b0 ee 08 47 10 f3 7f aa 73 2f fe
succeed to send
20 66 2e e0 d5 a7 00 00 06 2e 0f 80 02 5a 08 a9 06 f3 07 72
欧拉角：15.820176 39.681252 6.021484   欧拉角速度：2.217378 1.779517 1.906695
```

Figure 4-6 Running interface of the attitude control thread (output the sending and receiving data and attitude information)

## Results: 4.5 Processor In the Loop Test

The interface and working state of the inertial sensor are normal; The communication between the inertial sensor and the initial sample of the on-board computer is normal. The results obtained from the PIL test presented below.

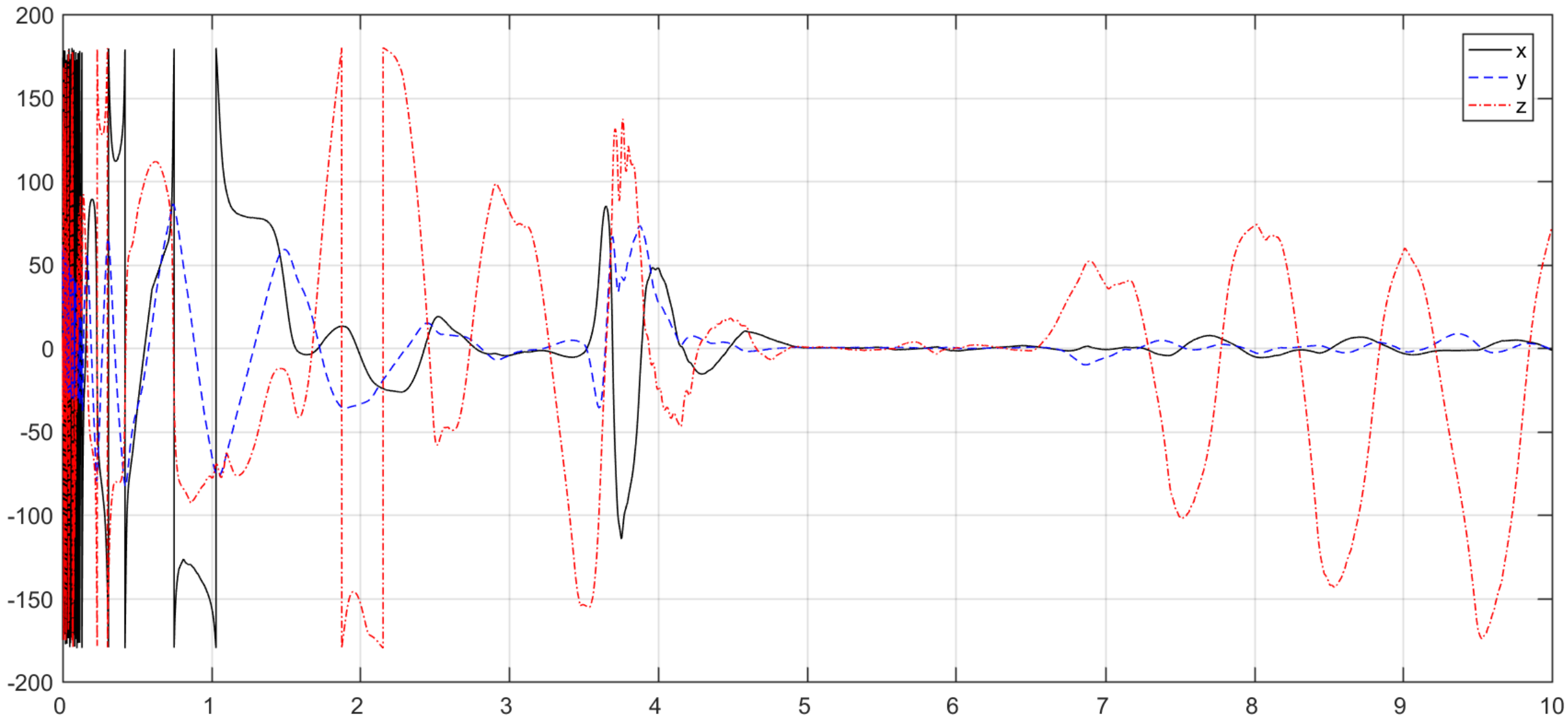

Figure 4-9 The Euler angles results from PIL test.

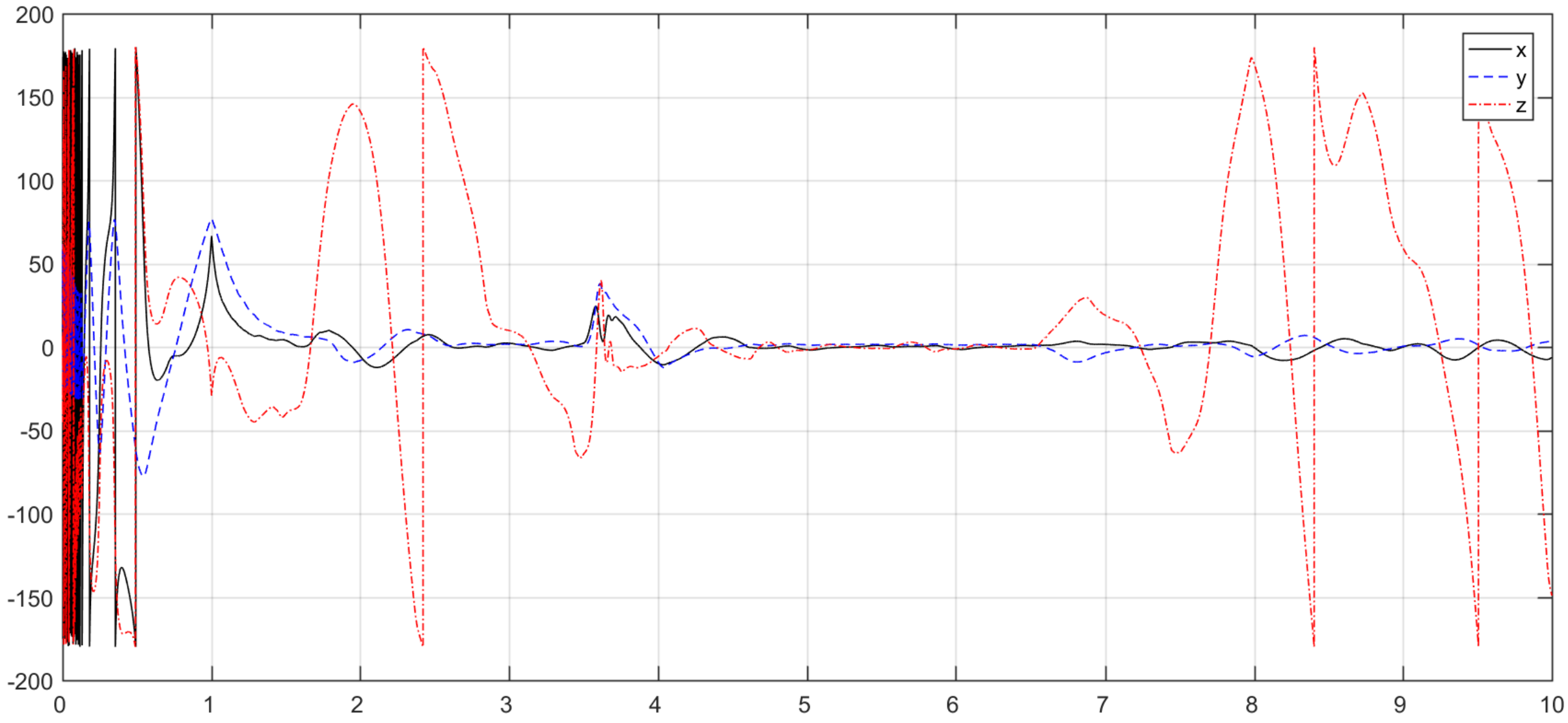

Figure 4-9 The Euler angles results from Simulink test.

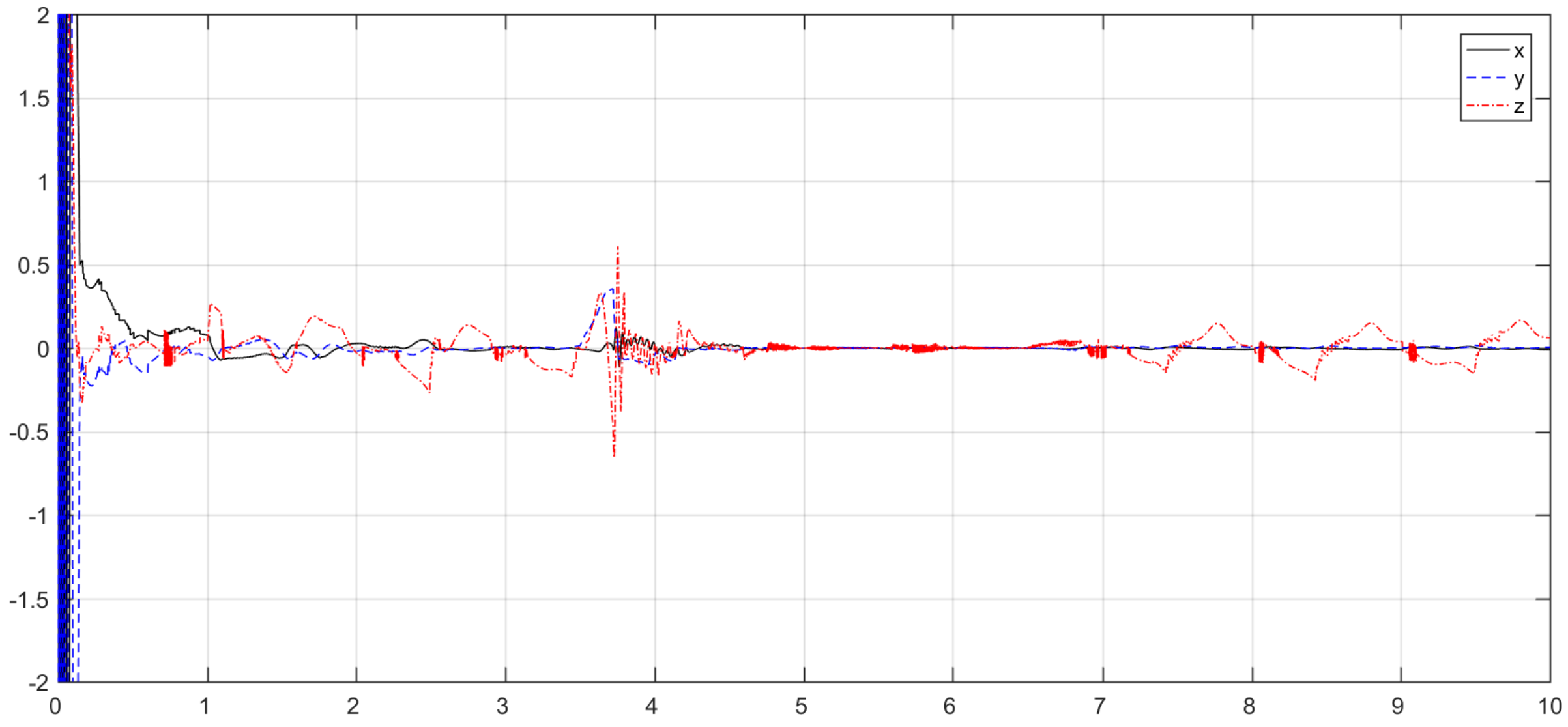

Figure 4-10 Angular velocities result from PIL test.

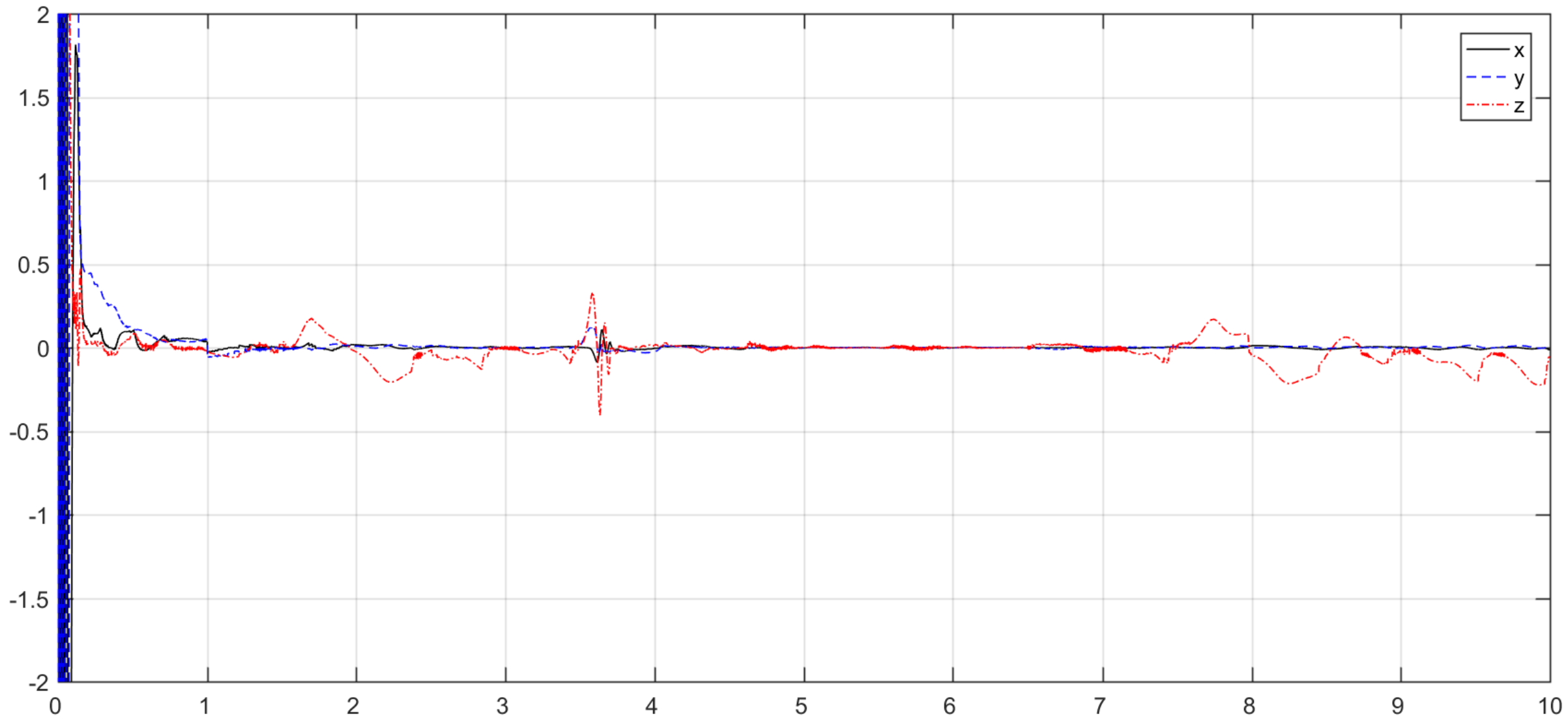

*Figure 4-11 Angular velocities result from Simulink test.*

The final results validate the performance, compatibility and reliability of the provided package. The minor difference between the Simulink and C code package in the obtained results can be explained by the lower accuracy in C code as fewer complex algorithms, less precision and several other factors will result in such an event.

However, this difference has been studied and it is proven that it doesn't affects the system performance and accuracy. The project requirements are meet and the test results can be considered as successful.

# 5. Conclusion

The objectives of the thesis have been met. The basic study on previous literatures is done. The mathematical model for satellite dynamics and kinematics, filters and controllers developed and designed. The designed software described with full detail. The Software-In-the-Loop test, Processor-In-the-Loop and several other tests have been performed. The system design results from Simulink presented in the table 2-4 obviously shows a better accuracy and stability for the system than the requirements.

The PIL test also explained and the results presented and compared with Simulink SIL test and the results are clearly the same with some small error which is negligible. The interfaces, drivers and all other sections are now complete.

The C code model of disturbances and orbital calculation and dynamic model, in addition to Simulink package for the same equivalents developed and tested. Such packages in C can be additionally added and uploaded on the on-board computer as the predictive packages in case of the failure of the sensors the predictive packages will offer an acceptable estimation of the parameters while some of them such as SGP4 model has the advantage of being usable all the time so less GPS data will be processed and retrieved which effects the power budget.

The provided package has a good quality in both performance and other parameters such as being user friendly and the code is also easy to read and understand in a combination of reading this thesis as all the API for all the functions are introduced.

Finally, for the future works, several optimizations on the designed algorithms can be applied. Some additional packages such as implementation of a voting algorithm can be added. The failure analysis and fault detection packages also can be improved at great extent. The filters and controllers also might be changing as the project evolving and receiving updates from the other member state countries cooperating in this project.

# 6. Acknowledgment

My deepest gratitude goes first and foremost to Professor Sun Liang, my supervisor, for his constant encouragement and guidance. He has walked me through all the stages of the writing of this thesis. Without his consistent and illuminating instruction, this thesis could not have reached its present form.

Last my thanks would go to my beloved family for their loving considerations and great confidence in me all through these years. I also owe my sincere gratitude to my friends and my fellow classmates who gave me their help and time in listening to me and helping me work out my problems during the difficult course of the thesis.

# 感谢

最深切的感谢首先是我的主管孙亮教授，感谢他不断的鼓励和指导。 他让我完成了本论文写作的各个阶段。 如果没有他的一贯和有启发性的指示，这篇论文就无法达到目前的形式。

最后，感谢我心爱的家人，感谢他们多年来的爱心和对我的信任。 我还要感谢我的朋友们和同学们，他们给了我们他们的帮助和时间，听我说话，帮助我在论文的艰难过程中解决我的问题。

# Appendices

```c
/*******************************************************************

    File Title:     main.c
    Created by:     Amir Hossein Alikhah Mishamandani
    Created at:     15:15:00, 14/2/2019
    Description:    The platform for integrated SIL test

*******************************************************************/

#include <stdio.h>
#include <inttypes.h>
#include "struct.h"
#include "orbit.h"
#include "Dynamic.h"
#include "AttitudeDeter.h"
#include "adcs_manager.h"

//Failure Flags
int FF = 0;

int main(){

    FILE *res;
    res = fopen("output.txt", "wb");

    Time CurrentTime;

    ADCS_DATA RunTimeData;

    FF = InitializationADCS(&RunTimeData,&CurrentTime);

    RunTimeData.CounterLoop = 0;

    for(int i = 0; i < 56761; i++){

    RunTimeData.CounterLoop = i;

    if(FF == 1){
        printf("\nError 0: The Initialization of ADCS failed ...!");
        FF = 0;
    }

    FF = Orbit(&RunTimeData,CurrentTime);

    if(FF == 1){
        printf("\nError 1: The calculation of orbital parameters failed ...!");
        FF = 0;
    }

    CurrentTime.sec = CurrentTime.sec + 1;

    TimeManager(&CurrentTime);
```

```c

    FF = AttitudeDetermination(&RunTimeData, CurrentTime);
    fprintf(res,"%20.20lf\t%20.20lf\t%20.20lf\n",
    RunTimeData.ECI_Pos[0], RunTimeData.ECI_Pos[1], RunTimeData.ECI_Pos[2]);
    if(FF == 1){
        printf("\nError 2: The Filters are failed ...!");
        FF = 0;
    }

    FF = MANAGER(RunTimeData, CurrentTime);



    if(FF == 1){
        printf("\nError 3: The ADCS manager failed ...!");
        FF = 0;
    }


    FF = Dynamic_Model(&RunTimeData);

    if(FF == 1){
        printf("\nError 4: The calculation of dynamic model failed ...!");
        FF = 0;
    }

    }

    fclose(res);

    return 0;
}

/* Test Print Output commands stored here

        fprintf(res,"6th Place: %20.20"PRIu16"\t%20.20"PRIu16"\t%20.20"PRIu16"\t%20.20"PRIu16"\t%20.20"PRIu16"\t%20.20lf\n",
                                                                        CurrentTime.year ,CurrentTime.mon ,CurrentTime.day ,CurrentTime.hr ,CurrentTime.min ,CurrentTime.sec);
        fprintf(res,"%20.20lf\t%20.20lf\t%20.20lf\t%20.20lf\t%20.20lf\t%20.20lf\n",
        RunTimeData.rsun_Inertia[0],RunTimeData.rsun_Inertia[1],RunTimeData.rsun_Inertia[2],
        RunTimeData.rmag_Inertia[0],RunTimeData.rmag_Inertia[1],RunTimeData.rmag_Inertia[2]
        );
        fprintf(res,"%20.20lf\t%20.20lf\t%20.20lf\t%20.20lf\t%20.20lf\t%20.20lf\n",
        RunTimeData.rsun_Body[0],RunTimeData.rsun_Body[1],RunTimeData.rsun_Body[2],
        RunTimeData.rmag_Body[0],RunTimeData.rmag_Body[1],RunTimeData.rmag_Body[2]
```

```
        );
        fprintf(res,"%20.20lf\t%20.20lf\t%20.20lf\t%20.20lf\t%20.20lf\t%20.20lf\n",
        RunTimeData.Omega_Orbit2Intertia[0], RunTimeData.Omega_Orbit2Intertia[1], RunTimeData.Omega_Orbit2Intertia[2],
        RunTimeData.Omega_B2I[0], RunTimeData.Omega_B2I[1], RunTimeData.Omega_B2I[2]
        );
        fprintf(res,"%20.20lf\t%20.20lf\t%20.20lf\n",
            RunTimeData.Tm[0],RunTimeData.Tm[1],RunTimeData.Tm[2]
        );


    fprintf(res,"%20.20lf\t%20.20lf\t%20.20lf\t%20.20lf\t%20.20lf\t%20.20lf\n",
    RunTimeData.Omega_Orbit2Intertia[0], RunTimeData.Omega_Orbit2Intertia[1], RunTimeData.Omega_Orbit2Intertia[2],
    RunTimeData.Omega_B2I[0], RunTimeData.Omega_B2I[1], RunTimeData.Omega_B2I[2]
    );

    fprintf(res,"%20.20lf\t%20.20lf\t%20.20lf\n",
        RunTimeData.Omega_B2I[0], RunTimeData.Omega_B2I[1], RunTimeData.Omega_B2I[2]
    );

    fprintf(res,"%20.20lf\t%20.20lf\t%20.20lf\t%20.20lf\t%20.20lf\t%20.20lf\n",
    RunTimeData.Quat_B2I[0], RunTimeData.Quat_B2I[1], RunTimeData.Quat_B2I[2],
    RunTimeData.Quat_B2I_prev[0], RunTimeData.Quat_B2I_prev[1], RunTimeData.Quat_B2I_prev[2]
    );

    fprintf(res,"%20.20d\n",RunTimeData.BZ1);

    fprintf(res,"%20.20lf\t%20.20lf\t%20.20lf\t%20.20lf\t%20.20lf\t%20.20lf\t%20.20lf\t%20.20lf\t%20.20lf\n",
    RunTimeData.DCM_B2I[0][0], RunTimeData.DCM_B2I[0][1], RunTimeData.DCM_B2I[0][2],
    RunTimeData.DCM_B2I[1][0], RunTimeData.DCM_B2I[1][1], RunTimeData.DCM_B2I[1][2],
    RunTimeData.DCM_B2I[2][0], RunTimeData.DCM_B2I[2][1], RunTimeData.DCM_B2I[2][2]
    );


    fprintf(res,"%20.20lf\t%20.20lf\t%20.20lf\t%20.20lf\t%20.20lf\t%20.20lf\n",
    RunTimeData.Quat_B2I[0], RunTimeData.Quat_B2I[1], RunTimeData.Quat_B2I[2],
    RunTimeData.Quat_B2I_prev[0], RunTimeData.Quat_B2I_prev[1],
```

```
        RunTimeData.Quat_B2I_prev[2]
    );


    fprintf(res,"%20.20lf\t%20.20lf\t%20.20lf\t%20.20lf\t%20.20lf\t%20.20lf\t%20.20lf\t%20.20lf\t%20.20lf\n",
    RunTimeData.DCM_B2I[0][0], RunTimeData.DCM_B2I[0][1], RunTimeData.DCM_B2I[0][2],
    RunTimeData.DCM_B2I[1][0], RunTimeData.DCM_B2I[1][1], RunTimeData.DCM_B2I[1][2],
    RunTimeData.DCM_B2I[2][0], RunTimeData.DCM_B2I[2][1], RunTimeData.DCM_B2I[2][2]
    );

fprintf(res,"%20.20lf\t%20.20lf\t%20.20lf\t%20.20lf\n",
    RunTimeData.Quat_B2I[0], RunTimeData.Quat_B2I[1], RunTimeData.Quat_B2I[2], RunTimeData.Quat_B2I[3]);
    );
    */
```

```c
/****************************************************************

    File Title:     struct.h
    Created by:     Amir Hossein Alikhah Mishamandani
    Created at:     15:19:00, 14/2/2019
    Description:    The unified data structure.

****************************************************************/

    #ifndef _STRUCT_H_
    #define _STRUCT_H_

    #include <stdio.h>
    #include <math.h>
    #include <stdint.h>

    typedef struct{

        //CounterLoop
        int32_t CounterLoop;
        //Satellite ECI Position
        double ECI_Pos[3];
        //Satellite ECI Position Previous value
        double ECI_Pos_Prev[3];
        //Satellite ECI Velocity
        double ECI_Vel[3];
        //Satellite ECI Velocity Previous value
        double ECI_Vel_Prev[3];
        //Normalized ECI Sun Position Vector
        double rsun_Inertia[3];
        //The Sun Position Vector in body frame
        double rsun_Body[3];
        //Geomagnetic field vector in nanotesla (nT) in eci frame
        double rmag_Inertia[3];
        //Geomagnetic field vector in nanotesla (nT) in body frame Delta
        double Delta_rmag_Body[3];
        //Geomagnetic field vector in nanotesla (nT) in body frame previous value
        double rmag_Body_prev[3];
        //Geomagnetic field vector in nanotesla (nT) in body frame
        double rmag_Body[3];
        //The rotation matrix of orbital system relative to the inertial system
        double DCM_B2I[3][3];
        //The Euler Rotation Matrix satellite body relative to orbital frame
        double DCM_B2O[3][3];
        //The Euler Rotation Matrix satellite body relative to orbital frame  previous value
        double DCM_B2O_Prev[3][3];
        //The quaternion of the orbital system relative to the inertial system
        double Quat_Orbit2Intertia[4];
        //The quaternion of the orbital system relative to the inertial
```

```
    system
        double Quat_B2O[4];
        //The quaternion of the orbital system relative to the inertial system previous value
        double Quat_B2O_prev[4];
        //The quaternion of the satellite body relative to the inertial system
        double Quat_B2I[4];
        //The quaternion of the satellite body relative to the inertial system previous value
        double Quat_B2I_prev[4];
        //The angular velocity of the orbital system relative to the inertial system
        double Omega_Orbit2Intertia[3];
        //The angular velocity of the satellite body relative to the inertial system
        double Omega_B2I[3];
        //The angular velocity of the satellite body relative to the orbital system
        double Omega_B2O[3];
        //The angular velocity of the satellite body relative to the inertial system previous value
        double Omega_B2I_prev[3];
        //Gravity gradient moment
        double Tg[3];
        //magnetic moment
        double Tc[3];
        //Angular Velocity Moment
        double Tw[3];
        //Gyro Out
        double GK[3];
        //sun sensor flag (True = 1 =  Sun, False = 0 = Shadow)
        int sun_sensor_flag;
        //Controller Output
        double Mc[3];
        //Controller output
        double Tm[3];
        //Flag
        int Hw;
        //Flag
        int Hw_set;
        //Flag
        int BZ6;
        //Flag
        int BZ1;
        //Flag (True = 1 =  not failed, False = 0 = failed)
        int FailureFlag;
        //Flag (True = 1 =  updated, False = 0 = no update)
        int ZT1UpdateFlag;
        //Flag
        int ZT1;
        //Kd gain of rate damping controller
        double Kd_damp[3][3];
```

```c
        //Kp gain of pointing controller
        double Kp_point[3][3];
        //Kd gain of pointing controller
        double Kd_point[3][3];
        //Kp gain of stabilization controller
        double Kp_stable[3][3];
        //Kd gain of stabilization controller
        double Kd_stable[3][3];
        //singular values of A
        double SVD[4];
        //Sampling time of Gyro
        double S_time;
        //GyroDrift
        double GyroDrift[3];
        //PK_1
        double PK1[7][7];
        //QK_1
        double QK1[7][7];
        //XK_1
        double XK1[7];
        //The measurement noise Covariance Matrix RK (diagonal)
        double RK[3][3];
        //XK
        double XK[7];
        //PK
        double PK[7][7];
    }ADCS_DATA;

    typedef struct{
    int16_t year;
    int16_t mon;
    int16_t day;
    int16_t hr;
    int16_t min;
    double sec;
    }Time;

    #define pi 3.1415926535897932384626433832795
    #define D2R (pi/180)
    #define R2D (180/pi)
    #define pi2 6.283185307179586476925286766559O
    #define ee 2.71828182845904553488480814849O3
    #define alpha_damp 1.0235987
    #define alpha_point 1.134464
    #define Mmax 12
    #define Hw_stand1 (4000*(0.067 / 6000))
    #define Hw_stand2 (5000*(0.067 / 6000))
    #define Hw_stand3 (2*(0.067 / 6000))

    extern ADCS_DATA RunTimeData;
    extern Time CurrentTime;
    static const double DEG2RAD = 3.14159265359 / 180.0;
    static const double u = 3.986004418e14;
```

```
/* Excerpt of constmath.m from Vallado (2013) Fundamentals of Astrodynamics and Applications */
static const double RAD2DEG = 180.0 / 3.14159265359;
static const double TWOPI = 2.0 * 3.14159265359;

/* Excerpt of constastro.m from Vallado (2013) Fundamentals of Astrodynamics and Applications */
// WGS-84/EGM-96 constants used here
static const double REQU      = 6378.137;         // km semimajor axis
static const double RPOL      = 6378.137*(1 - (1.0/298.257223563));  // km semiminor axis
static const double RMEAN     = 6371.2;           // km mean earth radius

// derived constants from the base values
static const double ECCSQRD = 2.0*(1.0/298.257223563) - (1.0/298.257223563)*(1.0/298.257223563);
static const double VELKMPS = sqrt(398600.4418 / 6378.137);

/* Excerpt fo getgravc.m from Vallado (2013) Fundamentals of Astrodynamics and Applications */
// WGS-84 constants
static const double XKE    = 1.0 / 13.4468520637;
static const double J2     =   0.00108262998905;
static const double J4     =  -0.00000161098761;
static const double J3OJ2  =  -0.00000253215306 / 0.00108262998905;

/* Excerpt of iau80in.m from Vallado (2013) Fundamentals of Astrodynamics and Applications */
static const int16_t IAR80[530] = {0, 0, 0, 0, 0, 1, 0, 0, 1, 0, 1, 0,
  -1, 1, 0, -1, -1, 1, 2, -2, 0, 2, 2, 1, 0, 0, -1, 0, 0, -1, 0, 1, 0,
  2, -1, 1, 0, 0, 1, 0, -2, 0, 2, 1, 1, 0, 0, 2, 1, 1, 0, 0, 1, 2, 0, 1,
  1, -1, 0, 1, 3, -2, 1, -1, 1, -1, 0, -2, 2, 3, 1, 0, 1, 1, 1, 0, 0,
  0, 1, 1, 1, 1, 2, 0, 0, -2, 2, 0, 0, 0, 0, 1, 3, -2, -1, 0, 0, -1, 2,
  2, 2, 2, 1, -1, -1, 0, 0, 0, 0, 0, 1, 0, 1, 0, 0, -1, 0, 0, 0, 0, 0,
  0, 0, 0, 0, 0, 0, 0, 0, 0, 0, 0, 0, 2, 2, 0, 1, 0, -1, 0, 0, 0, -1, 0,
  1, 1, 0, 0, 0, 0, 0, 0, -1, 0, -1, 0, 0, 1, 0, 0, 1, 1, -1, -1, -1,
  -1, 0, 0, 0, 0, 0, 0, -2, 0, 0, 0, 1, 0, 0, 0, 1, 1, 1, 1, 0, 0, 0, 0,
  0, 0, 0, 0, 0, -1, 0, 0, 1, 1, 0, 0, 0, 0, 1, 0, 1, 0, 0, 0, -1, 0,
  -1, 1, 0, 2, 2, 0, 0, 0, 2, 2, 2, 2, 0, 2, 2, 0, 0, 2, 0, 2, 0, 2, 2,
  2, 0, 2, 2, 2, 2, 0, 2, 0, 0, 0, 0, -2, 2, 2, 2, 2, 0, 2, 0, 0, 2, 0,
  2, 0, 2, 2, 0, 0, 0, 0, -2, 0, 2, 0, 0, 2, 2, 2, 2, 2, 2, 2, 0, 2, 2,
  0, 0, 0, 2, 2, 0, 2, 0, 0, 2, -2, -2, -2, 2, 0, 0, 2, 2, 2, 2, 2, -2,
  4, 0, 2, 2, 2, 0, -2, 2, 4, 0, 0, 2, -2, 0, 0, 0, 0, 0, -2, 0, 0, 0,
  0, -2, 0, 0, -2, -2, -2, 0, 0, 2, 2, 0, 0, -2, 0, 2, 0, 0, -2, 0, -2,
  0, 0, -2, 2, 0, -2, 0, 0, 2, 2, 0, 2, -2, 0, 2, 2, -2, 2, -2, -2, -2,
  0, 0, -1, 1, -2, 0, -2, -2, 0, -1, 2, 2, 0, 0, 0, 0, 4, 0, -2, -2, 0,
  0, 0, 0, 1, 2, 2, -2, 2, -2, 2, 2, -2, -2, -4, -4, 4, -1, 4, 2, 0, 0,
  -2, 0, -2, -2, 2, 0, 2, 0, 0, -2, 2, -2, 0, -2, 1, 2, 1, 1, 2, 2, 2,
  0, 0, 2, 1, 2, 2, 0, 1, 2, 1, 0, 2, 1, 1, 0, 1, 2, 2, 0, 2, 0, 0, 1,
  0, 2, 1, 1, 1, 1, 0, 1, 2, 2, 1, 0, 2, 1, 1, 2, 0, 1, 1, 1, 1, 0, 0,
  0, 0, 0, 1, 1, 0, 0, 2, 2, 2, 2, 2, 0, 2, 2, 1, 1, 1, 1, 0, 2, 2, 1,
  1, 1, 0, 0, 0, 0, 0, 0, 0, 0, 2, 2, 2, 2, 1, 1, 2, 2, 2, 2, 2, 2, 1,
```

```
        1, 2, 0, 0, 1, 1, 0, 1, 1, 0};
    static const double RAR80[424] = {-171996.0, -13187.0, -2274.0, 2062.0,
        1426.0, 712.0, -517.0, -386.0, -301.0, 217.0, -158.0, 129.0, 123.0,
        63.0, 63.0, -59.0, -58.0, -51.0, 48.0, 46.0, -38.0, -31.0, 29.0, 29.0,
         26.0, -22.0, 21.0, 17.0, -16.0, 16.0, -15.0, -13.0, -12.0, 11.0,
        -10.0, -8.0, -7.0, -7.0, -7.0, 7.0, -6.0, -6.0, 6.0, 6.0, 6.0, -5.0,
        -5.0, -5.0, 5.0, -4.0, -4.0, -4.0, 4.0, 4.0, 4.0, -3.0, -3.0, -3.0,
        -3.0, -3.0, -3.0, -3.0, 3.0, -2.0, -2.0, -2.0, -2.0, -2.0, 2.0, 2.0,
        2.0, 2.0, -1.0, -1.0, -1.0, -1.0, -1.0, -1.0, -1.0, -1.0, -1.0, -1.0,
        -1.0, -1.0, -1.0, -1.0, -1.0, -1.0, -1.0, 1.0, 1.0, 1.0, 1.0, 1.0,
        1.0, 1.0, 1.0, 1.0, 1.0, 1.0, 1.0, 1.0, 1.0, 1.0, 1.0, 1.0,  -174.2,
        -1.6, -0.2, 0.2, -3.4, 0.1, 1.2, -0.4, 0.0, -0.5, 0.0, 0.1, 0.0, 0.1,
        0.0, 0.0, -0.1, 0.0, 0.0, 0.0, 0.0, 0.0, 0.0, 0.0, 0.0, 0.0, 0.0,
        -0.1, 0.1, 0.0, 0.0, 0.0, 0.0, 0.0, 0.0, 0.0, 0.0, 0.0, 0.0, 0.0, 0.0,
         0.0, 0.0, 0.0, 0.0, 0.0, 0.0, 0.0, 0.0, 0.0, 0.0, 0.0, 0.0, 0.0, 0.0,
         0.0, 0.0, 0.0, 0.0, 0.0, 0.0, 0.0, 0.0, 0.0, 0.0, 0.0, 0.0, 0.0, 0.0,
         0.0, 0.0, 0.0, 0.0, 0.0, 0.0, 0.0, 0.0, 0.0, 0.0, 0.0, 0.0, 0.0, 0.0,
         0.0, 0.0, 0.0, 0.0, 0.0, 0.0, 0.0, 0.0, 0.0, 0.0, 0.0, 0.0, 0.0, 0.0,
         0.0, 0.0, 0.0, 0.0, 0.0, 0.0, 0.0, 0.0, 0.0, 92025.0, 5736.0, 977.0,
        -895.0, 54.0, -7.0, 224.0, 200.0, 129.0, -95.0, -1.0, -70.0, -53.0,
        -33.0, -2.0, 26.0, 32.0, 27.0, 1.0, -24.0, 16.0, 13.0, -1.0, -12.0,
        -1.0, 0.0, -10.0, 0.0, 7.0, -8.0, 9.0, 7.0, 6.0, 0.0, 5.0, 3.0, 3.0,
        3.0, 0.0, -3.0, 3.0, 3.0, -3.0, 0.0, -3.0, 3.0, 3.0, 3.0, 0.0, 0.0,
        0.0, 0.0, 0.0, -2.0, -2.0, 0.0, 0.0, 1.0, 1.0, 1.0, 1.0, 1.0, 0.0,
        1.0, 1.0, 1.0, 1.0, 1.0, -1.0, 0.0, -1.0, -1.0, 0.0, 1.0, 0.0, 0.0,
        0.0, 0.0, 0.0, 0.0, 0.0, 0.0, 0.0, 0.0, 0.0, 1.0, 0.0, 0.0, 0.0, 0.0,
        0.0, -1.0, 0.0, -1.0, -1.0, 0.0, 0.0, 0.0, 0.0, 0.0, -1.0, 0.0, 0.0,
        0.0, 0.0, 0.0, 8.9, -3.1, -0.5, 0.5, -0.1, 0.0, -0.6, 0.0, -0.1, 0.3,
        0.0, 0.0, 0.0, 0.0, 0.0, 0.0, 0.0, 0.0, 0.0, 0.0, 0.0, 0.0, 0.0, 0.0,
        0.0, 0.0, 0.0, 0.0, 0.0, 0.0, 0.0, 0.0, 0.0, 0.0, 0.0, 0.0, 0.0, 0.0,
        0.0, 0.0, 0.0, 0.0, 0.0, 0.0, 0.0, 0.0, 0.0, 0.0, 0.0, 0.0, 0.0, 0.0,
        0.0, 0.0, 0.0, 0.0, 0.0, 0.0, 0.0, 0.0, 0.0, 0.0, 0.0, 0.0, 0.0, 0.0,
        0.0, 0.0, 0.0, 0.0, 0.0, 0.0, 0.0, 0.0, 0.0, 0.0, 0.0, 0.0, 0.0, 0.0,
        0.0, 0.0, 0.0, 0.0, 0.0, 0.0, 0.0, 0.0, 0.0, 0.0, 0.0, 0.0, 0.0, 0.0,
        0.0, 0.0, 0.0, 0.0, 0.0, 0.0, 0.0, 0.0, 0.0, 0.0, 0.0, 0.0};

    /* Polar motion estimation coefficients */
    // Values taken from IERS Bulletin A 12 February 2015 Vol. XXVIII No.
        007 http://www.iers.org/IERS/EN/Publications/Bulletins/bulletins.html
    static const int16_t MJDPOLAR = 57065;
    static const double XPOLAR[5] = {0.1095, -0.0737, -0.0326, -0.0373,
        0.0080};
    static const double YPOLAR[5] = {0.3519, -0.0293, 0.0661, 0.0080,
        0.0373};

    /* IGRF12 coefficients */
    // These are used in a function originally written in MATLAB. MATLAB
        begins indexing with 1, C with 0. "0.0" was thus added as 0th element.
    static const double IGRFGH[196] = {0.0, -29442.0, -1501.0, 4797.1,
        -2445.1, 3012.9, -2845.6, 1676.7, -641.9, 1350.7, -2352.3, -115.3,
        1225.6, 244.9, 582.0, -538.4, 907.6, 813.7, 283.3, 120.4, -188.7,
        -334.9, 180.9, 70.4, -329.5, -232.6, 360.1, 47.3, 192.4, 197.0,
        -140.9, -119.3, -157.5, 16.0, 4.1, 100.2, 70.0, 67.7, -20.8, 72.7,
```

```
        33.2, -129.9, 58.9, -28.9, -66.7, 13.2, 7.3, -70.9, 62.6, 81.6, -76.1,
         -54.1, -6.8, -19.5, 51.8, 5.7, 15.0, 24.4, 9.4, 3.4, -2.8, -27.4,
        6.8, -2.2, 24.2, 8.8, 10.1, -16.9, -18.3, -3.2, 13.3, -20.6, -14.6,
        13.4, 16.2, 11.7, 5.7, -15.9, -9.1, -2.0, 2.1, 5.4, 8.8, -21.6, 3.1,
        10.8, -3.3, 11.8, 0.7, -6.8, -13.3, -6.9, -0.1, 7.8, 8.7, 1.0, -9.1,
        -4.0, -10.5, 8.4, -1.9, -6.3, 3.2, 0.1, -0.4, 0.5, 4.6, -0.5, 4.4,
        1.8, -7.9, -0.7, -0.6, 2.1, -4.2, 2.4, -2.8, -1.8, -1.2, -3.6, -8.7,
        3.1, -1.5, -0.1, -2.3, 2.0, 2.0, -0.7, -0.8, -1.1, 0.6, 0.8, -0.7,
        -0.2, 0.2, -2.2, 1.7, -1.4, -0.2, -2.5, 0.4, -2.0, 3.5, -2.4, -1.9,
        -0.2, -1.1, 0.4, 0.4, 1.2, 1.9, -0.8, -2.2, 0.9, 0.3, 0.1, 0.7, 0.5,
        -0.1, -0.3, 0.3, -0.4, 0.2, 0.2, -0.9, -0.9, -0.1, 0.0, 0.7, 0.0,
        -0.9, -0.9, 0.4, 0.4, 0.5, 1.6, -0.5, -0.5, 1.0, -1.2, -0.2, -0.1,
        0.8, 0.4, -0.1, -0.1, 0.3, 0.4, 0.1, 0.5, 0.5, -0.3, -0.4, -0.4, -0.3,
         -0.8};
    static const double IGRFGHSV[196] = {0.0, 10.3, 18.1, -26.6, -8.7, -3.3,
         -27.4, 2.1, -14.1, 3.4, -5.5, 8.2, -0.7, -0.4, -10.1, 1.8, -0.7, 0.2,
         -1.3, -9.1, 5.3, 4.1, 2.9, -4.3, -5.2, -0.2, 0.5, 0.6, -1.3, 1.7,
        -0.1, -1.2, 1.4, 3.4, 3.9, 0.0, -0.3, -0.1, 0.0, -0.7, -2.1, 2.1,
        -0.7, -1.2, 0.2, 0.3, 0.9, 1.6, 1.0, 0.3, -0.2, 0.8, -0.5, 0.4, 1.3,
        -0.2, 0.1, -0.3, -0.6, -0.6, -0.8, 0.1, 0.2, -0.2, 0.2, 0.0, -0.3,
        -0.6, 0.3, 0.5, 0.1, -0.2, 0.5, 0.4, -0.2, 0.1, -0.3, -0.4, 0.3, 0.3,
        0.0, 0.0, 0.0, 0.0, 0.0, 0.0, 0.0, 0.0, 0.0, 0.0, 0.0, 0.0, 0.0, 0.0,
        0.0, 0.0, 0.0, 0.0, 0.0, 0.0, 0.0, 0.0, 0.0, 0.0, 0.0, 0.0, 0.0, 0.0,
        0.0, 0.0, 0.0, 0.0, 0.0, 0.0, 0.0, 0.0, 0.0, 0.0, 0.0, 0.0, 0.0, 0.0,
        0.0, 0.0, 0.0, 0.0, 0.0, 0.0, 0.0, 0.0, 0.0, 0.0, 0.0, 0.0, 0.0, 0.0,
        0.0, 0.0, 0.0, 0.0, 0.0, 0.0, 0.0, 0.0, 0.0, 0.0, 0.0, 0.0, 0.0, 0.0,
        0.0, 0.0, 0.0, 0.0, 0.0, 0.0, 0.0, 0.0, 0.0, 0.0, 0.0, 0.0, 0.0, 0.0,
        0.0, 0.0, 0.0, 0.0, 0.0, 0.0, 0.0, 0.0, 0.0, 0.0, 0.0, 0.0, 0.0, 0.0,
        0.0, 0.0, 0.0, 0.0, 0.0, 0.0, 0.0, 0.0, 0.0, 0.0, 0.0, 0.0, 0.0, 0.0,
        0.0, 0.0, 0.0, 0.0};

    /* WMM2015 coefficients */
    // As above WMMG[0] etc. are just placeholders
    static const double WMMG[91] = {0.0, -29438.5, -1501.1, -2445.3, 3012.5,
         1676.6, 1351.1, -2352.3, 1225.6, 581.9, 907.2, 813.7, 120.3, -335.0,
        70.3, -232.6, 360.1, 192.4, -141.0, -157.4, 4.3, 69.5, 67.4, 72.8,
        -129.8, -29.0, 13.2, -70.9, 81.6, -76.1, -6.8, 51.9, 15.0, 9.3, -2.8,
        6.7, 24.0, 8.6, -16.9, -3.2, -20.6, 13.3, 11.7, -16.0, -2.0, 5.4, 8.8,
         3.1, -3.1, 0.6, -13.3, -0.1, 8.7, -9.1, -10.5, -1.9, -6.5, 0.2, 0.6,
        -0.6, 1.7, -0.7, 2.1, 2.3, -1.8, -3.6, 3.1, -1.5, -2.3, 2.1, -0.9,
        0.6, -0.7, 0.2, 1.7, -0.2, 0.4, 3.5, -2.0, -0.3, 0.4, 1.3, -0.9, 0.9,
        0.1, 0.5, -0.4, -0.4, 0.2, -0.9, 0.0};
    static const double WMMGSV[91] = {0.0, 10.7, 17.9, -8.6, -3.3, 2.4, 3.1,
         -6.2, -0.4, -10.4, -0.4, 0.8, -9.2, 4.0, -4.2, -0.2, 0.1, -1.4, 0.0,
        1.3, 3.8, -0.5, -0.2, -0.6, 2.4, -1.1, 0.3, 1.5, 0.2, -0.2, -0.4, 1.3,
         0.2, -0.4, -0.9, 0.3, 0.0, 0.1, -0.5, 0.5, -0.2, 0.4, 0.2, -0.4, 0.3,
         0.0, -0.1, -0.1, 0.4, -0.5, -0.2, 0.1, 0.0, -0.2, -0.1, 0.0, 0.0,
        -0.1, 0.3, -0.1, -0.1, -0.1, 0.0, -0.2, -0.1, -0.2, 0.0, 0.0, -0.1,
        0.1, 0.0, 0.0, 0.0, 0.0, 0.0, 0.0, -0.1, -0.1, 0.1, 0.0, 0.0, 0.1,
        -0.1, 0.0, 0.1, 0.0, 0.0, 0.0, 0.0, 0.0, 0.0};
    static const double WMMH[91] = {0.0, 0.0, 4796.2, 0.0, -2845.6, -642.0,
        0.0, -115.3, 245.0, -538.3, 0.0, 283.4, -188.6, 180.9, -329.5, 0.0,
        47.4, 196.9, -119.4, 16.1, 100.1, 0.0, -20.7, 33.2, 58.8, -66.5, 7.3,
```

```
        62.5, 0.0, -54.1, -19.4, 5.6, 24.4, 3.3, -27.5, -2.3, 0.0, 10.2,
        -18.1, 13.2, -14.6, 16.2, 5.7, -9.1, 2.2, 0.0, -21.6, 10.8, 11.7,
        -6.8, -6.9, 7.8, 1.0, -3.9, 8.5, 0.0, 3.3, -0.3, 4.6, 4.4, -7.9, -0.6,
         -4.1, -2.8, -1.1, -8.7, 0.0, -0.1, 2.1, -0.7, -1.1, 0.7, -0.2, -2.1,
        -1.5, -2.5, -2.0, -2.3, 0.0, -1.0, 0.5, 1.8, -2.2, 0.3, 0.7, -0.1,
        0.3, 0.2, -0.9, -0.2, 0.7};
    static const double WMMHSV[91] = {0.0, 0.0, -26.8, 0.0, -27.1, -13.3,
        0.0, 8.4, -0.4, 2.3, 0.0, -0.6, 5.3, 3.0, -5.3, 0.0, 0.4, 1.6, -1.1,
        3.3, 0.1, 0.0, 0.0, -2.2, -0.7, 0.1, 1.0, 1.3, 0.0, 0.7, 0.5, -0.2,
        -0.1, -0.7, 0.1, 0.1, 0.0, -0.3, 0.3, 0.3, 0.6, -0.1, -0.2, 0.3, 0.0,
        0.0, -0.2, -0.1, -0.2, 0.1, 0.1, 0.0, -0.2, 0.4, 0.3, 0.0, 0.1, -0.1,
        0.0, 0.0, -0.2, 0.1, -0.1, -0.2, 0.1, -0.1, 0.0, 0.0, 0.1, 0.0, 0.1,
        0.0, 0.0, 0.1, 0.0, -0.1, 0.0, -0.1, 0.0, 0.0, 0.0, -0.1, 0.0, 0.0,
        0.0, 0.0, 0.0, 0.0, 0.0, 0.0, 0.0};

    int InitializationADCS(ADCS_DATA *RunTimeData, Time *CurrentTime);



    #endif
```

```c
/*****************************************************************

    File Title:     struct.c
    Created by:     Amir Hossein Alikhah Mishamandani
    Created at:     15:19:00, 14/2/2019
    Description:    The unified data structure.

*****************************************************************/

#include "struct.h"

int InitializationADCS(ADCS_DATA *RunTimeData, Time *CurrentTime){

    RunTimeData->BZ1 = 2;
    RunTimeData->BZ6 = 1;
    RunTimeData->ZT1 = 2;
    RunTimeData->ZT1UpdateFlag = 0;
    RunTimeData->FailureFlag = 0;

    CurrentTime->year = 2018;
    CurrentTime->mon = 11;
    CurrentTime->day = 20;
    CurrentTime->hr = 0;
    CurrentTime->min = 0;
    CurrentTime->sec = 0;

    RunTimeData->ECI_Pos_Prev[0] = 5531956.71943065;
    RunTimeData->ECI_Pos_Prev[1] = 4087324.72958965;
    RunTimeData->ECI_Pos_Prev[2] = 0.0;
    RunTimeData->ECI_Vel_Prev[0] = 583.152304189998;
    RunTimeData->ECI_Vel_Prev[1] = -789.262787040461;
    RunTimeData->ECI_Vel_Prev[2] = 7549.09271648891;

    RunTimeData->GK[0]=0.174532925199433;
    RunTimeData->GK[1]=0.174532925199433;
    RunTimeData->GK[2]=0.174532925199433;

    RunTimeData->rmag_Body_prev[0] = 0;
    RunTimeData->rmag_Body_prev[1] = 0;
    RunTimeData->rmag_Body_prev[2] = 0;

    RunTimeData->Quat_B2I_prev[0] = 0.212208295572777;
    RunTimeData->Quat_B2I_prev[1] = -0.480140056392978;
    RunTimeData->Quat_B2I_prev[2] = 0.257986386781379;
    RunTimeData->Quat_B2I_prev[3] = 0.811095672391694;

    RunTimeData->sun_sensor_flag = 1;

    RunTimeData->Hw = 0 * (0.067 / 6000);


    RunTimeData->Omega_B2I_prev[0] = 10 * D2R;
    RunTimeData->Omega_B2I_prev[0] = 10 * D2R;
```

```
    RunTimeData->Omega_B2I_prev[0] = 10 * D2R;


    RunTimeData->Quat_B2O_prev[0] = 0.0150954361579326;
    RunTimeData->Quat_B2O_prev[1] = 0.233865102362646;
    RunTimeData->Quat_B2O_prev[2] = 0.0203567811834056;
    RunTimeData->Quat_B2O_prev[3] = 0.971938703398502;

    for(int i = 0; i < 3;i++){
        for(int j = 0; j < 3;j++){
            RunTimeData->Kd_damp[i][j] = 0;
        }
    }
    RunTimeData->Kd_damp[0][0] = 5.1101;
    RunTimeData->Kd_damp[1][1] = 5.0611;
    RunTimeData->Kd_damp[2][2] = 9.4984;

    for(int i = 0; i < 3;i++){
        for(int j = 0; j < 3;j++){
            RunTimeData->Kp_point[i][j] = 0;
        }
    }
    RunTimeData->Kp_point[0][0] = 0.0378;
    RunTimeData->Kp_point[1][1] = 0.03775;
    RunTimeData->Kp_point[2][2] = 0.00263;

    for(int i = 0; i < 3;i++){
        for(int j = 0; j < 3;j++){
            RunTimeData->Kd_point[i][j] = 0;
        }
    }
    RunTimeData->Kd_point[0][0] = 22.7481;
    RunTimeData->Kd_point[1][1] = 22.7184;
    RunTimeData->Kd_point[2][2] = 3.1661;

    for(int i = 0; i < 3;i++){
        for(int j = 0; j < 3;j++){
            RunTimeData->Kp_stable[i][j] = 0;
        }
    }
    RunTimeData->Kp_stable[0][0] = 0.0378;
    RunTimeData->Kp_stable[1][1] = 0.03775;
    RunTimeData->Kp_stable[2][2] = 0.00263;

    for(int i = 0; i < 3;i++){
        for(int j = 0; j < 3;j++){
            RunTimeData->Kd_stable[i][j] = 0;
        }
    }
    RunTimeData->Kd_stable[0][0] = 22.7481;
    RunTimeData->Kd_stable[1][1] = 22.7184;
    RunTimeData->Kd_stable[2][2] = 3.1661;

```

```
    RunTimeData->sun_sensor_flag = 1;

    RunTimeData->DCM_B2O_Prev[0][0] = RunTimeData->Quat_B2O_prev[0] *
      RunTimeData->Quat_B2O_prev[0] - RunTimeData->Quat_B2O_prev[1] *
      RunTimeData->Quat_B2O_prev[1] - RunTimeData->Quat_B2O_prev[2] *
      RunTimeData->Quat_B2O_prev[2] + RunTimeData->Quat_B2O_prev[3] *
      RunTimeData->Quat_B2O_prev[3];
    RunTimeData->DCM_B2O_Prev[0][1] = 2 * RunTimeData->Quat_B2O_prev[0] *
      RunTimeData->Quat_B2O_prev[1] + 2 * RunTimeData->Quat_B2O_prev[2] *
      RunTimeData->Quat_B2O_prev[3];
    RunTimeData->DCM_B2O_Prev[0][2] = 2 * RunTimeData->Quat_B2O_prev[0] *
      RunTimeData->Quat_B2O_prev[2] - 2 * RunTimeData->Quat_B2O_prev[1] *
      RunTimeData->Quat_B2O_prev[3];
    RunTimeData->DCM_B2O_Prev[1][0] = 2 * RunTimeData->Quat_B2O_prev[0] *
      RunTimeData->Quat_B2O_prev[1] - 2 * RunTimeData->Quat_B2O_prev[2] *
      RunTimeData->Quat_B2O_prev[3];
    RunTimeData->DCM_B2O_Prev[1][1] = -RunTimeData->Quat_B2O_prev[0] *
      RunTimeData->Quat_B2O_prev[0] + RunTimeData->Quat_B2O_prev[1] *
      RunTimeData->Quat_B2O_prev[1] - RunTimeData->Quat_B2O_prev[2] *
      RunTimeData->Quat_B2O_prev[2] + RunTimeData->Quat_B2O_prev[3] *
      RunTimeData->Quat_B2O_prev[3];
    RunTimeData->DCM_B2O_Prev[1][2] = 2 * RunTimeData->Quat_B2O_prev[0] *
      RunTimeData->Quat_B2O_prev[3] + 2 * RunTimeData->Quat_B2O_prev[1] *
      RunTimeData->Quat_B2O_prev[2];
    RunTimeData->DCM_B2O_Prev[2][0] = 2 * RunTimeData->Quat_B2O_prev[0] *
      RunTimeData->Quat_B2O_prev[2] + 2 * RunTimeData->Quat_B2O_prev[1] *
      RunTimeData->Quat_B2O_prev[3];
    RunTimeData->DCM_B2O_Prev[2][1] = 2 * RunTimeData->Quat_B2O_prev[1] *
      RunTimeData->Quat_B2O_prev[2] - 2 * RunTimeData->Quat_B2O_prev[0] *
      RunTimeData->Quat_B2O_prev[3];
    RunTimeData->DCM_B2O_Prev[2][2] = -RunTimeData->Quat_B2O_prev[0] *
      RunTimeData->Quat_B2O_prev[0] - RunTimeData->Quat_B2O_prev[1] *
      RunTimeData->Quat_B2O_prev[1] + RunTimeData->Quat_B2O_prev[2] *
      RunTimeData->Quat_B2O_prev[2] + RunTimeData->Quat_B2O_prev[3] *
      RunTimeData->Quat_B2O_prev[3];


    return 0;
}
```

```c
/*******************************************************************

    File Title:     utils.h
    Created by:     Amir Hossein Alikhah Mishamandani
    Created at:     15:23:00, 14/2/2019
    Description:    The unified math functions toolbox.

*******************************************************************/

#ifndef _UTILS_H_
#define _UTILS_H_

#include <stdio.h>
#include "struct.h"
#include "math.h"

#define MGet(p,i,j,n)  (p[(i)*(n)+(j)])

double norm(double *In, int len);
void rotmtx2quat(double In[3][3], double *q);
void Unit(double *In, double *out);
void MatProd(double *C1, double *C2, double C[3][3]);
double MaxArray(double *In, int len);
void printbuff(char *title, double *In, int len);
void printbuffMatrix(char *title, double In[3][3]);
#endif
```

```c
/***************************************************************

    File Title:     utils.c
    Created by:     Amir Hossein Alikhah Mishamandani
    Created at:     15:23:00, 14/2/2019
    Description:    The unified math functions toolbox.

***************************************************************/

#include "utils.h"

double norm(double *In, int len){
    double out_norm;
    for(int i = 0; i < len; i++){
        out_norm = out_norm + (In[i] * In[i]);
    }
    out_norm = sqrt(out_norm);
    return out_norm;
}

void Unit(double *In, double *out){
    double Cnorm = norm(In, 3);
    for (int i = 0; i<3; i++){
        out[i] = In[i]/Cnorm;
    }
}



void MatProd(double *C1, double *C2, double C[3][3]){
    int i,j;
    for (i=0;i<3;i++){
        for(j=0;j<3;j++){
            C[i][j] = 0.0;
        }
    }
    for (i=0;i<3;i++){
        for(j=0;j<3;j++){
            C[i][j] = C[i][j] + C2[i] * C1[j];
        }
    }
}


double MaxArray(double *In, int len){
    double MaxVal = 0;
    for (int c = 0; c < len; c++) {
        if (In[c] > MaxVal) {
           MaxVal = In[c];
        }
    }
    return MaxVal;
}
```

```c

void rotmtx2quat(double In[3][3], double *q){

    q[2] = 0;
    double sqtrp1;
    double sqtrp1x2;
    double d[3];
    double sqdip1;
    double rotmtx[9];

    rotmtx[0] = In[0][0];rotmtx[1] = In[0][1];rotmtx[2] = In[0][2];
    rotmtx[3] = In[1][0];rotmtx[4] = In[1][1];rotmtx[5] = In[1][2];
    rotmtx[6] = In[2][0];rotmtx[7] = In[2][1];rotmtx[8] = In[2][2];

    const double tr = rotmtx[0] + rotmtx[4] + rotmtx[8];
    if (tr > 1e-6f) {
        sqtrp1 = sqrt(tr + 1.0f);
        sqtrp1x2 = 2.0 * sqtrp1;

        q[3] = 0.5f * sqtrp1;
        q[0] = -(rotmtx[7] - rotmtx[5]) / sqtrp1x2;
        q[1] = -(rotmtx[2] - rotmtx[6]) / sqtrp1x2;
        q[2] = -(rotmtx[3] - rotmtx[1]) / sqtrp1x2;
    }
    else {
        d[0] = rotmtx[0];
        d[1] = rotmtx[4];
        d[2] = rotmtx[8];

        if ((d[1] > d[0]) && (d[1] > d[2])) {
            sqdip1 = sqrt(d[1] - d[0] - d[2] + 1.0f);
            q[1] = 0.5f * sqdip1;

            if (fabsf(sqdip1) > 1e-6f)
                sqdip1 = 0.5f / sqdip1;
            else
                sqdip1 = 0.0f;

            q[3] = -(rotmtx[2] - rotmtx[6]) * sqdip1;
            q[0] = (rotmtx[3] + rotmtx[1]) * sqdip1;
            q[2] = (rotmtx[7] + rotmtx[5]) * sqdip1;
        }
        else if (d[2] > d[0]) {
            //max value at R(3,3)
            sqdip1 = sqrt(d[2] - d[0] - d[1] + 1.0f);

            q[2] = 0.5f * sqdip1;

            if (fabsf(sqdip1) > 1e-6f)
                sqdip1 = 0.5f / sqdip1;
            else
                sqdip1 = 0.0f;

```

```c
            q[3] = -(rotmtx[3] - rotmtx[1]) * sqdip1;
            q[0] = (rotmtx[2] + rotmtx[6]) * sqdip1;
            q[1] = (rotmtx[7] + rotmtx[5]) * sqdip1;
        }
        else {
            // max value at R(1,1)
            sqdip1 = sqrt(d[0] - d[1] - d[2] + 1.0f);

            q[0] = 0.5f * sqdip1;

            if (fabsf(sqdip1) > 1e-6f)
                sqdip1 = 0.5f / sqdip1;
            else
                sqdip1 = 0.0f;

            q[3] = -(rotmtx[7] - rotmtx[5]) * sqdip1;
            q[1] = (rotmtx[3] + rotmtx[1]) * sqdip1;
            q[2] = (rotmtx[2] + rotmtx[6]) * sqdip1;
        }
    }

}

void printbuff(char *title, double *In, int len){
    printf("%s\t",title);
     double res;
    for(int i = 0; i < len; i++){
        res = 0;
        res = In[i];
        printf("%50.20lf\t", res);
    }
    printf("\n");
}

void printbuffMatrix(char *title, double In[3][3]){
    printf("%s",title);
    for(int i = 0; i < 3; i++){
        for(int j = 0; j < 3; j++){
            printf("%50.20lf\t", In[i][j]);
        }
        printf("\n");
    }
    printf("\n");
}
```

```c
/*****************************************************************

    File Title:     AttitudeDeter.h
    Created by:     Amir Hossein Alikhah Mishamandani
    Created at:     05:14:00, 20/2/2019
    Description:    Package for all control algorithms.
    Input:          ADCS data structure
    Outputs:        control mode

*****************************************************************/

#ifndef _ADCS_MANAGER_H_
#define _ADCS_MANAGER_H_

#include <stdio.h>
#include "struct.h"
#include "utils.h"
#include "control.h"

int MANAGER(ADCS_DATA RunTimeData, Time CurrentTime);
void TimeManager(Time *CurrentTime);
#endif
```

```c
/**********************************************************************

    File Title:     AttitudeDeter.h
    Created by:     Amir Hossein Alikhah Mishamandani
    Created at:     05:14:00, 20/2/2019
    Description:    Package for all control algorithms.
    Input:          ADCS data structure
    Outputs:        control mode

**********************************************************************/

#include "adcs_manager.h"
#include <stdint.h>

int numberOfDays (int16_t monthNumber, int16_t year){
    // January
    if (monthNumber == 0)
        return (31);

    // February
    if (monthNumber == 1)
    {
        // If the year is leap then February has
        // 29 days
        if (year % 400 == 0 ||
                (year % 4 == 0 && year % 100 != 0))
            return (29);
        else
            return (28);
    }

    // March
    if (monthNumber == 2)
        return (31);

    // April
    if (monthNumber == 3)
        return (30);

    // May
    if (monthNumber == 4)
        return (31);

    // June
    if (monthNumber == 5)
        return (30);

    // July
    if (monthNumber == 6)
        return (31);

    // August
    if (monthNumber == 7)
```

```c
        return (31);

    // September
    if (monthNumber == 8)
        return (30);

    // October
    if (monthNumber == 9)
        return (31);

    // November
    if (monthNumber == 10)
        return (30);

    // December
    if (monthNumber == 11)
        return (31);
}

void TimeManager(Time *CurrentTime){

    int NOD = 0;

        if(CurrentTime->sec > 60){
        CurrentTime->sec = 0;
        CurrentTime->min = CurrentTime->min + 1;
            if(CurrentTime->min > 60){
                CurrentTime->min = 0;
                CurrentTime->hr = CurrentTime->hr + 1;
                if(CurrentTime->hr > 12){
                    CurrentTime->hr = 0;
                    CurrentTime->day = CurrentTime->day + 1;
                    NOD = numberOfDays (CurrentTime->mon,  CurrentTime->year);
                    if(CurrentTime->day > NOD) {
                        CurrentTime->day = 0;
                        CurrentTime->mon = CurrentTime->mon + 1;
                        if(CurrentTime->mon > 12){
                            CurrentTime->mon = 0;
                            CurrentTime->year = CurrentTime->year + 1;
                        }
                    }
                }
            }
        }

}

int MANAGER(ADCS_DATA RunTimeData, Time CurrentTime) {
    double tV[3];
    double C;
    //if (RunTimeData.ZT1UpdateFlag == 0) {
    //  RunTimeData.BZ1 = RunTimeData.ZT1;
```

```c
    //}
    RunTimeData.FailureFlag = 0;
    switch (RunTimeData.BZ1) {
        case 2:
        {
            //printf("\nI'm here ...");
            Control_damp(RunTimeData);
            for (int i = 0; i<3; i++){
                RunTimeData.Tc[i] = RunTimeData.Tm[i];
                tV[i] = RunTimeData.Omega_B2I[i];
            }
            C = MaxArray(tV, 3);
            //printf("\n %lf",C);
            if (C < 0.2*(pi / 180)) {
                RunTimeData.BZ1 = 3;
            }
        }
            break;
        case 3:
        {
            Control_point(RunTimeData);
            for (int i = 0; i<3; i++){
                RunTimeData.Tc[i] = RunTimeData.Tm[i];
                tV[i] = RunTimeData.Omega_B2I[i];
            }
            C = MaxArray(tV, 3);
            if (C > 0.5*(pi / 180)) {
                RunTimeData.BZ1 = 2;
            }
        }
            break;
        case 5:
        {
            Control_point(RunTimeData);
            Control_MWspin(RunTimeData, CurrentTime);
            for (int i = 0; i<3; i++){
                RunTimeData.Tc[i] = RunTimeData.Tm[i] + RunTimeData.Tw[i];
            }

        }

            break;
        case 6:
        {
            Control_stable(RunTimeData);
            Control_MWdespin(RunTimeData, CurrentTime);
            for (int i = 0; i<3; i++){
                RunTimeData.Tc[i] = RunTimeData.Tm[i] + RunTimeData.Tw[i];
            }
            if (RunTimeData.BZ6 == 0) {
                RunTimeData.BZ1 = 3;
                RunTimeData.BZ6 = 1;
            }
```

```
        }

        break;
    }

    return 0;

}
```

```c
/*****************************************************************

    File Title:     orbit.h
    Created by:     Amir Hossein Alikhah Mishamandani
    Created at:     16:13:00, 14/2/2019
    Description:    Get the orbital param.
    Input:          time, position, velocity
    Outputs:        Position, Velocity, Angular Velocity, DCM, Quaternioun

*****************************************************************/

#ifndef _ORBIT_H_
#define _ORBIT_H_

#include "struct.h"
#include "utils.h"
#include "magPred.h"
#include "sunPred.h"

int Orbit(ADCS_DATA *,Time);

#endif
```

```c
/*****************************************************************

    File Title:     orbit.c
    Created by:     Amir Hossein Alikhah Mishamandani
    Created at:     16:13:00, 14/2/2019
    Description:    Get the orbital param.
    Input:          time, position, velocity
    Outputs:        Position, Velocity, Angular Velocity, DCM, Quaternioun

*****************************************************************/

#include "orbit.h"

int Orbit(ADCS_DATA *RunTimeData, Time CurrentTime){

    double reci[3] = {0};
    double dV[3] = {0};
    double norm_buff = 0;
    double  Buff = 0, normman = 0;

    sun(CurrentTime, RunTimeData->rsun_Inertia);

    for(int i = 0; i < 3; i++){
        RunTimeData->rsun_Body[i] = RunTimeData->rsun_Inertia[i] * 0.99;
    }

    reci[0] = RunTimeData->ECI_Pos_Prev[0] * 0.001;
    reci[1] = RunTimeData->ECI_Pos_Prev[1] * 0.001;
    reci[2] = RunTimeData->ECI_Pos_Prev[2] * 0.001;

    magWmm(reci, CurrentTime, RunTimeData->rmag_Inertia);

    for(int i = 0; i < 3; i++){
        RunTimeData->rmag_Body[i] = RunTimeData->rmag_Inertia[i] * 0.99;
    }

    normman = 1e-9;

    RunTimeData->rmag_Inertia[0] = RunTimeData->rmag_Inertia[0] * normman;
    RunTimeData->rmag_Inertia[1] = RunTimeData->rmag_Inertia[1] * normman;
    RunTimeData->rmag_Inertia[2] = RunTimeData->rmag_Inertia[2] * normman;

    double X1[3], X[3], Y[3], Z[3];
    for (int i = 0; i < 3; i++) {
        X1[i] = RunTimeData->ECI_Vel_Prev[i];
        Z[i] = 0.0 - RunTimeData->ECI_Pos_Prev[i];
    }

    norm_buff = norm(X1, 3);

    X1[0] = X1[0] /norm_buff;
    X1[1] = X1[1] /norm_buff;
    X1[2] = X1[2] /norm_buff;
```

```

    norm_buff = norm(Z, 3);

    Z[0] = Z[0] / norm_buff;
    Z[1] = Z[1] / norm_buff;
    Z[2] = Z[2] / norm_buff;

    Y[0] = Z[1] * X1[2] - Z[2] * X1[1];
    Y[1] = Z[2] * X1[0] - Z[0] * X1[2];
    Y[2] = Z[0] * X1[1] - Z[1] * X1[0];

    X[0] = Y[1] * Z[2] - Y[2] * Z[1];
    X[1] = Y[2] * Z[0] - Y[0] * Z[2];
    X[2] = Y[0] * Z[1] - Y[1] * Z[0];

    RunTimeData->DCM_B2I[0][0] = X[0];
    RunTimeData->DCM_B2I[0][1] = X[1];
    RunTimeData->DCM_B2I[0][2] = X[2];
    RunTimeData->DCM_B2I[1][0] = Y[0];
    RunTimeData->DCM_B2I[1][1] = Y[1];
    RunTimeData->DCM_B2I[1][2] = Y[2];
    RunTimeData->DCM_B2I[2][0] = Z[0];
    RunTimeData->DCM_B2I[2][1] = Z[1];
    RunTimeData->DCM_B2I[2][2] = Z[2];

    RunTimeData->Quat_Orbit2Intertia[3] = -0.5 * sqrt(1.0 + RunTimeData->DCM_B2I[0][0] + RunTimeData->DCM_B2I[1][1] + RunTimeData->DCM_B2I[2][2]);
    RunTimeData->Quat_Orbit2Intertia[0] = 0.25 / RunTimeData->Quat_Orbit2Intertia[3] * (RunTimeData->DCM_B2I[1][2] - RunTimeData->DCM_B2I[2][1]);
    RunTimeData->Quat_Orbit2Intertia[1] = 0.25 / RunTimeData->Quat_Orbit2Intertia[3] * (RunTimeData->DCM_B2I[2][0] - RunTimeData->DCM_B2I[0][2]);
    RunTimeData->Quat_Orbit2Intertia[2] = 0.25 / RunTimeData->Quat_Orbit2Intertia[3] * (RunTimeData->DCM_B2I[0][1] - RunTimeData->DCM_B2I[1][0]);

    RunTimeData->Omega_Orbit2Intertia[0] = 0;
    RunTimeData->Omega_Orbit2Intertia[1] = -sqrt(u / (norm_buff*norm_buff*norm_buff));;
    RunTimeData->Omega_Orbit2Intertia[2] = 0;

    norm_buff = norm(RunTimeData->ECI_Pos_Prev, 3);

    Buff = - u / (norm_buff * norm_buff * norm_buff);

    dV[0] = RunTimeData->ECI_Pos_Prev[0] * Buff;
    dV[1] = RunTimeData->ECI_Pos_Prev[1] * Buff;
    dV[2] = RunTimeData->ECI_Pos_Prev[2] * Buff;

    RunTimeData->ECI_Pos[0] = RunTimeData->ECI_Pos_Prev[0] + RunTimeData->ECI_Vel_Prev[0];
```

```c
    RunTimeData->ECI_Pos[1] = RunTimeData->ECI_Pos_Prev[1] + RunTimeData->ECI_Vel_Prev[1];
    RunTimeData->ECI_Pos[2] = RunTimeData->ECI_Pos_Prev[2] + RunTimeData->ECI_Vel_Prev[2];

    RunTimeData->ECI_Vel[0] = dV[0] + RunTimeData->ECI_Vel_Prev[0];
    RunTimeData->ECI_Vel[1] = dV[1] + RunTimeData->ECI_Vel_Prev[1];
    RunTimeData->ECI_Vel[2] = dV[2] + RunTimeData->ECI_Vel_Prev[2];

    RunTimeData->ECI_Pos_Prev[0] = RunTimeData->ECI_Pos[0];
    RunTimeData->ECI_Pos_Prev[1] = RunTimeData->ECI_Pos[1];
    RunTimeData->ECI_Pos_Prev[2] = RunTimeData->ECI_Pos[2];

    RunTimeData->ECI_Vel_Prev[0] = RunTimeData->ECI_Vel[0];
    RunTimeData->ECI_Vel_Prev[1] = RunTimeData->ECI_Vel[1];
    RunTimeData->ECI_Vel_Prev[2] = RunTimeData->ECI_Vel[2];

    return 0;
}
```

```c
/****************************************************************

    File Title:     AttitudeDeter.h
    Created by:     Amir Hossein Alikhah Mishamandani
    Created at:     02:44:00, 20/2/2019
    Description:    Package for all control algorithms.
    Input:          Magnetic Vector Body, Magnetic Vector Inertia, OmegaO2I,
                    quaternion
    Outputs:        control torque, Control Moment

****************************************************************/

#ifndef _CONTROL_H_
#define _CONTROL_H_

#include <stdio.h>
#include "struct.h"
#include "utils.h"

void Control_damp(ADCS_DATA RunTimeData);
void Control_point(ADCS_DATA RunTimeData);
void Control_stable(ADCS_DATA RunTimeData);
void Control_MWspin(ADCS_DATA RunTimeData, Time CurrentTime);
void Control_MWdespin(ADCS_DATA RunTimeData, Time CurrentTime);
#endif
```

```c
/****************************************************************

    File Title:     AttitudeDeter.h
    Created by:     Amir Hossein Alikhah Mishamandani
    Created at:     02:44:00, 20/2/2019
    Description:    Package for all control algorithms.
    Input:          Magnetic Vector Body, Magnetic Vector Inertia, OmegaO2I,
                    quaternion
    Outputs:        control torque, Control moment

****************************************************************/

#include "control.h"

/*=================================Control_damp=====================================*/
void Control_damp(ADCS_DATA RunTimeData) {

    double tV[3] = {0.0, 0.0, 0.0};
    double tV2[3] = {0.0, 0.0, 0.0};
    double tV3[3] = {0.0, 0.0, 0.0};
    double tV4[3] = {0.0, 0.0, 0.0};
    double C1 = 0.0;
    double C2 = 0.0;
    double C3 = 0.0;

    //tV = Te = -Kd*Wbi
    for(int i=0;i<3;i++){
        for(int j=0;j<3;j++){
        tV[i] = RunTimeData.Kd_damp[i][j]*RunTimeData.Omega_B2I[0] + RunTimeData.Kd_damp[i][j]*RunTimeData.Omega_B2I[1] + RunTimeData.Kd_damp[i][j]*RunTimeData.Omega_B2I[2];
        tV[i] = -tV[i];
        }
    }
    //tV2 = Me = cross(Bm,Te)/square(||Bm||)
    tV2[0] = RunTimeData.rmag_Body[1] * tV[2] - RunTimeData.rmag_Body[2] * tV[1];
    tV2[1] = RunTimeData.rmag_Body[2] * tV[0] - RunTimeData.rmag_Body[0] * tV[2];
    tV2[2] = RunTimeData.rmag_Body[0] * tV[1] - RunTimeData.rmag_Body[1] * tV[0];

    C1 = norm(RunTimeData.rmag_Body, 3);

    for (int i = 0; i<3; i++){
        tV2[i] = tV2[i] / C1;
    }

    //tV3 = Mr = min(Mmax,max(|Me|))/max(|Me|) * Me
    for (int i = 0; i<3; i++){
        tV3[i] = fabs(tV2[i]);
```

```c
    }

    C1 = MaxArray(tV3, 3);

    if(Mmax > C1){
        C2 = C1;
    }else if(Mmax <= C1){
        C2 = Mmax;
    }

    C1 = C2 / C1;

    for (int i = 0; i<3; i++){
        tV3[i] = tV2[i] * C1;
    }

    //tV2 = Tr = cross(Mr,Bm)
    tV2[0] = tV3[1] * RunTimeData.rmag_Body[2] - tV3[2] * RunTimeData.rmag_Body[1];
    tV2[1] = tV3[2] * RunTimeData.rmag_Body[0] - tV3[0] * RunTimeData.rmag_Body[2];
    tV2[2] = tV3[0] * RunTimeData.rmag_Body[1] - tV3[1] * RunTimeData.rmag_Body[0];

    //C1 = alpha = arccos(Tr*Te/(||Tr||*||Te||))
    C1 = tV[0] * tV2[0] + tV[1] * tV2[1] + tV[2] * tV2[2];
    C2 = norm(tV2, 3);
    C3 = norm(tV, 3);
    C2 = C2 * C3;
    C3 = C1 / C2;
    C1 = acos(C3);
    //printf("\n %lf ? 1.0235987",C1);
    if (C1 < alpha_damp){

        for (int i = 0; i<3; i++){
            RunTimeData.Mc[i] = tV3[i];
        }
    }
    else{
        for (int i = 0; i<3; i++){
            RunTimeData.Mc[i] = 0;
        }
    }
    //printf("%20.20lf\t%20.20lf\t%20.20lf\n",RunTimeData.Mc[0],RunTimeData.Mc[1],RunTimeData.Mc[2]);
    //tV2 = Tm = cross(Mc,Bm)
    RunTimeData.Tm[0] = RunTimeData.Mc[1] * RunTimeData.rmag_Body[2] - RunTimeData.Mc[2] * RunTimeData.rmag_Body[1];
    RunTimeData.Tm[1] = RunTimeData.Mc[2] * RunTimeData.rmag_Body[0] - RunTimeData.Mc[0] * RunTimeData.rmag_Body[2];
    RunTimeData.Tm[2] = RunTimeData.Mc[0] * RunTimeData.rmag_Body[1] - RunTimeData.Mc[1] * RunTimeData.rmag_Body[0];

```

```c
}


/*=================================Control_point===========================================*/
void Control_point(ADCS_DATA RunTimeData) {

    double tV[3] = { 0.0f, 0.0f, 0.0f };
    double tV2[3] = { 0.0f, 0.0f, 0.0f };
    double tV3[3] = { 0.0f, 0.0f, 0.0f };
    double C1 = 0.0f;
    double C2 = 0.0f;
    double C3 = 0.0f;

    for (int i = 0; i<3; i++){
        tV[i] = RunTimeData.Quat_B2O[i];
    }

    for (int i=0;i<3;i++){
        tV2[i] = RunTimeData.Kp_point[i][0]*tV[0] + RunTimeData.Kp_point[i][1]*tV[1] + RunTimeData.Kp_point[i][2]*tV[2];
    }

    for (int i=0;i<3;i++){
        tV[i] = RunTimeData.Kd_point[i][0]*RunTimeData.Omega_B2O[0] + RunTimeData.Kd_point[i][1]*RunTimeData.Omega_B2O[1] + RunTimeData.Kd_point[i][2]*RunTimeData.Omega_B2O[2];
    }

    for (int i = 0; i<3; i++){
        tV3[i] = tV[i] + tV2[i];
        tV3[i] = -tV3[i];
    }

    tV2[0] = RunTimeData.rmag_Body[1] * tV[2] - RunTimeData.rmag_Body[2] * tV[1];
    tV2[1] = RunTimeData.rmag_Body[2] * tV[0] - RunTimeData.rmag_Body[0] * tV[2];
    tV2[2] = RunTimeData.rmag_Body[0] * tV[1] - RunTimeData.rmag_Body[1] * tV[0];

    C1 = 1 / norm(RunTimeData.rmag_Body, 3);
    C1 = 1 / C1;

    for (int i = 0; i<3; i++){
        tV2[i] = tV2[i] * C1;
    }

    for (int i = 0; i<3; i++){
        tV3[i] = fabs(tV2[i]);
    }

```

```c
    C1 = MaxArray(tV3, 3);

    if(Mmax > C1){
        C2 = C1;
    }else if(Mmax <= C1){
        C2 = Mmax;
    }

    C1 = C2 / C1;

    for (int i = 0; i<3; i++){
        tV3[i] = tV2[i] * C1;
    }

    tV2[0] = tV3[1] * RunTimeData.rmag_Body[2] - tV3[2] * RunTimeData.rmag_Body[1];
    tV2[1] = tV3[2] * RunTimeData.rmag_Body[0] - tV3[0] * RunTimeData.rmag_Body[2];
    tV2[2] = tV3[0] * RunTimeData.rmag_Body[1] - tV3[1] * RunTimeData.rmag_Body[0];

    C1 = tV[0] * RunTimeData.rmag_Body[0] + tV[1] * RunTimeData.rmag_Body[1] + tV[2] * RunTimeData.rmag_Body[2];
    C2 = norm(RunTimeData.rmag_Body, 3);
    C3 = norm(tV, 3);
    C2 = C2 * C3;
    C3 = fabs(C1) / C2;
    C1 = acos(C3);

    if (C1 < alpha_point)
        for (int i = 0; i<3; i++){
            RunTimeData.Mc[i] = 0;
        }
    else{
        for (int i = 0; i<3; i++){
            RunTimeData.Mc[i] = tV3[i];
        }
    }

    RunTimeData.Tm[0] = RunTimeData.Mc[1] * RunTimeData.rmag_Body[2] - RunTimeData.Mc[2] * RunTimeData.rmag_Body[1];
    RunTimeData.Tm[1] = RunTimeData.Mc[2] * RunTimeData.rmag_Body[0] - RunTimeData.Mc[0] * RunTimeData.rmag_Body[2];
    RunTimeData.Tm[2] = RunTimeData.Mc[0] * RunTimeData.rmag_Body[1] - RunTimeData.Mc[1] * RunTimeData.rmag_Body[0];

}


/*=================================Control_stable======================================*/
void Control_stable(ADCS_DATA RunTimeData) {
```

```c
    double tV[3] = { 0.0, 0.0, 0.0 };
    double tV2[3] = { 0.0, 0.0, 0.0 };
    double tV3[3] = { 0.0, 0.0, 0.0 };
    double C1 = 0.0;
    double C2 = 0.0;
    double C3 = 0.0;

    for (int i = 0; i<3; i++){
        tV[i] = RunTimeData.Quat_B2O[i];
    }

    for (int i=0;i<3;i++){
        tV2[i] = RunTimeData.Kp_stable[i][0]*tV[0] + RunTimeData.Kp_stable[i][1]*tV[1] + RunTimeData.Kp_stable[i][2]*tV[2];
    }

    for (int i=0;i<3;i++){
        tV[i] = RunTimeData.Kd_stable[i][0]*RunTimeData.Omega_B2O[0] + RunTimeData.Kd_stable[i][1]*RunTimeData.Omega_B2O[1] + RunTimeData.Kd_stable[i][2]*RunTimeData.Omega_B2O[2];
    }

    for (int i = 0; i<3; i++){
        tV3[i] = tV[i] + tV2[i];
        tV3[i] = -tV3[i];
    }

    tV2[0] = RunTimeData.rmag_Body[1] * tV[2] - RunTimeData.rmag_Body[2] * tV[1];
    tV2[1] = RunTimeData.rmag_Body[2] * tV[0] - RunTimeData.rmag_Body[0] * tV[2];
    tV2[2] = RunTimeData.rmag_Body[0] * tV[1] - RunTimeData.rmag_Body[1] * tV[0];

    C1 = 1 / norm(RunTimeData.rmag_Body, 3);
    C1 = 1 / C1;

    for (int i = 0; i<3; i++){
        tV2[i] = tV2[i] * C1;
    }

    C2 = -atan(RunTimeData.DCM_B2I[2][1] / RunTimeData.DCM_B2I[2][2]);
    C3 = -0.1*RunTimeData.rmag_Body[0] * C2 - 10 * RunTimeData.Delta_rmag_Body[1];
    tV2[1] = C3;


    for (int i = 0; i<3; i++){
        tV3[i] = fabs(tV2[i]);
    }

    C1 = MaxArray(tV3, 3);

```

```c
    if(Mmax > C1){
        C2 = C1;
    }else if(Mmax <= C1){
        C2 = Mmax;
    }

    C1 = C2 / C1;

    for (int i = 0; i<3; i++){
        tV3[i] = tV2[i] * C1;
    }

    tV2[0] = tV3[1] * RunTimeData.rmag_Body[2] - tV3[2] * RunTimeData.rmag_Body[1];
    tV2[1] = tV3[2] * RunTimeData.rmag_Body[0] - tV3[0] * RunTimeData.rmag_Body[2];
    tV2[2] = tV3[0] * RunTimeData.rmag_Body[1] - tV3[1] * RunTimeData.rmag_Body[0];

    C1 = tV[0] * RunTimeData.rmag_Body[0] + tV[1] * RunTimeData.rmag_Body[1] + tV[2] * RunTimeData.rmag_Body[2];
    C2 = norm(RunTimeData.rmag_Body, 3);
    C3 = norm(tV, 3);
    C2 = C2 * C3;
    C3 = fabs(C1) / C2;
    C1 = acos(C3);

    if (C1 < alpha_point)
        for (int i = 0; i<3; i++){
            RunTimeData.Mc[i] = 0;
        }
    else{
        for (int i = 0; i<3; i++){
            RunTimeData.Mc[i] = tV3[i];
        }
    }

    RunTimeData.Tm[0] = tV2[0];
    RunTimeData.Tm[1] = tV2[1];
    RunTimeData.Tm[2] = tV2[2];

}

/*======================== Wheel_spin ====================================*/
void Control_MWspin(ADCS_DATA RunTimeData, Time CurrentTime) {
    int i;
    double dHw1 = 4000 * (0.067 / 6000) / 1000;
    double dHw2 = -4000 * (0.067 / 6000) / 2000;
    double dHw = 0;
    double tV[3], tV2[3], tV3[3];

    if (RunTimeData.BZ6 == 1){
```

```c
        dHw = dHw1;
        if (RunTimeData.Hw > Hw_stand1) {
            RunTimeData.BZ6 = 4;
        }
    }

    if (RunTimeData.BZ6 == 2){
        dHw = dHw2;
        if (RunTimeData.Hw < Hw_stand1) {
            RunTimeData.BZ6 = 4;
        }
    }

    if (RunTimeData.BZ6 == 4){
        for (i = 0; i<3; i++){
            tV[i] = RunTimeData.Quat_B2O[i];
        }

    for (int i=0;i<3;i++){
        tV2[i] = RunTimeData.Kp_stable[i][0]*tV[0] + RunTimeData.Kp_stable[i][1]*tV[1] + RunTimeData.Kp_stable[i][2]*tV[2];
    }

    for (int i=0;i<3;i++){
        tV3[i] = RunTimeData.Kd_stable[i][0]*RunTimeData.Omega_B2O[0] + RunTimeData.Kd_stable[i][1]*RunTimeData.Omega_B2O[1] + RunTimeData.Kd_stable[i][2]*RunTimeData.Omega_B2O[2];
    }

    for (int i = 0; i<3; i++){
        tV[i] = tV3[i] + tV2[i];
        tV[i] = -tV3[i];
    }

    for (int i = 0; i<3; i++){
        tV[i] = tV3[i] / 30000;
    }
    dHw = tV[1];

    if (RunTimeData.Hw > Hw_stand2)
            RunTimeData.BZ6 = 2;
    }
    RunTimeData.Hw_set = RunTimeData.Hw + dHw * CurrentTime.sec;
    double Hw_real = ceilf(RunTimeData.Hw_set * (6000 / 0.067)) * (0.067 / 6000);
    double dHw_real = Hw_real - ceilf(RunTimeData.Hw * (6000 / 0.067)) * (0.067 / 6000);
    RunTimeData.Tw[0] = -RunTimeData.Omega_B2I[2] * Hw_real;
    RunTimeData.Tw[1] = dHw_real;
    RunTimeData.Tw[2] =  RunTimeData.Omega_B2I[0] * Hw_real;

}

```

```
/*======================== Wheel_despin ====================================*/
void Control_MWdespin(ADCS_DATA RunTimeData, Time CurrentTime) {
    double dHw3 = -4000 * (0.067 / 6000) / 2000;

    if (RunTimeData.Hw < Hw_stand3) {
        RunTimeData.BZ6 = 0;
    }

    RunTimeData.Hw_set = RunTimeData.Hw + dHw3 * CurrentTime.sec;

    double Hw_real = ceilf(RunTimeData.Hw_set * (6000 / 0.067)) * (0.067 / 6000);
    double dHw_real = Hw_real - ceilf(RunTimeData.Hw * (6000 / 0.067)) * (0.067 / 6000);
    RunTimeData.Tw[0] = -RunTimeData.Omega_B2I[2] * Hw_real;
    RunTimeData.Tw[1] = dHw_real;
    RunTimeData.Tw[2] =  RunTimeData.Omega_B2I[0] * Hw_real;


}
```

```c
/******************************************************************

    File Title:     AttitudeDeter.h
    Created by:     Amir Hossein Alikhah Mishamandani
    Created at:     11:14:00, 16/2/2019
    Description:    Get the orbital param.
    Input:          Gyro Angular Velocity, Sun vector body, Sun vector Inertia
                    Magnetic Vector Body, Magnetic Vector Inertia, OmegaO2I,
                    Previous State Vector, previous quaternion
    Outputs:        State Vector, QuaternionB2I, QuaternionB2O,
                    OmegaB2I, OmegaB2O, Eurler Angels, Phi_K

******************************************************************/

#ifndef _ATTITUDEDETER_H_
#define _ATTITUDEDETER_H_

#include <stdio.h>
#include "struct.h"
#include "utils.h"
#include "solver/solver.h"
#include "solver/solver_initialize.h"
#include "solver/svd.h"
void TRIAD(ADCS_DATA *RunTimeData);
void EKF(ADCS_DATA *RunTimeData, Time CurrentTime);
int AttitudeDetermination(ADCS_DATA *RunTimeData, Time CurrentTime);

#endif
```

```c
/*****************************************************************

    File Title:     AttitudeDeter.c
    Created by:     Amir Hossein Alikhah Mishamandani
    Created at:     11:14:00, 16/2/2019
    Description:    Get the orbital param.
    Input:          Gyro Angular Velocity, Sun vector body, Sun vector Inertia
                    Magnetic Vector Body, Magnetic Vector Inertia, OmegaO2I,
                    Previous State Vector, previous quaternion
    Outputs:        State Vector, QuaternionB2I, QuaternionB2O,
                    OmegaB2I, OmegaB2O, Eurler Angels, Phi_K

*****************************************************************/

#include "AttitudeDeter.h"

int AttitudeDetermination(ADCS_DATA *RunTimeData, Time CurrentTime){

    if (RunTimeData->sun_sensor_flag == 1) {

        TRIAD(RunTimeData);
        EKF(RunTimeData, CurrentTime);

    }
    else{

        printf("\nEKF Filter is required ...!");
        EKF(RunTimeData, CurrentTime);

    }


    double Quat[4] = {0};

    rotmtx2quat(RunTimeData->DCM_B2I, Quat);

    for (int i = 0; i < 4; i++){
        RunTimeData->Quat_B2I[i] = Quat[i];
    }

    double a, b, c, d;
    a = fabsf(RunTimeData->Quat_B2I[0] - RunTimeData->Quat_B2I_prev[0]);
    b = fabsf(RunTimeData->Quat_B2I[1] - RunTimeData->Quat_B2I_prev[1]);
    c = fabsf(RunTimeData->Quat_B2I[2] - RunTimeData->Quat_B2I_prev[2]);
    d = fabsf(RunTimeData->Quat_B2I[3] - RunTimeData->Quat_B2I_prev[3]);

    if ((a > 0.1) || (b > 0.1) || (c > 0.1) || (d > 0.1))    {
        RunTimeData->Quat_B2I[0] = -RunTimeData->Quat_B2I[0];
        RunTimeData->Quat_B2I[1] = -RunTimeData->Quat_B2I[1];
        RunTimeData->Quat_B2I[2] = -RunTimeData->Quat_B2I[2];
        RunTimeData->Quat_B2I[3] = -RunTimeData->Quat_B2I[3];
    }
```

```c


    double dQuat_B2I[4];
    double X[3] = {0, 0, 0};

    solver_initialize();

    solver(dQuat_B2I[0], dQuat_B2I[1], dQuat_B2I[2], dQuat_B2I[3], RunTimeData->Quat_B2I[0], RunTimeData->Quat_B2I[1], RunTimeData->Quat_B2I[2], RunTimeData->Quat_B2I[3], X);


    double cross_pos[3] = {0};
    double E_S_radius =  6371 + 600;

    cross_pos[0] = (RunTimeData->ECI_Pos[1]/1000) * (RunTimeData->ECI_Vel[2]/1000) - (RunTimeData->ECI_Pos[2]/1000) * (RunTimeData->ECI_Vel[1]/1000);
    cross_pos[1] = (RunTimeData->ECI_Pos[2]/1000) * (RunTimeData->ECI_Vel[0]/1000) - (RunTimeData->ECI_Pos[0]/1000) * (RunTimeData->ECI_Vel[2]/1000);
    cross_pos[2] = (RunTimeData->ECI_Pos[0]/1000) * (RunTimeData->ECI_Vel[1]/1000) - (RunTimeData->ECI_Pos[1]/1000) * (RunTimeData->ECI_Vel[0]/1000);

    RunTimeData->Omega_B2I[0] = cross_pos[0] / (E_S_radius*E_S_radius);
    RunTimeData->Omega_B2I[1] = cross_pos[1] / (E_S_radius*E_S_radius);
    RunTimeData->Omega_B2I[2] = cross_pos[2] / (E_S_radius*E_S_radius);


    double Cio[3][3], Cbo[3][3];

    for(int i=0;i<3;i++){
        for(int j=0;j<3;j++){
            Cio[i][j] = RunTimeData->DCM_B2I[j][i];
        }
    }

    for (int i=0;i<3;i++){
        for(int j=0;j<3;j++){
            Cbo[i][j] = 0.0;
            for (int k=0;k<3;k++){
                Cbo[i][j] = Cbo[i][j] + RunTimeData->DCM_B2I[i][k]*Cio[k][j];
            }
        }
    }

    rotmtx2quat(Cbo, RunTimeData->Quat_B2O);

    a = fabsf(RunTimeData->Quat_B2O[0] - RunTimeData->Quat_B2O_prev[0]);
    b = fabsf(RunTimeData->Quat_B2O[1] - RunTimeData->Quat_B2O_prev[1]);
    c = fabsf(RunTimeData->Quat_B2O[2] - RunTimeData->Quat_B2O_prev[2]);
```

```c
    d = fabsf(RunTimeData->Quat_B2O[3] - RunTimeData->Quat_B2O_prev[3]);

    if ((a > 0.1) || (b > 0.1) || (c > 0.1) || (d > 0.1))
    {
        RunTimeData->Quat_B2O[0] = -RunTimeData->Quat_B2O[0];
        RunTimeData->Quat_B2O[1] = -RunTimeData->Quat_B2O[1];
        RunTimeData->Quat_B2O[2] = -RunTimeData->Quat_B2O[2];
        RunTimeData->Quat_B2O[3] = -RunTimeData->Quat_B2O[3];
    }

    double tV[3];

    for (int i=0;i<3;i++){
        tV[i]=0.0;
        for(int j=0;j<3;j++){
            tV[i] = tV[i] + Cbo[i][j] * RunTimeData->Omega_Orbit2Intertia[j];
        }
    }

    for (int i = 0; i<3; i++){
        RunTimeData->Omega_B2O[i] = RunTimeData->Omega_B2I[i] - tV[i];
    }

    return 0;

}

void TRIAD(ADCS_DATA *RunTimeData){

    double w1[3], w2[3], v1[3], v2[3], Ou1[3], Ou2[3], Ou3[3], R1[3], R2[3], R3[3];
    double crossw1w2[3],crossw1crossw1w2[3], crossv1v2[3], crossv1crossv1v2[3];

    Unit(RunTimeData->rmag_Body, w1);
    Unit(RunTimeData->rsun_Body, w2);

    Ou1[0] = w1[0];
    Ou1[1] = w1[1];
    Ou1[2] = w1[2];

    crossw1w2[0] = w1[1] * w2[2] - w1[2] * w2[1];
    crossw1w2[1] = w1[2] * w2[0] - w1[0] * w2[2];
    crossw1w2[2] = w1[0] * w2[1] - w1[1] * w2[0];

    Ou2[0] = crossw1w2[0] / norm(crossw1w2, 3);
    Ou2[1] = crossw1w2[1] / norm(crossw1w2, 3);
    Ou2[2] = crossw1w2[2] / norm(crossw1w2, 3);

    crossw1crossw1w2[0] = w1[1] * crossw1w2[2] - w1[2] * crossw1w2[1];
    crossw1crossw1w2[1] = w1[2] * crossw1w2[0] - w1[0] * crossw1w2[2];
    crossw1crossw1w2[2] = w1[0] * crossw1w2[1] - w1[1] * crossw1w2[0];
```

```
    Ou3[0] = crossw1crossw1w2[0] / norm(crossw1w2, 3);
    Ou3[1] = crossw1crossw1w2[1] / norm(crossw1w2, 3);
    Ou3[2] = crossw1crossw1w2[2] / norm(crossw1w2, 3);


    v1[0] = RunTimeData->rmag_Inertia[0];
    v1[1] = RunTimeData->rmag_Inertia[1];
    v1[2] = RunTimeData->rmag_Inertia[2];

    v2[0] = RunTimeData->rsun_Inertia[0];
    v2[1] = RunTimeData->rsun_Inertia[1];
    v2[2] = RunTimeData->rsun_Inertia[2];

    R1[0] = v1[0];
    R1[1] = v1[1];
    R1[2] = v1[2];

    crossv1v2[0] = v1[1] * v2[2] - v1[2] * v2[1];
    crossv1v2[1] = v1[2] * v2[0] - v1[0] * v2[2];
    crossv1v2[2] = v1[0] * v2[1] - v1[1] * v2[0];

    R2[0] = crossv1v2[0] / norm(crossv1v2, 3);
    R2[1] = crossv1v2[1] / norm(crossv1v2, 3);
    R2[2] = crossv1v2[2] / norm(crossv1v2, 3);

    crossv1crossv1v2[0] = v1[1] * crossv1v2[2] - v1[2] * crossv1v2[1];
    crossv1crossv1v2[1] = v1[2] * crossv1v2[0] - v1[0] * crossv1v2[2];
    crossv1crossv1v2[2] = v1[0] * crossv1v2[1] - v1[1] * crossv1v2[0];

    R3[0] = crossv1crossv1v2[0] / norm(crossv1v2, 3);
    R3[1] = crossv1crossv1v2[1] / norm(crossv1v2, 3);
    R3[2] = crossv1crossv1v2[2] / norm(crossv1v2, 3);

    RunTimeData->DCM_B2I[0][0] = R1[0] * Ou1[0] + R1[1] * Ou1[1] + R1[2] * 
      Ou1[2];
    RunTimeData->DCM_B2I[0][1] = R2[0] * Ou1[0] + R2[1] * Ou1[1] + R2[2] * 
      Ou1[2];
    RunTimeData->DCM_B2I[0][2] = R3[0] * Ou1[0] + R3[1] * Ou1[1] + R3[2] * 
      Ou1[2];
    RunTimeData->DCM_B2I[1][0] = R1[0] * Ou2[0] + R1[1] * Ou2[1] + R1[2] * 
      Ou2[2];
    RunTimeData->DCM_B2I[1][1] = R2[0] * Ou2[0] + R2[1] * Ou2[1] + R2[2] * 
      Ou2[2];
    RunTimeData->DCM_B2I[1][2] = R3[0] * Ou2[0] + R3[1] * Ou2[1] + R3[2] * 
      Ou2[2];
    RunTimeData->DCM_B2I[2][0] = R1[0] * Ou3[0] + R1[1] * Ou3[1] + R1[2] * 
      Ou3[2];
    RunTimeData->DCM_B2I[2][1] = R2[0] * Ou3[0] + R2[1] * Ou3[1] + R2[2] * 
      Ou3[2];
    RunTimeData->DCM_B2I[2][2] = R3[0] * Ou3[0] + R3[1] * Ou3[1] + R3[2] * 
      Ou3[2];

```

```c
    //DCM to Quat
    double Quat[4];
    rotmtx2quat(RunTimeData->DCM_B2I, Quat);

    for(int i =0; i < 4;i++){
    RunTimeData->Quat_B2I[i] = Quat[i];
    }
    //SVD
    double A[4][1];
    double v[1][4];
    /*printbuff("\nQuat: ", Quat, 4);
    for(int i = 0;i < 4; i++){
        A[i][0] = Quat[i];
    }*/

    dsvd(A, RunTimeData->SVD, v);

    //printbuff("\nSVD: ", RunTimeData->SVD, 4);
}

void EKF(ADCS_DATA *RunTimeData, Time CurrentTime){

    int n = 7;
    int m = 3;

    double TPhik[7][7], Phik[7][7], tCn[7][7], tCn3[7][7], tCn4[7][7], tCn5[7][7];
    double Hk_T[7][3], tCnm[7][3], tCnm2[7][3], tCm[3][3], tCm2[3][3], tCm3[3][3], tCmn[3][7], tCmn2[3][7];
    double tV[7], tV2[7], tV3[7], tVm[3];
    double Xk[4], normm;
    double q[4], g[3], c[3];
    double b[3] = { RunTimeData->rmag_Inertia[0], RunTimeData->rmag_Inertia[1], RunTimeData->rmag_Inertia[2] };
    //State Vector XK_1 = [Quat_B2I,OmegaB2I,GyroDrift]
    q[0] = RunTimeData->Quat_B2I[0];
    q[1] = RunTimeData->Quat_B2I[1];
    q[2] = RunTimeData->Quat_B2I[2];
    q[3] = RunTimeData->Quat_B2I[3];
    g[0] = RunTimeData->Omega_B2I[0];
    g[1] = RunTimeData->Omega_B2I[1];
    g[2] = RunTimeData->Omega_B2I[2];
    c[0] = RunTimeData->GyroDrift[0];
    c[1] = RunTimeData->GyroDrift[1];
    c[2] = RunTimeData->GyroDrift[2];
    //f(x)
    double fx[7] = { (q[2] * (c[1] - g[1])) / 2.0f - (q[3] * (c[0] - g[0])) / 2.0f - (q[1] * (c[2] - g[2])) / 2.0f,\
        (q[0] * (c[2] - g[2])) / 2.0f - (q[3] * (c[1] - g[1])) / 2.0f - (q[2] * (c[0] - g[0])) / 2.0f,\
        (q[1] * (c[0] - g[0])) / 2.0f - (q[0] * (c[1] - g[1])) / 2.0f - (q[3] * (c[2] - g[2])) / 2.0f,\
        (q[0] * (c[0] - g[0])) / 2.0f + (q[1] * (c[1] - g[1])) / 2.0f + (q
```

```c
            [2] * (c[2] - g[2])) / 2.0f,\
        0, 0, 0 };
    //F_k
    float Fk[7][7] = { { 0, g[2] / 2 - c[2] / 2, c[1] / 2 - g[1] / 2, g[0] / 2 - c[0] / 2, -q[3] / 2, q[2] / 2, -q[1] / 2 },\
        { c[2] / 2 - g[2] / 2, 0, g[0] / 2 - c[0] / 2, g[1] / 2 - c[1] / 2, -q[2] / 2, -q[3] / 2, q[0] / 2 },\
        { g[1] / 2 - c[1] / 2, c[0] / 2 - g[0] / 2, 0, g[2] / 2 - c[2] / 2, q[1] / 2, -q[0] / 2, -q[3] / 2 },\
        { c[0] / 2 - g[0] / 2, c[1] / 2 - g[1] / 2, c[2] / 2 - g[2] / 2, 0, q[0] / 2, q[1] / 2, q[2] / 2 },\
        { 0, 0, 0, 0, 0, 0, 0 },{ 0, 0, 0, 0, 0, 0, 0 },{ 0, 0, 0, 0, 0, 0, 0 } };
    //F_k * T_Sample
    for(int i = 0;i < 7;i++){
        for(int j = 0;j < 7;j++){
            Fk[i][j] = Fk[i][j] * RunTimeData->S_time;
        }
    }
    //Phik_ = I + Fk*ts
    for(int i = 0;i < 7;i++){
        for(int j = 0;j < 7;j++){
            Phik[i][j] = Fk[i][j];
            if(i = j){
                Phik[i][j] = Phik[i][j] + 1;
            }
        }
    }
    //Phik_' (Transpose)
    for(int i = 0;i < 7;i++){
        for(int j = 0;j < 7;j++){
            TPhik[i][j] = Phik[j][i];
        }
    }
    //tCn3 = Pk_1 * Phik_'
    for (int i=0;i<7;i++){
        for(int j=0;j<7;j++){
            tCn3[i][j] = 0.0;
            for (int k=0;k<7;k++){
                tCn3[i][j] = tCn3[i][j] + RunTimeData->PK1[i][j] * TPhik[i][j];
            }
        }
    }
    //tCn  = Phik_*tCn3 = Phik_ * Pk_1 * Phik_'
    for (int i=0;i<7;i++){
        for(int j=0;j<7;j++){
            tCn[i][j] = 0.0;
            for (int k=0;k<7;k++){
                tCn[i][j] = tCn[i][j] + Phik[i][j] * tCn3[i][j];
            }
        }
    }
```

```c
    //tCn  = tCn+Qk_1 = Phik_*Pk_1*Phik_'+Qk_1
    for(int i = 0;i < 7;i++){
        for(int j = 0;j < 7;j++){
            tCn[i][j] = tCn[i][j] + RunTimeData->QK1[i][j];
        }
    }
    //tV = T_Sample * fx
    for(int i = 0;i < 7;i++){
        tV[i] = fx[i] * RunTimeData->S_time;
    }
    //tV = ts * fx + Xk_1
    for(int i = 0;i < 7;i++){
        tV[i] = tV[i] * RunTimeData->XK1[i];
    }
    //copy first 4 element of tV to quaternion
    for(int i = 0;i < 4;i++){
        RunTimeData->Quat_B2I[i] = tV[i];
    }
    //h
    float h_[3] = { b[0] * (q[0] * q[0] - q[1] * q[1] - q[2] * q[2] + q[3] * q[3]) + b[1] * (2 * q[0] * q[1] + 2 * q[2] * q[3]) + b[2] * (2 * q[0] * q[2] - 2 * q[1] * q[3]), \
        b[0] * (2 * q[0] * q[1] - 2 * q[2] * q[3]) - b[1] * (q[0] * q[0] - q[1] * q[1] + q[2] * q[2] - q[3] * q[3]) + b[2] * (2 * q[0] * q[3] + 2 * q[1] * q[2]),\
        b[0] * (2 * q[0] * q[2] + 2 * q[1] * q[3]) - b[2] * (q[0] * q[0] + q[1] * q[1] - q[2] * q[2] - q[3] * q[3]) - b[1] * (2 * q[0] * q[3] - 2 * q[1] * q[2]) };
    //H_k
    float Hk[3][7] = { { 2 * b[0] * q[0] + 2 * b[1] * q[1] + 2 * b[2] * q[2], 2 * b[1] * q[0] - 2 * b[0] * q[1] - 2 * b[2] * q[3], 2 * b[1] * q[3] - 2 * b[0] * q[2] + 2 * b[2] * q[0], 2 * b[0] * q[3] + 2 * b[1] * q[2] - 2 * b[2] * q[1], 0, 0, 0 },\
        { 2 * b[0] * q[1] - 2 * b[1] * q[0] + 2 * b[2] * q[3], 2 * b[0] * q[0] + 2 * b[1] * q[1] + 2 * b[2] * q[2], 2 * b[2] * q[1] - 2 * b[1] * q[2] - 2 * b[0] * q[3], 2 * b[1] * q[3] - 2 * b[0] * q[2] + 2 * b[2] * q[0], 0, 0, 0 },\
        { 2 * b[0] * q[2] - 2 * b[1] * q[3] - 2 * b[2] * q[0], 2 * b[0] * q[3] + 2 * b[1] * q[2] - 2 * b[2] * q[1], 2 * b[0] * q[0] + 2 * b[1] * q[1] + 2 * b[2] * q[2], 2 * b[0] * q[1] - 2 * b[1] * q[0] + 2 * b[2] * q[3], 0, 0, 0 } };
    // Hk_T = Hk'
    for(int i=0;i<7;i++){
        for(int j=0;j<3;j++){
            Hk_T[i][j] = Hk[j][i];
        }
    }
    //tCnm = tCn * Hk_T
    for (int i=0;i<7;i++){
        for(int j=0;j<3;j++){
            tCnm[i][j] = 0.0;
            for (int k=0;k<7;k++){
                tCnm[i][j] = tCnm[i][j] + tCn[i][k] * Hk_T[k][j];
```

```
            }
        }
    }
    //tCm = Hk * Pk_ * Hk'
    for (int i=0;i<3;i++){
        for(int j=0;j<3;j++){
            tCm[i][j] = 0.0;
            for (int k=0;k<7;k++){
                tCm[i][j] = tCm[i][j] + Hk[i][k] * tCnm[k][j];
            }
        }
    }
    //tCm2 = Hk*Pk_*Hk' + RK
    for (int i=0;i<3;i++){
        for(int j=0;j<3;j++){
            tCm2[i][j] = tCm[i][j] + RunTimeData->RK[i][j];
        }
    }
    //Inverse tCm2
    double det = tCm2[0][0] * tCm2[1][1] * tCm2[2][2] +      tCm2[0][1] * tCm2[1][2] * tCm2[2][0] +    tCm2[1][0] * tCm2[2][1] * tCm2[0][2] - tCm2[0][2] * tCm2[1][1] * tCm2[2][0] -  tCm2[0][1] * tCm2[1][0] * tCm2[2][2] -    tCm2[2][1] * tCm2[1][2] * tCm2[0][0];
    tCm3[0][0] =  (tCm2[1][1]*tCm2[2][2] - tCm2[2][1]*tCm2[1][2])/det;
    tCm3[0][1] = -(tCm2[0][1]*tCm2[2][2] - tCm2[2][1]*tCm2[0][2])/det;
    tCm3[0][2] =  (tCm2[0][1]*tCm2[1][2] - tCm2[1][1]*tCm2[0][2])/det;
    tCm3[1][0] = -(tCm2[1][0]*tCm2[2][2] - tCm2[2][0]*tCm2[1][2])/det;
    tCm3[1][1] =  (tCm2[0][0]*tCm2[2][2] - tCm2[2][0]*tCm2[0][2])/det;
    tCm3[1][2] = -(tCm2[0][0]*tCm2[1][2] - tCm2[1][0]*tCm2[0][2])/det;
    tCm3[2][0] =  (tCm2[1][0]*tCm2[2][1] - tCm2[2][0]*tCm2[1][1])/det;
    tCm3[2][1] = -(tCm2[0][0]*tCm2[2][1] - tCm2[2][0]*tCm2[0][1])/det;
    tCm3[2][2] =  (tCm2[0][0]*tCm2[1][1] - tCm2[0][1]*tCm2[1][0])/det;
    //tCnm = Hk' / (Hk*Pk_*Hk' + RK)
    for (int i=0;i<7;i++){
        for(int j=0;j<3;j++){
            tCnm[i][j] = 0.0;
            for (int k=0;k<3;k++){
                tCnm[i][j] = tCnm[i][j] + Hk_T[i][k] *tCm3[k][j];
            }
        }
    }
    //tCnm2 = Pk_*Hk' / (Hk*Pk_*Hk' + RK)
    for (int i=0;i<7;i++){
        for(int j=0;j<3;j++){
            tCnm2[i][j] = 0.0;
            for (int k=0;k<7;k++){
                tCnm2[i][j] = tCnm2[i][j] + tCn[i][k] *tCnm[k][j];
            }
        }
    }
    //tCn3 = Kk*Hk
    for (int i=0;i<7;i++){
        for(int j=0;j<7;j++){
```

```
                tCn3[i][j] = 0.0;
                for (int k=0;k<3;k++){
                    tCn3[i][j] = tCn3[i][j] + tCnm2[i][k] * Hk[k][j];
                }
            }
        }
        //tCn3 = I-Kk*Hk
        for (int i=0;i<7;i++){
            for(int j=0;j<7;j++){
                if(i = j){
                    tCn3[i][j] = 1 - tCn3[i][j];
                }
            }
        }
        //tCn4 = (I-Kk*Hk)'
        for(int i=0;i<7;i++){
            for(int j=0;j<7;j++){
                tCn4[i][j] = tCn3[j][i];
            }
        }
        //tCn5 = tCn * tCn4
        for (int i=0;i<7;i++){
            for(int j=0;j<7;j++){
                tCn5[i][j] = 0.0;
                for (int k=0;k<7;k++){
                    tCn5[i][j] = tCn5[i][j] + tCn[i][k] * tCn4[k][j];
                }
            }
        }
        //tCn4 = (I-Kk*Hk)*Pk_*(I-Kk*Hk)'
        for (int i=0;i<7;i++){
            for(int j=0;j<7;j++){
                tCn4[i][j] = 0.0;
                for (int k=0;k<7;k++){
                    tCn4[i][j] = tCn4[i][j] + tCn3[i][k] * tCn5[k][j];
                }
            }
        }
        //tCnm2'
        for(int i=0;i<3;i++){
            for(int j=0;j<7;j++){
                tCmn[i][j] = tCnm2[j][i];
            }
        }
        //tCmn2 = RK * tCmn
        for(int i=0;i<3;i++){
            for(int j=0;j<7;j++){
                tCmn2[i][j] = 0.0;
                for (int k=0;k<3;k++){
                    tCmn2[i][j] = tCmn2[i][j] + RunTimeData->RK[i][k] * tCmn[k]
                      [j];
                }
            }
```

```
    }
    //tCn5 = Kk*RK*Kk'
    for(int i=0;i<7;i++){
        for(int j=0;j<7;j++){
            tCn5[i][j] = 0.0;
            for (int k=0;k<3;k++){
                tCn5[i][j] = tCn5[i][j] + tCnm2[i][k] * tCmn2[k][j];
            }
        }
    }
    //tCn3 = (I-Kk*Hk)*Pk_*(I-Kk*Hk)' + Kk*RK*Kk'
    for(int i=0;i<7;i++){
        for(int j=0;j<7;j++){
            tCn3[i][j] = tCn4[i][j] + tCn5[i][j];
        }
    }
    //tVm = Zk - h_
    for(int i = 0; i<3; i++){
        tVm[i] = RunTimeData->rmag_Body[i] - h_[i];
    }
    //tV3 = Kk*(Zk-h_)
    for(int i=0;i<7;i++){
        tV3[i]=0.0;
        for(int j=0;j<3;j++){
            tV3[i] = tV3[i]+ tCnm2[i][j] * tVm[j];
        }
    }
    //tV2 = Xk_+Kk*(Zk-h_)
    for(int i=0;i<7;i++){
        tV2[i]=tV[i]+tV3[i];
    }
    //Copy Xk <- tV2
    for(int i=0;i<4;i++){
        Xk[i] = tV2[i];
    }
    //norm(XK)
    normm = norm(Xk, 4);
    //XK = XK / norm
    for(int i=0;i<4;i++){
        Xk[i] = Xk[i] / normm;
    }
    //Copy Xk -> tV2
    for(int i=0;i<4;i++){
        tV2[i] = Xk[i];
    }

    for(int i=0;i<7;i++){
        RunTimeData->XK[i] = tV2[i];
    }
    for(int i=0;i<7;i++){
        for(int j=0;j<7;j++){
            RunTimeData->PK[i][j] = tCn3[i][j];
        }
```

```
    }
}
```

```c
/*******************************************************************

    File Title:     Dynamic.h
    Created by:     Amir Hossein Alikhah Mishamandani
    Created at:     14:59:00, 16/2/2019
    Description:    the Dynamic Model.
    Input:          Tc, Tg, ECI Position, ECI Sun Position, ECI Geomagnetic field vector,
                    Prev val OmegaB2I, Prev val QuaternionB2I, Prev val QuaternionB2O,
                    Prev val Euler rotation matrix, Rotational matrix Inertial to Orbit
    Outputs:        OmegaB2I, OmegaB2O, QuaternionB2I, QuaternionB2O, Gyro Outputs,
                    Magnetometer Output, Sun sensor Output, Euler rotation matrix

*******************************************************************/

#ifndef _DYNAMIC_H_
#define _DYNAMIC_H_

#include <stdio.h>
#include "struct.h"
#include "utils.h"

int Dynamic_Model(ADCS_DATA *);

#endif
```

```c
/****************************************************************

    File Title:     Dynamic.c
    Created by:     Amir Hossein Alikhah Mishamandani
    Created at:     14:59:016/2/2019
    Description:    the Dynamic Model.
    Input:          Tc, Tg, ECI Position, ECI Sun Position, ECI Geomagnetic field vector,
                    Prev val OmegaB2I, Prev val QuaternionB2I, Prev val QuaternionB2O,
                    Prev val Euler rotation matrix, Rotational matrix Inertial to Orbit
    Outputs:        OmegaB2I, OmegaB2O, QuaternionB2I, QuaternionB2O, Gyro Outputs,
                    Magnetometer Output, Sun sensor Output, Euler rotation matrix

****************************************************************/
#include <stdio.h>
#include "Dynamic.h"


int Dynamic_Model(ADCS_DATA *RunTimeData){

    double Ib[9] = {7.6590, -0.0020, 0.0380, -0.0020, 7.6490, -0.0170, 0.0380, -0.0170, 0.5330};
    double norm_buff, r, index;
    double tV[3] = {0}, tV2[3] = {0}, tV3[4] = {0}, tV4[4] = {0}, Ro[3] = {0, 0 ,0}, Rb[3] = {0}, IR[3] = {0}, RI[3] = {0}, T[3] = {0};
    double tM[3][3];

    for(int i = 0; i < 100; i++) {

        for (int i=0;i<3;i++){
            tV[i] = Ib[i*3+0]*RunTimeData->Omega_B2I_prev[0] + Ib[i*3+1]*RunTimeData->Omega_B2I_prev[1] + Ib[i*3+2]*RunTimeData->Omega_B2I_prev[2];
        }

        //printbuff("TV\t",tV, 3);

        tV2[0] = RunTimeData->Omega_B2I_prev[1] * tV[2] - RunTimeData->Omega_B2I_prev[2] * tV[1];
        tV2[1] = RunTimeData->Omega_B2I_prev[2] * tV[0] - RunTimeData->Omega_B2I_prev[0] * tV[2];
        tV2[2] = RunTimeData->Omega_B2I_prev[0] * tV[1] - RunTimeData->Omega_B2I_prev[1] * tV[0];

        //printbuff("TV2\t",tV2, 3);

        Ro[0] = RunTimeData->DCM_B2I[0][0] * RunTimeData->ECI_Pos[0] + RunTimeData->DCM_B2I[0][1] * RunTimeData->ECI_Pos[1] + RunTimeData->DCM_B2I[0][2] * RunTimeData->ECI_Pos[2];
```

```c
        Ro[1] = RunTimeData->DCM_B2I[1][0] * RunTimeData->ECI_Pos[0] + RunTimeData->DCM_B2I[1][1] * RunTimeData->ECI_Pos[1] + RunTimeData->DCM_B2I[1][2] * RunTimeData->ECI_Pos[2];
        Ro[2] = RunTimeData->DCM_B2I[2][0] * RunTimeData->ECI_Pos[0] + RunTimeData->DCM_B2I[2][1] * RunTimeData->ECI_Pos[1] + RunTimeData->DCM_B2I[2][2] * RunTimeData->ECI_Pos[2];


        //printbuffMatrix("DCM O2I", RunTimeData->DCM_B2I);
        //printbuff("ECI_Pos\t",RunTimeData->ECI_Pos, 3);
        //printbuff("Ro\t",Ro, 3);

        Rb[0] = RunTimeData->DCM_B2O_Prev[0][0] * Ro[0] + RunTimeData->DCM_B2O_Prev[0][1] * Ro[1] + RunTimeData->DCM_B2O_Prev[0][2] * Ro[2];
        Rb[1] = RunTimeData->DCM_B2O_Prev[1][0] * Ro[0] + RunTimeData->DCM_B2O_Prev[1][1] * Ro[1] + RunTimeData->DCM_B2O_Prev[1][2] * Ro[2];
        Rb[2] = RunTimeData->DCM_B2O_Prev[2][0] * Ro[0] + RunTimeData->DCM_B2O_Prev[2][1] * Ro[1] + RunTimeData->DCM_B2O_Prev[2][2] * Ro[2];

        //printbuff("Rb\t",Rb, 3);

        r = norm(Rb, 3);

        index = 3 * u / (r*r*r*r*r);

        //printf("the r = %lf, the index = %lf",r,index);

        for (int i=0;i<3;i++){
            IR[i] = Ib[i*3+0]*Rb[0] + Ib[i*3+1]*Rb[1] + Ib[i*3+2]*Rb[2];
        }

        //printbuff("IR\t",IR, 3);

        RI[0] = Rb[1] * IR[2] - Rb[2] * IR[1];
        RI[1] = Rb[2] * IR[0] - Rb[0] * IR[2];
        RI[2] = Rb[0] * IR[1] - Rb[1] * IR[0];

        //printbuff("RI\t",RI, 3);

        for (int i = 0; i<3; i++){
            RunTimeData->Tg[i] = RI[i] * index;
        }

        //printbuff("Tg\t",RunTimeData->Tg, 3);
        //printbuff("Tc\t",RunTimeData->Tc, 3);

        for (int i = 0; i<3; i++){
            T[i] = RunTimeData->Tg[i] + RunTimeData->Tc[i];
        }

```

```c
        //printbuff("T\t",T, 3);

        for (int i = 0; i<3; i++){
            tV[i] = T[i] - tV2[i];
        }

        //printbuff("tV _ below\t",tV, 3);

        double det = 0;

        det=Ib[0]*Ib[1 * 3 + 1]*Ib[2 * 3 + 2]+Ib[0 * 3 + 1]*Ib[1 * 3 + 2]*Ib[2 * 3 + 0]+Ib[1 * 3 + 0]*Ib[2 * 3 + 1]*Ib[0 * 3 + 2]-Ib[0 * 3 + 2]*Ib[1 * 3 + 1]*Ib[2 * 3 + 0]-Ib[0 * 3 + 1]*Ib[1 * 3 + 0]*Ib[2 * 3 + 2]-Ib[2 * 3 + 1]*Ib[1 * 3 + 2]*Ib[0];
        /*
        if (fabs(det)<2.2204e-16){
            return NULL;
        }
        */
        tM[0][0] =  (Ib[1 * 3 + 1]*Ib[2 * 3 + 2] - Ib[2 * 3 + 1]*Ib[1 * 3 + 2])/det;
        tM[0][1] = -(Ib[0 * 3 + 1]*Ib[2 * 3 + 2] - Ib[2 * 3 + 1]*Ib[0 * 3 + 2])/det;
        tM[0][2] =  (Ib[0 * 3 + 1]*Ib[1 * 3 + 2] - Ib[1 * 3 + 1]*Ib[0 * 3 + 2])/det;
        tM[1][0] = -(Ib[1 * 3 + 0]*Ib[2 * 3 + 2] - Ib[2 * 3 + 0]*Ib[1 * 3 + 2])/det;
        tM[1][1] =  (Ib[0 * 3 + 0]*Ib[2 * 3 + 2] - Ib[2 * 3 + 0]*Ib[0 * 3 + 2])/det;
        tM[1][2] = -(Ib[0 * 3 + 0]*Ib[1 * 3 + 2] - Ib[1 * 3 + 0]*Ib[0 * 3 + 2])/det;
        tM[2][0] =  (Ib[1 * 3 + 0]*Ib[2 * 3 + 1] - Ib[2 * 3 + 0]*Ib[1 * 3 + 1])/det;
        tM[2][1] = -(Ib[0 * 3 + 0]*Ib[2 * 3 + 1] - Ib[2 * 3 + 0]*Ib[0 * 3 + 1])/det;
        tM[2][2] =  (Ib[0 * 3 + 0]*Ib[1 * 3 + 1] - Ib[0 * 3 + 1]*Ib[1 * 3 + 0])/det;


        tV2[0] = (tM[0][0] * tV[0] + tM[0][1] * tV[1] + tM[0][2] * tV[2]) * 0.01f;
        tV2[1] = (tM[1][0] * tV[0] + tM[1][1] * tV[1] + tM[1][2] * tV[2]) * 0.01f;
        tV2[2] = (tM[2][0] * tV[0] + tM[2][1] * tV[1] + tM[2][2] * tV[2]) * 0.01f;

        //printbuff("tV2 _ below\t",tV2, 3);

        RunTimeData->Omega_B2I[0] = RunTimeData->Omega_B2I_prev[0] + tV2[0];
        RunTimeData->Omega_B2I[1] = RunTimeData->Omega_B2I_prev[1] + tV2[1];
        RunTimeData->Omega_B2I[2] = RunTimeData->Omega_B2I_prev[2] + tV2[2];

        //printbuff("Omega_B2I\t",RunTimeData->Omega_B2I, 3);
```

```c

        RunTimeData->Omega_B2I_prev[0] = RunTimeData->Omega_B2I[0];
        RunTimeData->Omega_B2I_prev[1] = RunTimeData->Omega_B2I[1];
        RunTimeData->Omega_B2I_prev[2] = RunTimeData->Omega_B2I[2];

        double Wo[3] = {0, -sqrt(u/(r*r*r)), 0};

        tV[0] = RunTimeData->DCM_B2O_Prev[0][0] * Wo[0] + RunTimeData->DCM_B2O_Prev[0][1] * Wo[1] + RunTimeData->DCM_B2O_Prev[0][2] * Wo[2];
        tV[1] = RunTimeData->DCM_B2O_Prev[1][0] * Wo[0] + RunTimeData->DCM_B2O_Prev[1][1] * Wo[1] + RunTimeData->DCM_B2O_Prev[1][2] * Wo[2];
        tV[2] = RunTimeData->DCM_B2O_Prev[2][0] * Wo[0] + RunTimeData->DCM_B2O_Prev[2][1] * Wo[1] + RunTimeData->DCM_B2O_Prev[2][2] * Wo[2];

        //printbuff("tV _ below 2\t",tV, 3);

        for (int i = 0; i<3; i++){
            tV2[i] = RunTimeData->Omega_B2I[i] - tV[i];
            RunTimeData->Omega_B2O[i] = tV2[i];
        }

        double Mulmatrix[4][3] = { { RunTimeData->Quat_B2O_prev[3], -RunTimeData->Quat_B2O_prev[2], RunTimeData->Quat_B2O_prev[1] }, { RunTimeData->Quat_B2O_prev[2], RunTimeData->Quat_B2O_prev[3], -RunTimeData->Quat_B2O_prev[0] },{ -RunTimeData->Quat_B2O_prev[1], RunTimeData->Quat_B2O_prev[0], RunTimeData->Quat_B2O_prev[3] },{ -RunTimeData->Quat_B2O_prev[0], -RunTimeData->Quat_B2O_prev[1], -RunTimeData->Quat_B2O_prev[2] } };
        float W[3] = {tV2[0] * 0.5, tV2[1] * 0.5, tV2[2] *0.5};

        for (int i=0;i<4;i++){
            tV3[i]=0.0;
            for(int j=0;j<3;j++){
                tV3[i] = tV3[i]+Mulmatrix[i][j]*W[j];
            }
            tV3[i] = tV3[i] * 0.01f;
        }

        for(int i=0;i<4;i++){
            tV4[i] = tV3[i] + RunTimeData->Quat_B2O_prev[i];
        }

        norm_buff = norm(tV4, 4);

        for(int i=0;i<4;i++){
            RunTimeData->Quat_B2O[i] = tV4[i] * (1/norm_buff) ;
            RunTimeData->Quat_B2O_prev[i] = RunTimeData->Quat_B2O[i];
        }

        RunTimeData->DCM_B2O[0][0] = RunTimeData->Quat_B2O[0] * RunTimeData-
```

```
                >Quat_B2O[0] - RunTimeData->Quat_B2O[1] * RunTimeData->Quat_B2O[1] - RunTimeData->Quat_B2O[2] * RunTimeData->Quat_B2O[2] + RunTimeData->Quat_B2O[3] * RunTimeData->Quat_B2O[3];
            RunTimeData->DCM_B2O[0][1] = 2 * RunTimeData->Quat_B2O[0] * RunTimeData->Quat_B2O[1] + 2 * RunTimeData->Quat_B2O[2] * RunTimeData->Quat_B2O[3];
            RunTimeData->DCM_B2O[0][2] = 2 * RunTimeData->Quat_B2O[0] * RunTimeData->Quat_B2O[2] - 2 * RunTimeData->Quat_B2O[1] * RunTimeData->Quat_B2O[3];
            RunTimeData->DCM_B2O[1][0] = 2 * RunTimeData->Quat_B2O[0] * RunTimeData->Quat_B2O[1] - 2 * RunTimeData->Quat_B2O[2] * RunTimeData->Quat_B2O[3];
            RunTimeData->DCM_B2O[1][1] = -RunTimeData->Quat_B2O[0] * RunTimeData->Quat_B2O[0] + RunTimeData->Quat_B2O[1] * RunTimeData->Quat_B2O[1] - RunTimeData->Quat_B2O[2] * RunTimeData->Quat_B2O[2] + RunTimeData->Quat_B2O[3] * RunTimeData->Quat_B2O[3];
            RunTimeData->DCM_B2O[1][2] = 2 * RunTimeData->Quat_B2O[0] * RunTimeData->Quat_B2O[3] + 2 * RunTimeData->Quat_B2O[1] * RunTimeData->Quat_B2O[2];
            RunTimeData->DCM_B2O[2][0] = 2 * RunTimeData->Quat_B2O[0] * RunTimeData->Quat_B2O[2] + 2 * RunTimeData->Quat_B2O[1] * RunTimeData->Quat_B2O[3];
            RunTimeData->DCM_B2O[2][1] = 2 * RunTimeData->Quat_B2O[1] * RunTimeData->Quat_B2O[2] - 2 * RunTimeData->Quat_B2O[0] * RunTimeData->Quat_B2O[3];
            RunTimeData->DCM_B2O[2][2] = -RunTimeData->Quat_B2O[0] * RunTimeData->Quat_B2O[0] - RunTimeData->Quat_B2O[1] * RunTimeData->Quat_B2O[1] + RunTimeData->Quat_B2O[2] * RunTimeData->Quat_B2O[2] + RunTimeData->Quat_B2O[3] * RunTimeData->Quat_B2O[3];

            for(int i = 0; i < 3; i++){
                for(int j = 0;j < 3;j++){
                    RunTimeData->DCM_B2O_Prev[i][j] = RunTimeData->DCM_B2O[i][j];
                }
            }

        }

        for (int i=0;i<3;i++){
            for(int j=0;j<3;j++){
                tM[i][j] = 0.0;
                for (int k=0;k<3;k++){
                    tM[i][j] = tM[i][j] + RunTimeData->DCM_B2O[i][k]*RunTimeData->DCM_B2I[k][j];
                }
            }
        }

        rotmtx2quat(tM, RunTimeData->Quat_B2I);

        float a, b, c, d;
        a = fabsf(RunTimeData->Quat_B2I[0] - RunTimeData->Quat_B2I_prev[0]);
```

```c
    b = fabsf(RunTimeData->Quat_B2I[1] - RunTimeData->Quat_B2I_prev[1]);
    c = fabsf(RunTimeData->Quat_B2I[2] - RunTimeData->Quat_B2I_prev[2]);
    d = fabsf(RunTimeData->Quat_B2I[3] - RunTimeData->Quat_B2I_prev[3]);

    if ((a > 0.1) || (b > 0.1) || (c > 0.1) || (d > 0.1))
    {
        RunTimeData->Quat_B2I[0] = -RunTimeData->Quat_B2I[0];
        RunTimeData->Quat_B2I[1] = -RunTimeData->Quat_B2I[1];
        RunTimeData->Quat_B2I[2] = -RunTimeData->Quat_B2I[2];
        RunTimeData->Quat_B2I[3] = -RunTimeData->Quat_B2I[3];
    }

    //double Gyroc[3] = {5e-5,5e-5,5e-5};

    for (int i=0;i<3;i++){
        RunTimeData->GK[i] = GRC[RunTimeData->CounterLoop][i] + RunTimeData->Omega_B2I[i];
    }
    /*
    RunTimeData->rsun_Body[0] = RunTimeData->rsun_Inertia[0] * tM[0][0] + RunTimeData->rsun_Inertia[1] * tM[0][1] + RunTimeData->rsun_Inertia[2] * tM[0][2];
    RunTimeData->rsun_Body[1] = RunTimeData->rsun_Inertia[0] * tM[1][0] + RunTimeData->rsun_Inertia[1] * tM[1][1] + RunTimeData->rsun_Inertia[2] * tM[1][2];
    RunTimeData->rsun_Body[2] = RunTimeData->rsun_Inertia[0] * tM[2][0] + RunTimeData->rsun_Inertia[1] * tM[2][1] + RunTimeData->rsun_Inertia[2] * tM[2][2];

    RunTimeData->rmag_Body[0] = RunTimeData->rmag_Inertia[0] * tM[0][0] + RunTimeData->rmag_Inertia[1] * tM[0][1] + RunTimeData->rmag_Inertia[2] * tM[0][2];
    RunTimeData->rmag_Body[1] = RunTimeData->rmag_Inertia[0] * tM[1][0] + RunTimeData->rmag_Inertia[1] * tM[1][1] + RunTimeData->rmag_Inertia[2] * tM[1][2];
    RunTimeData->rmag_Body[2] = RunTimeData->rmag_Inertia[0] * tM[2][0] + RunTimeData->rmag_Inertia[1] * tM[2][1] + RunTimeData->rmag_Inertia[2] * tM[2][2];
    */

    return 0;
}
```

```c
/*
 * File: solver.h
 *
 * MATLAB Coder version            : 4.0
 * C/C++ source code generated on  : 24-Mar-2019 11:42:47
 */

#ifndef SOLVER_H
#define SOLVER_H

/* Include Files */
#include <stddef.h>
#include <stdlib.h>
#include "rtwtypes.h"
#include "solver_types.h"

/* Function Declarations */
extern void solver(double dq1, double dq2, double dq3, double dq4, double q1,
                   double q2, double q3, double q4, double X[3]);

#endif

/*
 * File trailer for solver.h
 *
 * [EOF]
 */
```

```c
/*
 * File: solver.c
 *
 * MATLAB Coder version            : 4.0
 * C/C++ source code generated on  : 24-Mar-2019 11:42:47
 */

/* Include Files */
#include <math.h>
#include <string.h>
#include "rt_nonfinite.h"
#include "solver.h"

/* Function Definitions */

/*
 * Arguments    : double dq1
 *                double dq2
 *                double dq3
 *                double dq4
 *                double q1
 *                double q2
 *                double q3
 *                double q4
 *                double X[3]
 * Return Type  : void
 */
void solver(double dq1, double dq2, double dq3, double dq4, double q1, double q2,
            double q3, double q4, double X[3])
{
  double A[12];
  int p1;
  double x[9];
  double y[9];
  int p2;
  int p3;
  double absx11;
  double absx21;
  double absx31;
  int itmp;
  double b_dq1[4];
  double b_y[12];
  A[0] = q4;
  A[4] = -q3;
  A[8] = q2;
  A[1] = q3;
  A[5] = q4;
  A[9] = -q1;
  A[2] = -q2;
  A[6] = q1;
  A[10] = -q4;
  A[3] = -q1;
```

```c
  A[7] = -q2;
  A[11] = -q3;
  for (p1 = 0; p1 < 3; p1++) {
    for (p2 = 0; p2 < 3; p2++) {
      y[p1 + 3 * p2] = 0.0;
      for (p3 = 0; p3 < 4; p3++) {
        y[p1 + 3 * p2] += A[p3 + (p1 << 2)] * A[p3 + (p2 << 2)];
      }
    }
  }

  memcpy(&x[0], &y[0], 9U * sizeof(double));
  p1 = 0;
  p2 = 3;
  p3 = 6;
  absx11 = fabs(y[0]);
  absx21 = fabs(y[1]);
  absx31 = fabs(y[2]);
  if ((absx21 > absx11) && (absx21 > absx31)) {
    p1 = 3;
    p2 = 0;
    x[0] = y[1];
    x[1] = y[0];
    x[3] = y[4];
    x[4] = y[3];
    x[6] = y[7];
    x[7] = y[6];
  } else {
    if (absx31 > absx11) {
      p1 = 6;
      p3 = 0;
      x[0] = y[2];
      x[2] = y[0];
      x[3] = y[5];
      x[5] = y[3];
      x[6] = y[8];
      x[8] = y[6];
    }
  }

  absx11 = x[1] / x[0];
  x[1] /= x[0];
  absx21 = x[2] / x[0];
  x[2] /= x[0];
  x[4] -= absx11 * x[3];
  x[5] -= absx21 * x[3];
  x[7] -= absx11 * x[6];
  x[8] -= absx21 * x[6];
  if (fabs(x[5]) > fabs(x[4])) {
    itmp = p2;
    p2 = p3;
    p3 = itmp;
    x[1] = absx21;
```

```
    x[2] = absx11;
    absx11 = x[4];
    x[4] = x[5];
    x[5] = absx11;
    absx11 = x[7];
    x[7] = x[8];
    x[8] = absx11;
  }

  absx11 = x[5] / x[4];
  x[5] /= x[4];
  x[8] -= absx11 * x[7];
  absx11 = (x[5] * x[1] - x[2]) / x[8];
  absx21 = -(x[1] + x[7] * absx11) / x[4];
  y[p1] = ((1.0 - x[3] * absx21) - x[6] * absx11) / x[0];
  y[p1 + 1] = absx21;
  y[p1 + 2] = absx11;
  absx11 = -x[5] / x[8];
  absx21 = (1.0 - x[7] * absx11) / x[4];
  y[p2] = -(x[3] * absx21 + x[6] * absx11) / x[0];
  y[p2 + 1] = absx21;
  y[p2 + 2] = absx11;
  absx11 = 1.0 / x[8];
  absx21 = -x[7] * absx11 / x[4];
  y[p3] = -(x[3] * absx21 + x[6] * absx11) / x[0];
  y[p3 + 1] = absx21;
  y[p3 + 2] = absx11;
  b_dq1[0] = dq1;
  b_dq1[1] = dq2;
  b_dq1[2] = dq3;
  b_dq1[3] = dq4;
  for (p1 = 0; p1 < 3; p1++) {
    X[p1] = 0.0;
    for (p2 = 0; p2 < 4; p2++) {
      b_y[p1 + 3 * p2] = 0.0;
      for (p3 = 0; p3 < 3; p3++) {
        b_y[p1 + 3 * p2] += y[p1 + 3 * p3] * A[p2 + (p3 << 2)];
      }

      X[p1] += b_y[p1 + 3 * p2] * b_dq1[p2];
    }
  }
}

/*
 * File trailer for solver.c
 *
 * [EOF]
 */
```